\pdfoutput=1 
\documentclass[10pt]{article}
\usepackage[numbers, sort&compress]{natbib}
\usepackage{geometry}
\usepackage{amsmath}
\usepackage{amssymb}
\usepackage{multirow}
\usepackage{wrapfig}
\usepackage{url}
\usepackage{xcolor}
\usepackage{enumitem}
\usepackage{capt-of}

\newcommand{\eg}{{\it e.g.}, }
\newcommand{\ie}{{\it i.e.}, }

\newcommand{\topk}{\textsc{Top-$K$}}
\newcommand{\mstopk}{\textsc{MSTop-$K$}}
\newcommand{\cifart}{\textsc{Cifar-10}}

\usepackage[utf8]{inputenc}
\usepackage[T1]{fontenc}
\usepackage[colorlinks = true,urlcolor=blue, citecolor=black, linkcolor=blue]{hyperref}
\usepackage[noabbrev]{cleveref}
\usepackage{booktabs}
\usepackage{amsfonts}
\usepackage{nicefrac}
\usepackage{microtype}
\usepackage{balance}

\usepackage{subcaption}
\usepackage{graphicx}
\usepackage{algorithm}
\usepackage{algorithmic}
\usepackage{textcomp}
\usepackage{xfrac}
\usepackage{xspace}

\usepackage{bigstrut}
\newsavebox{\algleft}
\newsavebox{\algright}
\usepackage{pifont}

\newcommand{\signsgd}{\textsc{sign}SGD}
\newcommand{\powersgd}{\textsc{PowerSGD}}
\newcommand{\cmark}{\ding{51}}%
\newcommand{\xmark}{\ding{55}}%

\usepackage{amsthm}

\newtheorem{remark}{Takeaway}

\title{On the Utility of Gradient Compression in Distributed Training Systems}

\author{Saurabh Agarwal, Hongyi Wang, Shivaram Venkataraman, Dimitris Papailiopoulos
\\University of Wisconsin-Madison}
\date{}
\begin{document}
\maketitle

\begin{abstract}

A rich body of prior work has highlighted the existence of communication bottlenecks in synchronous data-parallel training. To alleviate these bottlenecks, a long line of recent work proposes gradient and model compression methods.  In this work, we evaluate the efficacy of gradient compression methods and compare their scalability with optimized implementations of synchronous data-parallel SGD across more than 200 different setups. Surprisingly, we observe that only in 6 cases out of more than 200,  gradient compression methods provide speedup over optimized synchronous data-parallel training in the typical data-center setting. We conduct an extensive investigation to identify the root causes of this phenomenon, and offer a performance model that can be used to identify the benefits of gradient compression for a variety of system setups. Based on our analysis, we propose a list of desirable properties that gradient compression methods should satisfy, in order for them to provide a meaningful end-to-end speedup.

\end{abstract}

\section{Introduction}

Synchronous data parallel training using stochastic gradient descent (SGD) is one of the most widely adopted approaches for distributed learning~\cite{li2020pytorch,narayanan2019pipedream, dean2012comm}.
One iteration of distributed data parallel SGD comprises two main phases: gradient computation and gradient aggregation. During the computation phase the gradient of the model is typically computed  using backpropagation. This is followed by an aggregation phase, where gradients are synchronously averaged among all participating nodes~\cite{iandola2016firecaffe, goyal2017accurate}. During this second phase, for state-of-the-art neural network models, millions to billions of parameters are communicated among nodes~\cite{gpt-3}, which has been shown to lead to communication bottlenecks~\cite{dean2012large, seide20141, qi17paleo, grubic2018synchronous, alistarh2017qsgd}. 

Alleviating communication bottlenecks in distributed training has been an active area of research in recent years. 
A long line of work has focused on lossy gradient compression methods to mitigate communication costs. Lossy gradient compression methods typically use techniques such as low-precision training~\cite{seide20141, alistarh2017qsgd, bernstein2018signsgd, wen2017terngrad}, sparsification~\cite{aji2017sparse, lin2017deep}, or low-rank updates~\cite{wang2018atomo, vogels2019powersgd}, with the common goal of reduced communication. Although these methods require significant effort to integrate in deep learning frameworks and often introduce extra hyper-parameters, they promise significant reduction in communication, \eg \powersgd~\cite{vogels2019powersgd} provides greater than $100\times$ reduction in communication with minimal effect on accuracy on certain tasks.

\begin{figure*}[t]
    \begin{center}
    \begin{minipage}[t]{0.57\textwidth}
    \includegraphics[width=\linewidth]{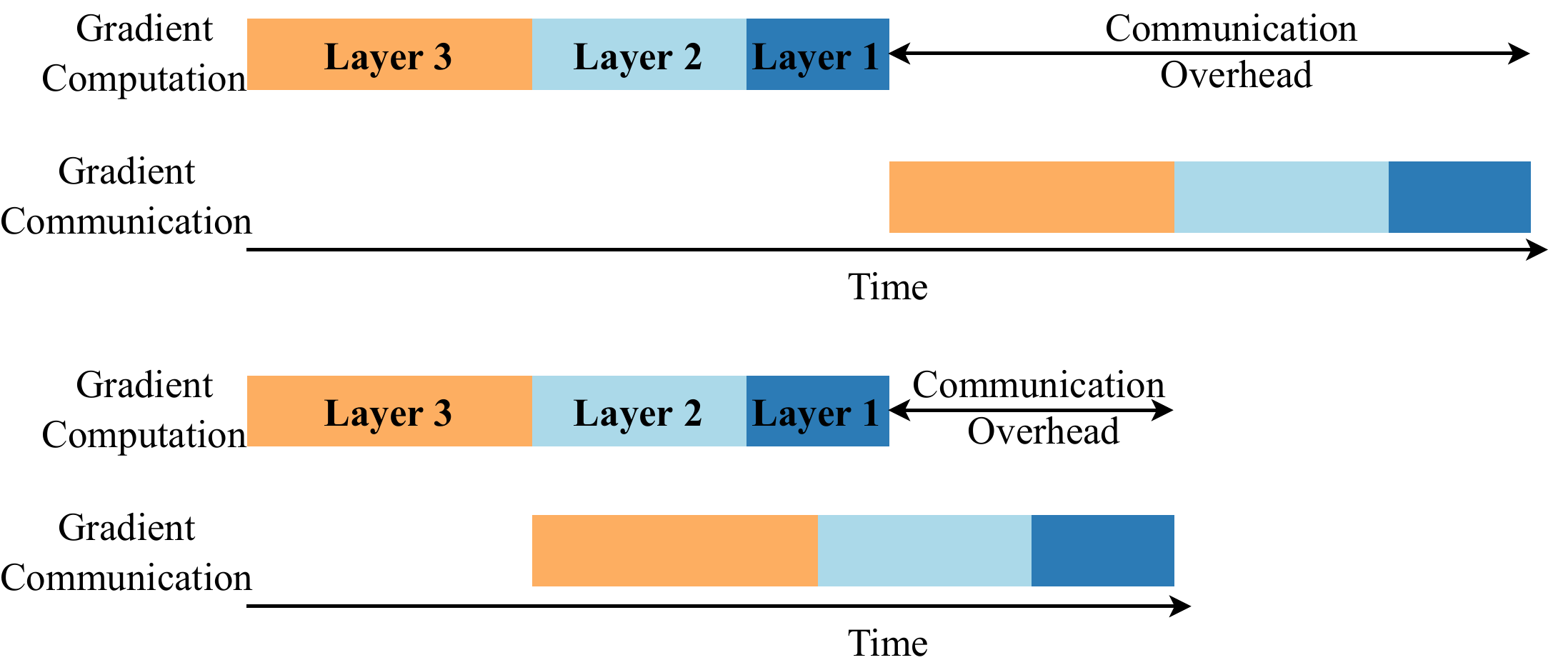}
    \caption{\small{\textbf{Illustration of how overlapping can reduce the total iteration time.} (Above) Gradient computation and communication done serially. (Below) Gradient computation and communication being overlapped, \ie when the gradient of a layer is computed, it is communicated right after the gradient of the previous layer.}}
      \label{fig:overlap}
    \end{minipage}\quad
    \begin{minipage}[t]{0.4\textwidth}
    \includegraphics[width=\linewidth]{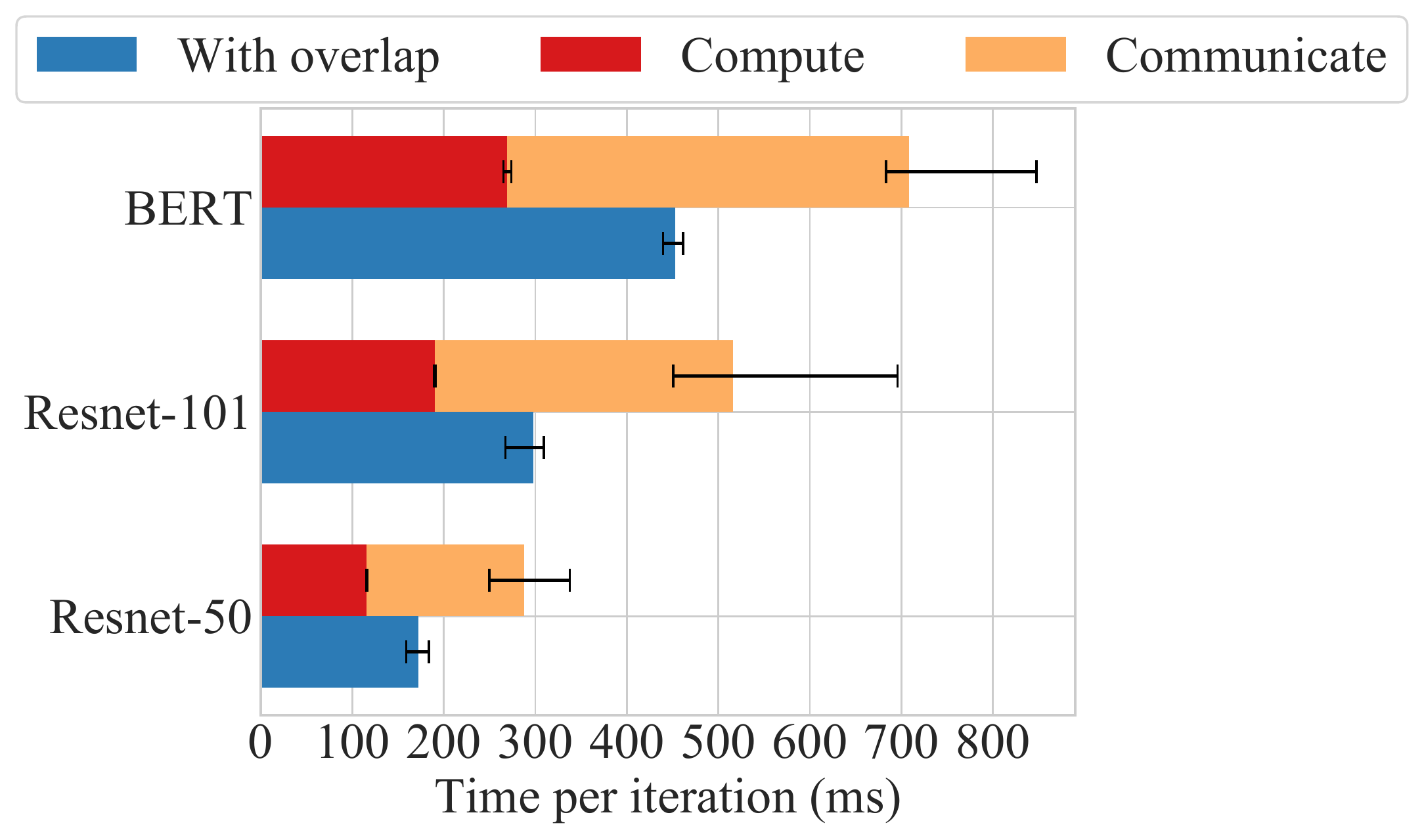}
    \caption{\small{\textbf{Effect of Overlap:} We plot the time for backward computation and gradient synchronization for 64 GPUs, both with and without overlap. In case of Resnet-50 overlapping reduces iteration time by upto 46\%. 
    }}
     \label{fig:serial_overlap}
    \end{minipage}
\end{center}
\vspace{-20pt}
\end{figure*}  
Concurrent to the work on gradient compression,
 a number of systems optimizations have been proposed to speed up distributed data-parallel synchronous SGD (syncSGD). Techniques like ring-reduce~\cite{thakur2005optimization} and tree-reduce~\cite{sanders2009two} have been implemented in several high performance communication libraries (\eg NCCL and Gloo) which in turn are tightly integrated in popular deep learning libraries like PyTorch~\cite{paszke2019pytorch, li2020pytorch} and Tensorflow~\cite{abadi2016tensorflow}. Both ring- and tree-reduce, are bandwidth efficient and have a constant, and logarithmic dependence on the number of nodes, respectively, \ie the number of bytes communicated remains sublinear in the number of machines used for training. To further reduce the observed overhead of communication, systems implement overlapping between the gradient computation and communication phases~\cite{li2020pytorch, sergeev2018horovod}. Figure~\ref{fig:overlap} shows an illustration of how overlapping the backward pass and the communication phases is implemented and Figure~\ref{fig:serial_overlap} shows the extent to which overlapping helps
 improves the scalability of distributed training. For PyTorch DPP~\cite{li2020pytorch} with ResNet-50, we observe almost 46\% reduction in time when overlapping communication with backward pass. These system optimizations are transparent to the user, \ie there is no requirement for additional hyper-parameters and the user need not worry about accuracy degradation. 

Given the above trends, our objective in this work is to measure  the utility of gradient compression in distributed training. We empirically compare three popular gradient compression methods against of-the-shelf implementations of syncSGD using new functionality~\cite{pytorchcommhooks} provided in Pytorch v1.8 for efficiently integrating gradient compression methods. 
The compression methods we compare against are, \signsgd{}~\cite{bernstein2018signsgd,bernstein2018signsgdtol}, \mstopk{}~\cite{shi2021towards} and \powersgd{}~\cite{vogels2019powersgd}. We test these methods on three popular models, \ie ResNet-50, ResNet-101 and BERT~\cite{devlin2018bert}, and conduct a large scale evaluation with up to 96 GPUs. Overall, we test across more than 200 experimental settings accounting for all different models, compression algorithms, compression ratios, batch sizes, network bandwidths, etc. 
\vspace{-2mm}
\paragraph{Our Contributions:} We observe that due to aforementioned systems optimizations to speedup syncSGD, at typical data-center bandwidths, there is \emph{limited opportunity for gradient compression} to provide speedup. We observe that even for communication heavy models like $\text{BERT}_{\text{BASE}}$~\cite{devlin2018bert}, the difference between linear scaling and observed per iteration time for off-the-shelf (PyTorch DDP) implementation of syncSGD is approximately 200 milliseconds when using 96 GPUs. If gradient compression methods have to provide speedups then they need to perform encode-decode and communication within this limited time-frame. However, existing gradient compression methods have high encode-decode times (upwards of 50 milliseconds), which significantly limits the ability of gradient compression to provide significant speedups. 

Further, we observe that gradient compression methods cannot fully utilize system optimizations like overlapping compression and backward pass. This is because both gradient compression and backward pass are compute intensive and compete for GPU resources leading to an overall slowdown.  We also show the extent to which large batch sizes further reduce the benefits of gradient compression.
Large batch sizes increase the time spent in computation providing more opportunity to ``hide'' communication overheads.

Finally, we also observe that, as reported by previous works~\cite{vogels2019powersgd, cho2019gradzip}, gradient compression methods that are compatible with the {\it all-reduce} collective scale better. For instance, we observe that \signsgd{}, which is not compatible with {\it all-reduce}, takes 1042ms for a single iteration of ResNet-101 on 96 GPUs, while \powersgd{}, which is compatible with {\it all-reduce}, takes only 470ms, while syncSGD which is also compatible with {\it all-reduce} takes only 262ms.   

To understand the regimes in which gradient compression can be helpful, we develop an analytical performance model and verify its accuracy. 

\begin{figure}[t]
\begin{minipage}{0.48\textwidth}
\captionof{table}{\small{\textbf{Comparing aggregation schemes:} We show how latency and bandwidth term scale for different aggregation strategies. $\alpha$ is the latency, $\beta$ is the inverse of bandwidth, and $n$ is the size of vector communicated. $p$ is the number of machines}}
\label{tab:latency_bandwidth}
\resizebox{\linewidth}{!}{
\begin{tabular}{@{}lll@{}}
\toprule
Algorithm          & Latency         & Bandwidth               \\ \midrule
Ring Reduce        & $2(p-1)\alpha$  & $2\beta \frac{(p-1)}{p}n$ \\ 
Tree Reduce & $2\alpha \log p$ & $ 2\beta (\log p) n$       \\ \midrule
Parameter Server   & $2\alpha$             & $2 \beta (p-1) n$             \\ \bottomrule
\end{tabular}}
\end{minipage}\quad
  \begin{minipage}[t]{0.5\textwidth}
    \centering
     \includegraphics[width=0.9\linewidth]{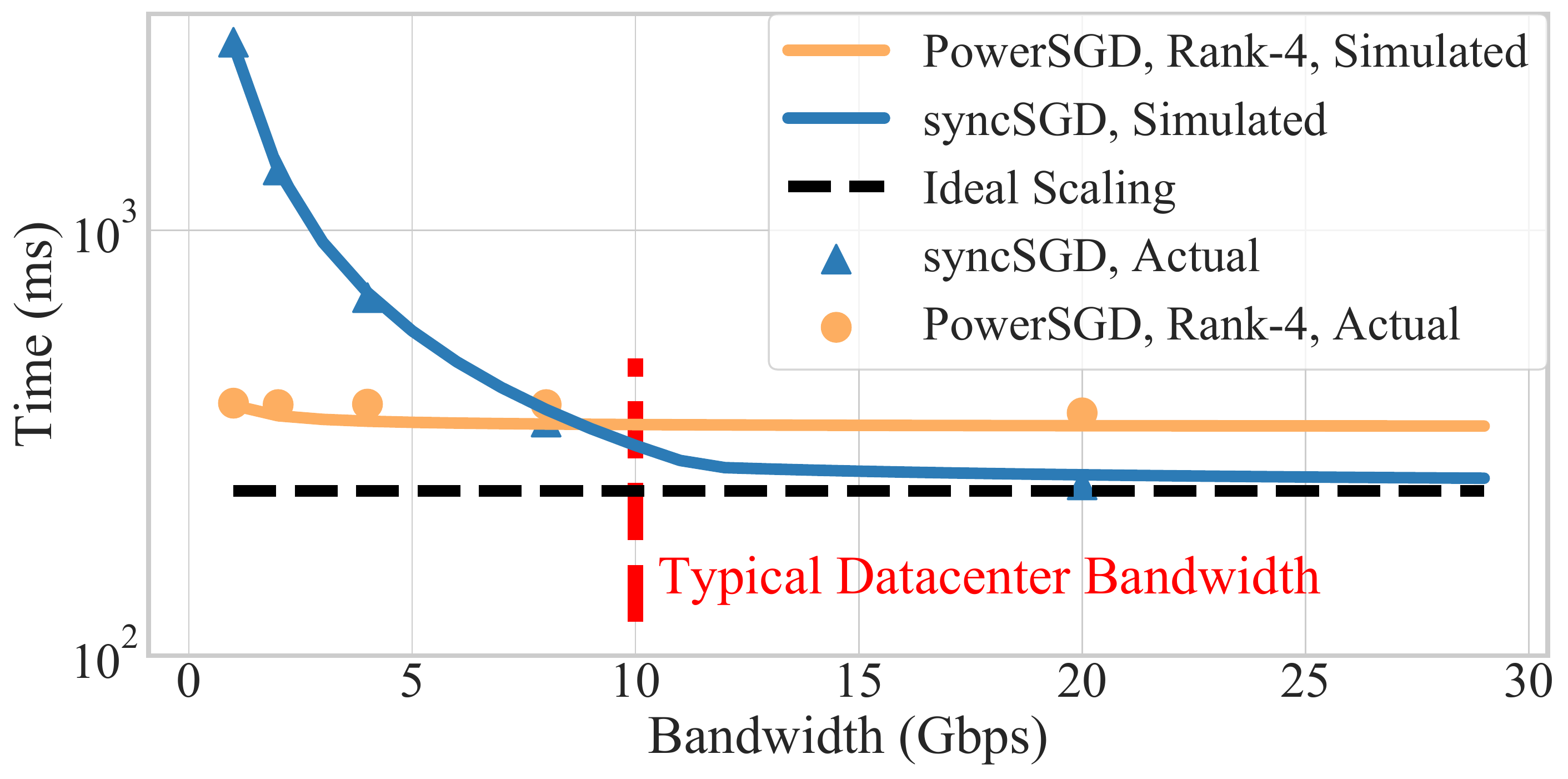}
    \caption{\small{\textbf{Evaluating effect of network bandwidth (simulated):} Above curve is for Resnet-101, batch size 64 on 64 GPUs. We observe that at bandwidth lower than 8.2 Gbps, PowerSGD Rank-4 can provide speedups but above that syncSGD performs better.}}
    \label{fig:network_resnet101_bsize64_main}
    \end{minipage}
    \vspace{-5pt}
\end{figure}

Using the performance model we investigate how various factors like network bandwidth and compute availability affect the scalability of distributed training and discuss scenarios where gradient compression can be effective. For \eg in Figure~\ref{fig:network_resnet101_bsize64_main} with the aid of our performance model we show that at lower bandwidths gradient compression can provide significant benefits. The markers in Figure~\ref{fig:network_resnet101_bsize64_main} are measurements on actual hardware, showing how close our performance model tracks the observed values in actual experiments.  From the performance model we find, that the focus of compression algorithm designers should be on reducing the overhead of encoding rather than trying to achieve high compression ratios. This is because at typical data-center bandwidths ($>$ 10Gbps) we only need a compression ratio of $4\times$ or less even for large models like ResNet-101 and $\text{BERT}_{\text{BASE}}$ to achieve linear scalability. However we find that in other settings where bandwidth is scant, \eg wide area learning~\cite{wang2021network}, existing gradient compression algorithms can provide meaningful speedups.

We would like to point out that our results are derived from analyzing per-iteration times and do not account for any loss in accuracy incurred by gradient compression. In that sense our analysis is {\it generous}  to gradient compression methods, as many lead to some small accuracy loss. This loss typically requires a larger number of iterations to overcome, or  mitigation techniques with additional computation or memory footprint (\eg the error feedback scheme~\cite{karimireddy2019error, stich2018sparsified}).

In summary, our analysis establishes that in a datacenter setting, for popular models, gradient compression methods do not provide promised speedups once we account for system level optimizations in syncSGD. To identify regimes where gradient compression can provide benefits, we develop a performance model that can be used by both practitioners and researchers to predict performance at large scale without the need of performing any real experiments. Based on our empirical analysis and performance model we provide guidelines for building future gradient compression algorithms.
\section{Background and Related Work}
We first provide a brief background of several different threads of prior work that aim at enabling faster distributed machine learning.    
\label{sec:related_work}
\vspace{-1mm}
\subsection{Gradient Compression}
\label{subsec:background_gradient}
Several lossy gradient compression methods based on quantization~\cite{alistarh2017qsgd, bernstein2018signsgd, karimireddy2019error, dettmers20158, seide20141, wen2017terngrad, bernstein2018signsgdtol, yu2019exploring, li2018network, horvath2019natural, tang2019doublesqueeze, dryden2016communication, strom2015scalable, gandikota2021vqsgd, shuai2019comm, zhang2017zipml, wu2018error, tang2018communication}, sparsification~\cite{ stich2018sparsified, lin2018deep, aji2017sparse, alistarh2018convergence, lin2018deep, shi2019understanding, shi2019distributed, fang2019redsync, m2021efficient, shi2021towards, wangni2018gradient, tang2018communication, sattler2019robust, sattler2019sparse}, low rank decomposition~\cite{ wang2018atomo, vogels2019powersgd, wang2021pufferfish}, and other approaches~\cite{acharya2019distributed, suresh2017distributed,ivkin2019communication} have been proposed in literature. 
Recent surveys~\cite{10754/662495, tang2020communication} describe these methods in detail.

In this work we study three popular gradient compression schemes, quantization based \signsgd{}~\cite{bernstein2018signsgd,bernstein2018signsgdtol}, low-rank decomposition based \powersgd{}~\cite{vogels2019powersgd} and sparsification based \mstopk{}~\cite{shi2021towards}. We compare these schemes and evaluate if these schemes provide any benefit over optimized implementations of syncSGD~\cite{li2020pytorch}.

\vspace{-1mm}
\subsection{System Advances}
\label{sec:system_advances}

Next, we provide a brief overview of several system advances which have been applied to syncSGD to improve the performance of distributed training.


\paragraph{All-reduce.}
In recent years, systems have shifted from using a parameter server based topology to an all-reduce topology for gradient aggregation. For example, we observe that all submissions to DawnBench~\cite{coleman2019analysis} use all-reduce for performing distributed training. 

Communication costs are typically modeled using a cost model~\cite{sarvotham2001connection} where cost of sending/receiving a vector of size $n$ is computed as the sum of latency and bandwidth requirements. There are several optimizations~\cite{rabenseifner2004optimization, thakur2005optimization,hoefler2011performance, sanders2009two} for all-reduce based collectives like ring-reduce~\cite{barnett1994interprocessor}, tree-reduce~\cite{sanders2009two}, recursive doubling~\cite{yuichiro2019exhaustive}, 2D-Torus ~\cite{mikami2018massively, jouppi2017datacenter} etc. These optimizations explore the trade-off between the latency and bandwidth terms. We list latency and bandwidth terms for a few aggregation strategies in Table~\ref{tab:latency_bandwidth} for synchronizing a vector of size $n$ among $p$ machines. In Table~\ref{tab:latency_bandwidth}, $\alpha$ represents latency (typically between 0.5 to 1ms in public clouds) and $\beta$ represents bandwidth. We would like to point out that the bandwidth requirement for ring reduce stays almost constant even with an increase in the number of machines $p$.

High performance implementations like Nvidia-NCCL~\cite{nccl} dynamically chooses between tree and ring reduce based on several factors like number of machines, bandwidth, interconnect, communication size to list a few.  In this work for simplicity, we analyze our results with the communication model of ring-reduce.\looseness=-1

\vspace{-1mm}
\paragraph{Communication and Computation Overlap.}
\label{sec:comm_comp_over}
 Gradients for DNN's are calculated layerwise,  therefore, gradients of later layers are available before initial layers. Instead of waiting for the availability of all the gradients, popular deep learning frameworks~\cite{li2020pytorch, paszke2019pytorch, abadi2016tensorflow} start gradient communication when some of the gradients are available. This leads to overlapping gradient computation with communication, hiding the time spent in communication. Figure~\ref{fig:overlap} illustrates how overlap can provide speedups.

In Figure~\ref{fig:serial_overlap}, we observe that overlapping can provide speedups of almost 46\% for Resnet-50.

\vspace{-1mm}
\paragraph{Bucketing Gradients.}
\label{sec:bucket}
Calling the all-reduce collective per layer can often lead to large overheads. To amortize the overheads of calling allreduce optimized implementation of syncSGD~\cite{li2020pytorch, sergeev2018horovod} create fixed size buckets. Once the gradients for a bucket are calculated then {\it all-reduce} is called on the entire bucket. Bucket sizes are typically large (25 MB by default in PyTorch). 

In this paper, we benchmark the runtime of the systems with the aforementioned optimizations to compare against gradient compression methods on real-world computer vision and natural language processing tasks.

\vspace{-1mm}
\subsection{Other Related Work}
\label{sec:related}
Several works have looked at using Gossip based protocols~\cite{lian2017can, tang2018d, pmlr-v97-koloskova19a, koloskova2019decentralized} to improve communication efficiency. Other methods have looked into improving efficiency of distributed training by enabling use of large batch sizes~\cite{you2019large, you2017scaling, smith2017don, devarakonda2017adabatch} or lower precision~\cite{micikevicius2017mixed} without accuracy loss. Other works have also looked at different forms of parallelism~\cite{narayanan2019pipedream, jia2018beyond, jia2018exploring, rasley2020deepspeed} for speeding up distributed training. MLperf~\cite{mattson2019mlperf} and DawnBench~\cite{coleman2019analysis} are two well known industry supported efforts to perform periodic benchmarking on training and inference speed at scale. 

Our findings about scalability of all-reduce based compression scheme has also been reported by prior works~\cite{vogels2019powersgd, cho2019gradzip}. A recent survey~\cite{10754/662495} quantitatively compares several gradient compression methods. However unlike our work it does not account for systems optimization like overlap of communication and computation.  In ~\cite{zhang2020network} authors study whether network is the bottleneck in distributed training. Unlike~\cite{zhang2020network} and other listed works,  our study focuses on the utility of gradient compression methods in several different settings and analyzes others aspects like compute availability, batch size, model size, system advances etc. beyond just focusing on network bandwidth. Further, our performance model allows practitioners to reason about performance of distributed training and predict expected performance gains without running
large scale experiments.

\section{ Evaluating Gradient Compression Schemes}
\label{sec:eval_grad_compression}
In this section we perform a detailed experimental evaluation comparing the scalability of gradient compression with an optimized syncSGD implementation.  We start by analyzing the effects of overlapping gradient compression techniques with gradient computation. Next we run large scale experiments to study how  gradient compression methods scale across a range of models. 
\vspace{-1.5mm}
\paragraph{Methodology.} We choose three popular gradient compression schemes to compare with syncSGD, \signsgd{}~\cite{bernstein2018signsgd,bernstein2018signsgdtol} which only communicates the sign of the gradient providing $32\times$ compression, \mstopk{}~\cite{shi2021towards} an extremely scalable \topk method and \powersgd{}, a low overhead method with compression ratios of around $100\times$. For syncSGD we use Pytorch-DDP module~\cite{li2020pytorch}.

We would like to point out that we use optimistic compression ratios; \eg for \powersgd{} we use Rank-4, 8 and 16. Such high compression ratios have been shown to work~\cite{vogels2019powersgd} for small datasets like \cifart\ and \textsc{WikiText-2} but can lead to accuracy loss for large datasets~\cite{vogels2019powersgd, ramesh2021zeroshot}.  While for \mstopk{} we are again being optimistic and consider dropping 99.9\% gradients and assuming no loss in accuracy. We chose these since we wanted to consider a best case scenario 
for gradient compression methods when used on large datasets. 

We use ResNet-50 (97MB), ResNet-101(170MB) and $\text{BERT}_{\text{BASE}}$(418MB) as the models to study given their very different sizes. For all our timing measurements on vision models we used  the ImageNet dataset~\cite{deng2009imagenet} and we fine-tune the $\text{BERT}_{\text{BASE}}$ model on Sogou News dataset~\cite{sun2019fine}. For the timing measurements we run $60$ iterations for each setup and discard the first 10. We plot the mean of the remaining 50. The error bars in the figure correspond to minimum and maximum values.

For experiments we use \textit{p3.8xlarge} instances on Amazon EC2. Each instance is equipped with 4 V100 GPUs and provides around 10Gpbs of bandwidth. We scale our experiments up to 96 GPUs (24 \textit{p3.8xlarge} instances) and consider weak scaling, \ie the number of inputs per worker is kept constant as the number of workers increase. This is a commonly used scenario for evaluating the scalability of deep learning training~\cite{coleman2019analysis,narayanan2019pipedream}.
 Thus, when we refer to a particular batch size, it is the batch size at each worker.



\subsection{Overlapping Compression and Computation}
\label{sec:overlap_failure}
\begin{wrapfigure}[19]{r}{0.4\textwidth}
    \centering
    \includegraphics[width=\linewidth]{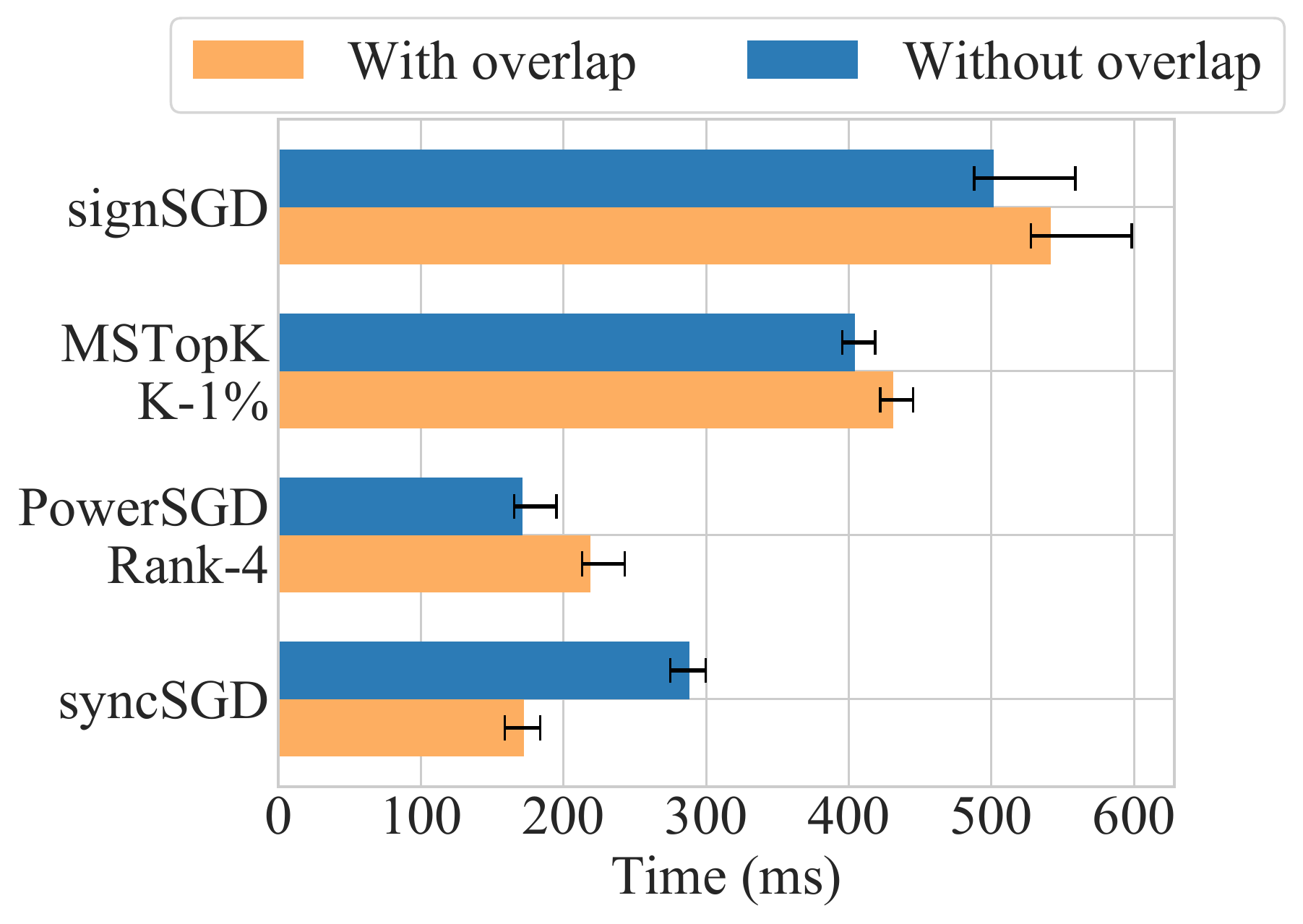}
    \caption{\small{\textbf{Overlapping Gradient Compression with Computation:} Overlapping compression leads to requiring more time per iteration than performing it sequentially, due to resource contention for compute resources. The results are for 64 GPUs. }}
    \label{fig:effect_overlap}
\end{wrapfigure}
We integrate \mstopk{} and \signsgd{}  gradient compression methods to overlap compression with backward pass using the new distributed communication hooks functionality provided in Pytorch v1.8~\cite{pytorchcommhooks}. \powersgd{} is already integrated in PyTorch with overlap~\cite{pytorchpsgd}.

We observe that when gradient compression is performed in parallel with backward computation it is slower than 
performing gradient compression after completing backward pass. Figure~\ref{fig:effect_overlap} depicts this phenomenon on ResNet-50 using PowerSGD Rank-4, \mstopk{}-1\%, and \signsgd{}. 
Since both gradient compression and gradient computation are compute-heavy steps, when performed in parallel  they end up competing for compute resources on the GPU leading to an overall slow down. 
On the other hand, syncSGD only performs {\it all-reduce} operation which is communication heavy with very little compute, thus efficiently utilizing the communication resources on the GPU without affecting the backward pass. Since we consistently observe that compression schemes perform better when not overlapped, for the next set of experiments we use {\bf non-overlapped versions of compression}. For more results with compression overlapped, we refer the reader to Appendix~\ref{app:extra_overlap_results}.
In summary we find:
\begin{remark}
Gradient Compression methods are poor candidates for overlap with gradient computations since both gradient compression and gradient computation are compute heavy processes leading to an overall slowdown.
\end{remark}

\subsection{Comparing Gradient Compression with Optimized syncSGD}


 \begin{figure*}[t]
    \begin{center}
    \includegraphics[width=0.8\textwidth]{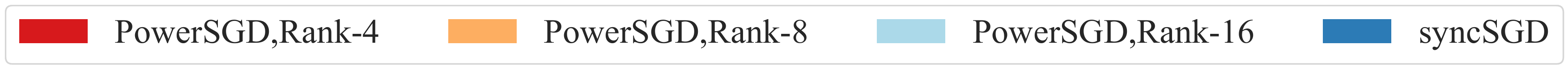}\\
    \vspace{-2pt}
    \begin{subfigure}[b]{0.32\textwidth}
    \includegraphics[width=\textwidth]{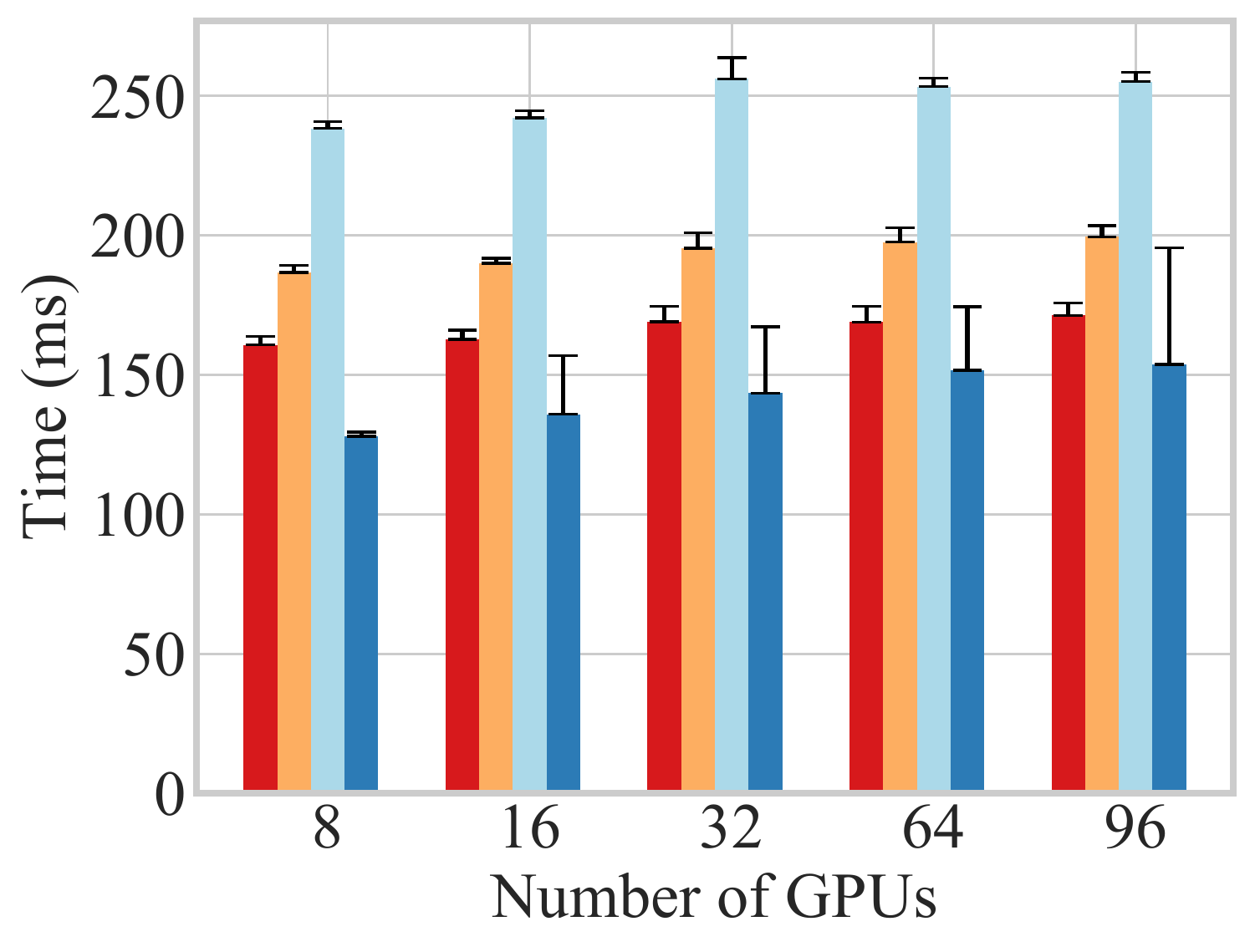}
        \vspace{-16pt}

    \caption{ResNet 50: Batch Size 64}
    \label{fig:psgd_resnet50_bsize64}
    \end{subfigure}
    \begin{subfigure}[b]{0.32\textwidth}
        \includegraphics[width=\textwidth]{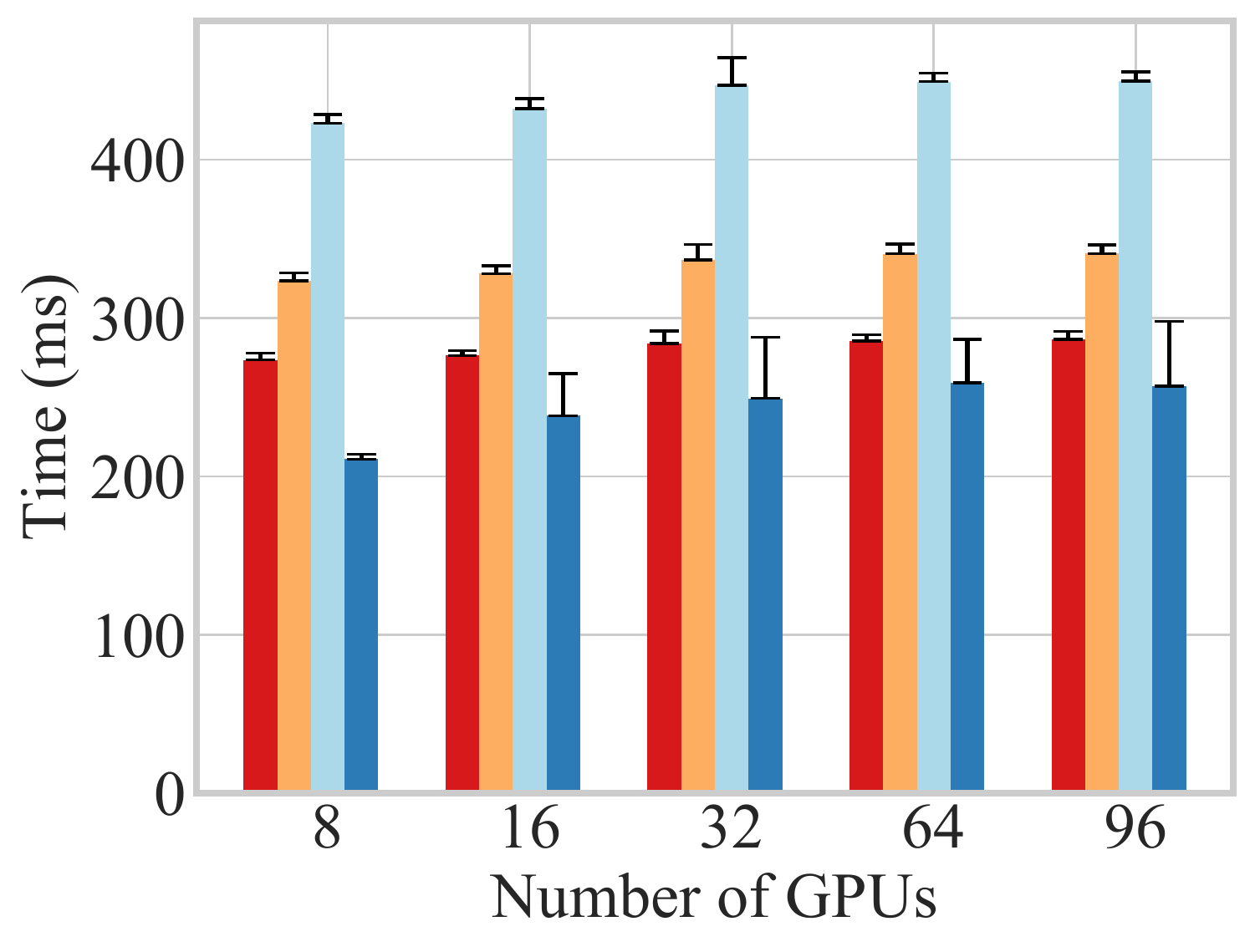}
            \vspace{-16pt}

    \caption{ResNet 101: Batch Size 64}
    \label{fig:psgd_resnet101_bsize64}
    \end{subfigure}
    \begin{subfigure}[b]{0.32\textwidth}
    \includegraphics[width=\textwidth]{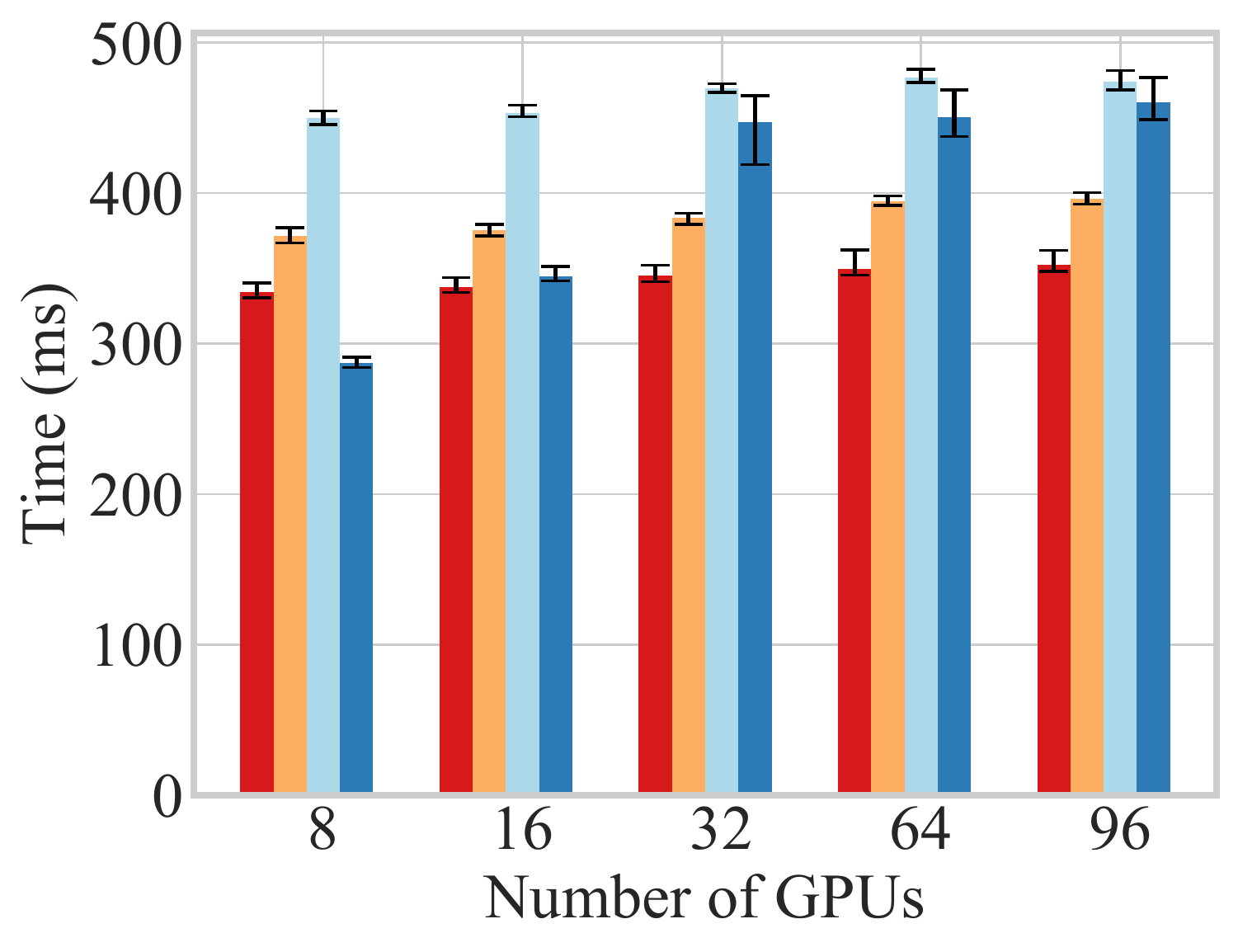}
            \vspace{-16pt}
    \caption{BERT: Batch Size 12}
    \label{fig:psgd_bert_bsize12}
    \end{subfigure}
    \end{center}
    \vspace{-10pt}
    \caption{\small{\textbf{Scalability of \powersgd:} When compared against an optimized implementation of syncSGD, \powersgd{} provides speedups only in case of $\text{BERT}_{\text{BASE}}$ when using Rank-4 and Rank-8 above 32 GPUs. In other cases it has a high per iteration time.}}
    \label{fig:psgd_all_models}
    \vspace{-0.1in}
\end{figure*}

\begin{figure*}[t]
    \vspace{-7pt}
    \begin{center}
    \includegraphics[width=0.5\textwidth]{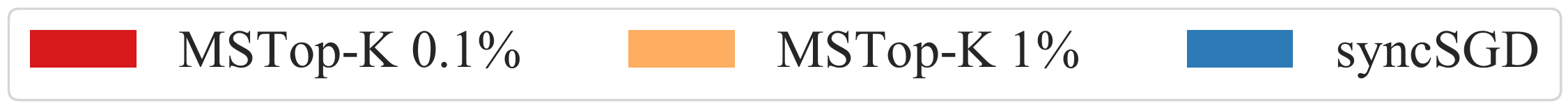}\\
    \vspace{-2pt}
    \begin{subfigure}[b]{0.32\textwidth}
    \includegraphics[width=\linewidth]{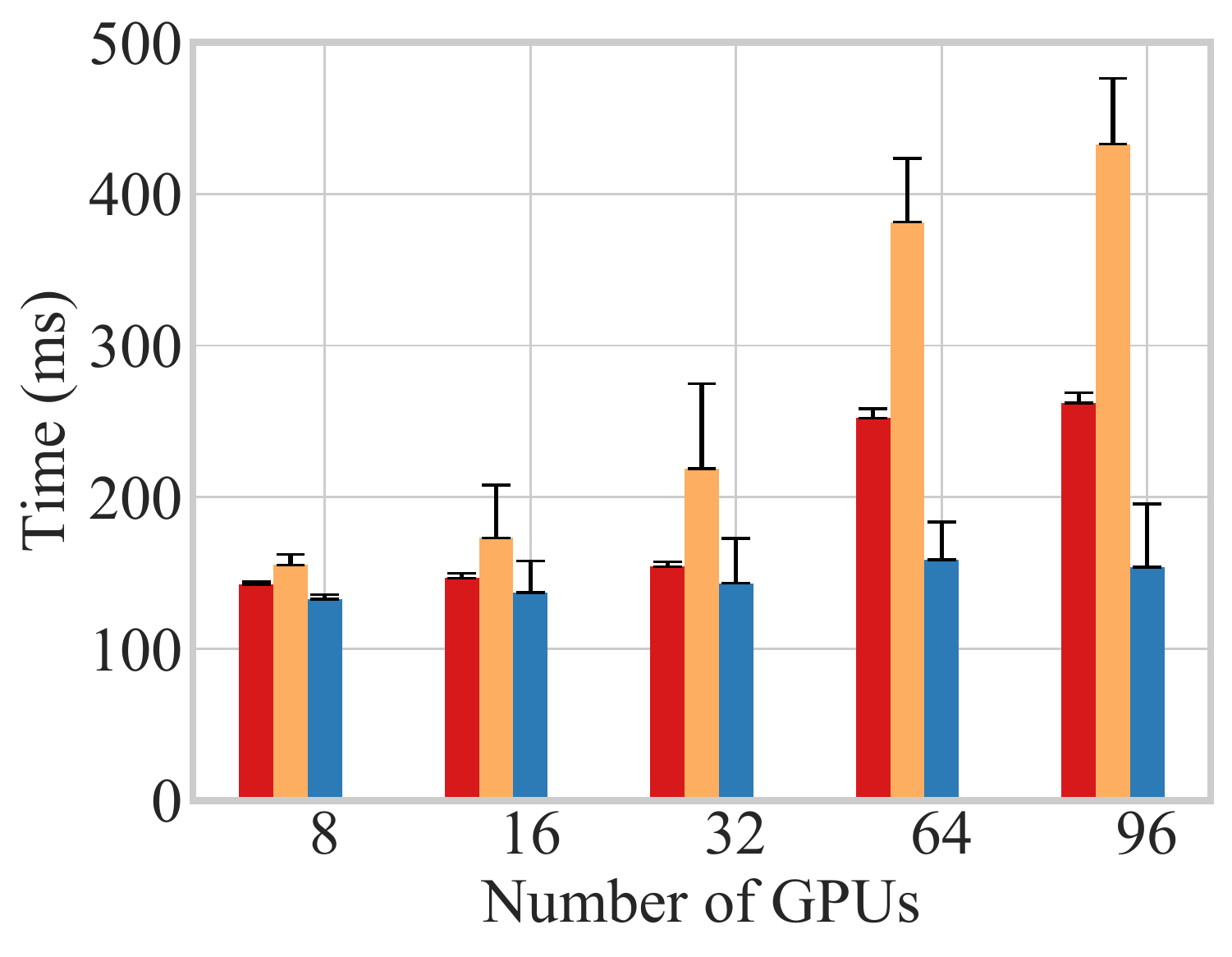}
        \vspace{-16pt}
    \caption{ResNet50: Batch Size 64}
    \end{subfigure}
    \begin{subfigure}[b]{0.32\textwidth}
    \includegraphics[width=\linewidth]{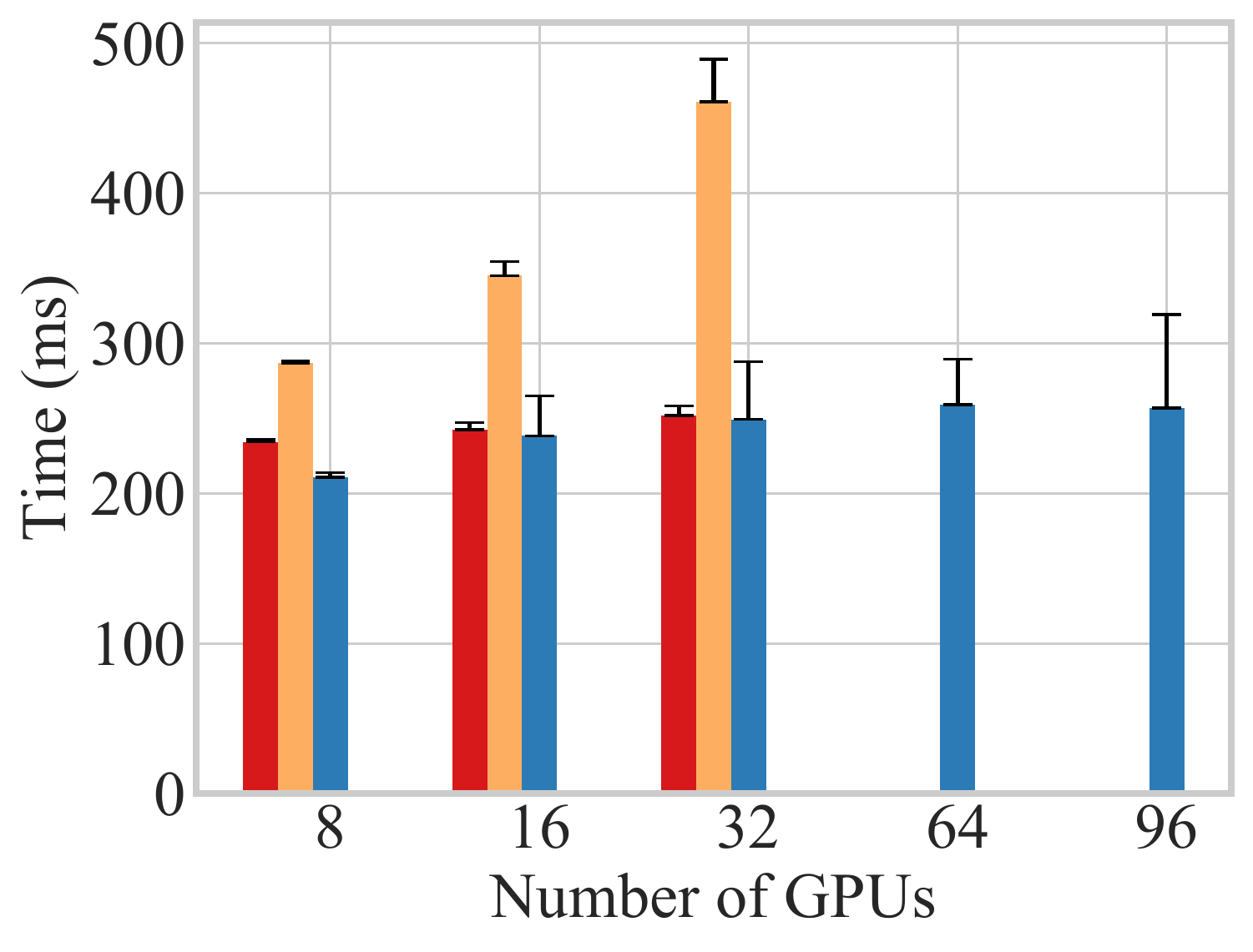}
        \vspace{-16pt}
    \caption{ResNet 101: Batch Size 64}
    \end{subfigure}
    \begin{subfigure}[b]{0.32\textwidth}
    \includegraphics[width=\linewidth]{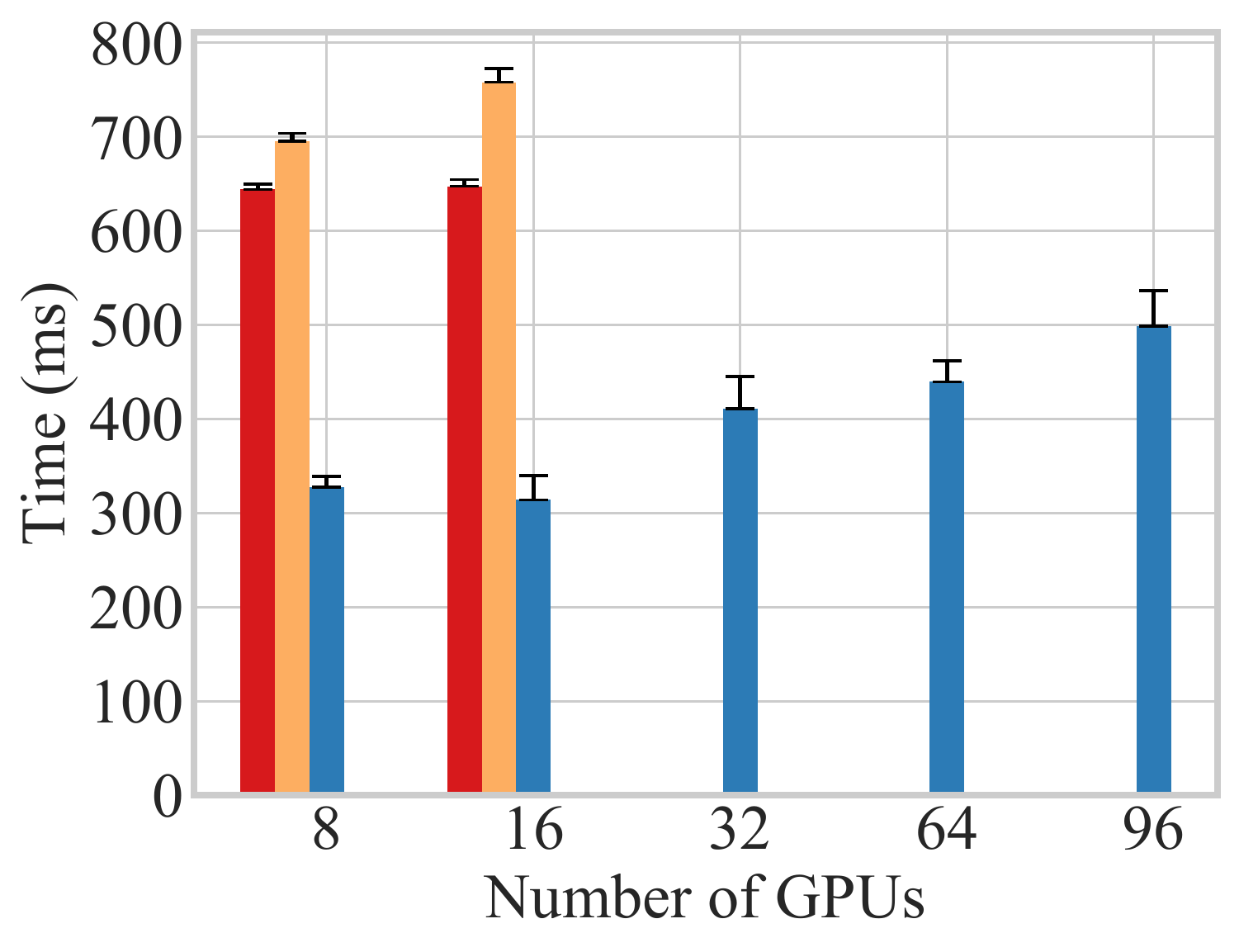}
        \vspace{-16pt}
    \caption{BERT: Batch Size 12}
    \end{subfigure}
    \end{center}
    \vspace{-10pt}
    \caption{\small{\textbf{Scalability of \mstopk{}:} Comparing \mstopk{} against syncSGD we observe due to lack of compatibility with {\it all-reduce} \mstopk\ performs slower than or comparable to syncSGD . For ResNet-101 and BERT we could not scale \topk\ beyond 16 and 32 GPUs respectively, due to running out of memory as  memory requirement increasing linearly with number of machines.}}
    \label{fig:mstopk_all_models}
    \vspace{-0.1in}
    \end{figure*}

\begin{figure*}[t]
    \begin{center}
    \includegraphics[width=0.3\textwidth]{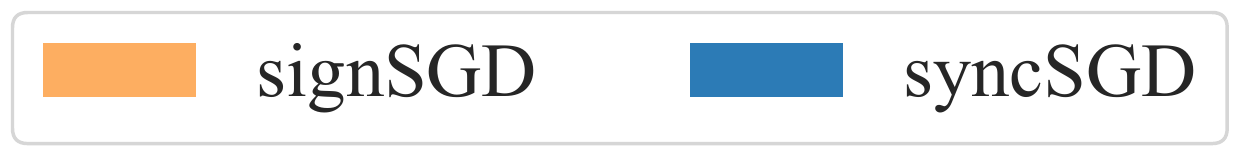}\\
    \vspace{-3pt}
    \begin{subfigure}[b]{0.3\textwidth}
    \includegraphics[width=\textwidth]{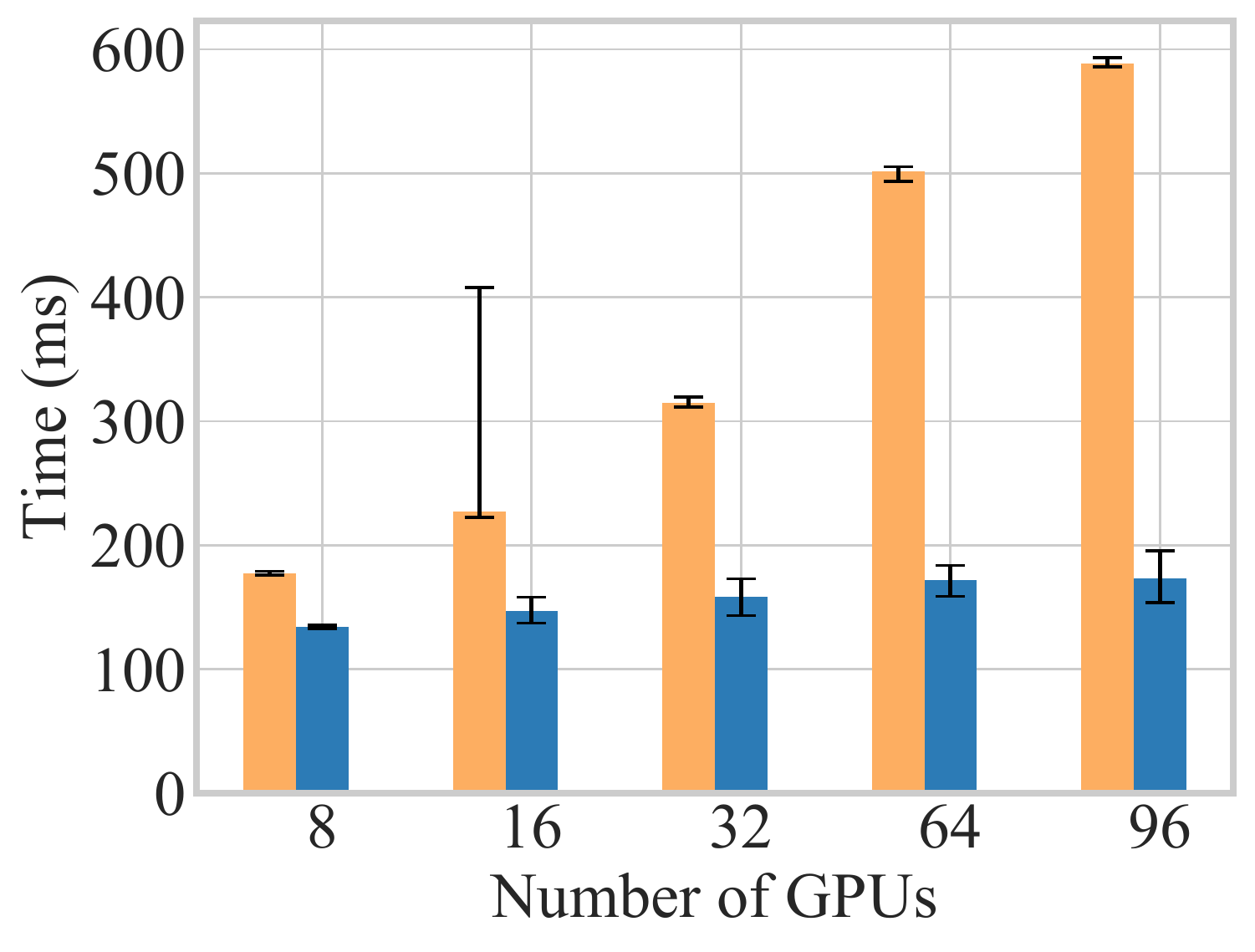}
    \vspace{-16pt}
    \caption{ResNet50: Batch Size 64}
    \end{subfigure}
    \begin{subfigure}[b]{0.3\textwidth}
    \includegraphics[width=\textwidth]{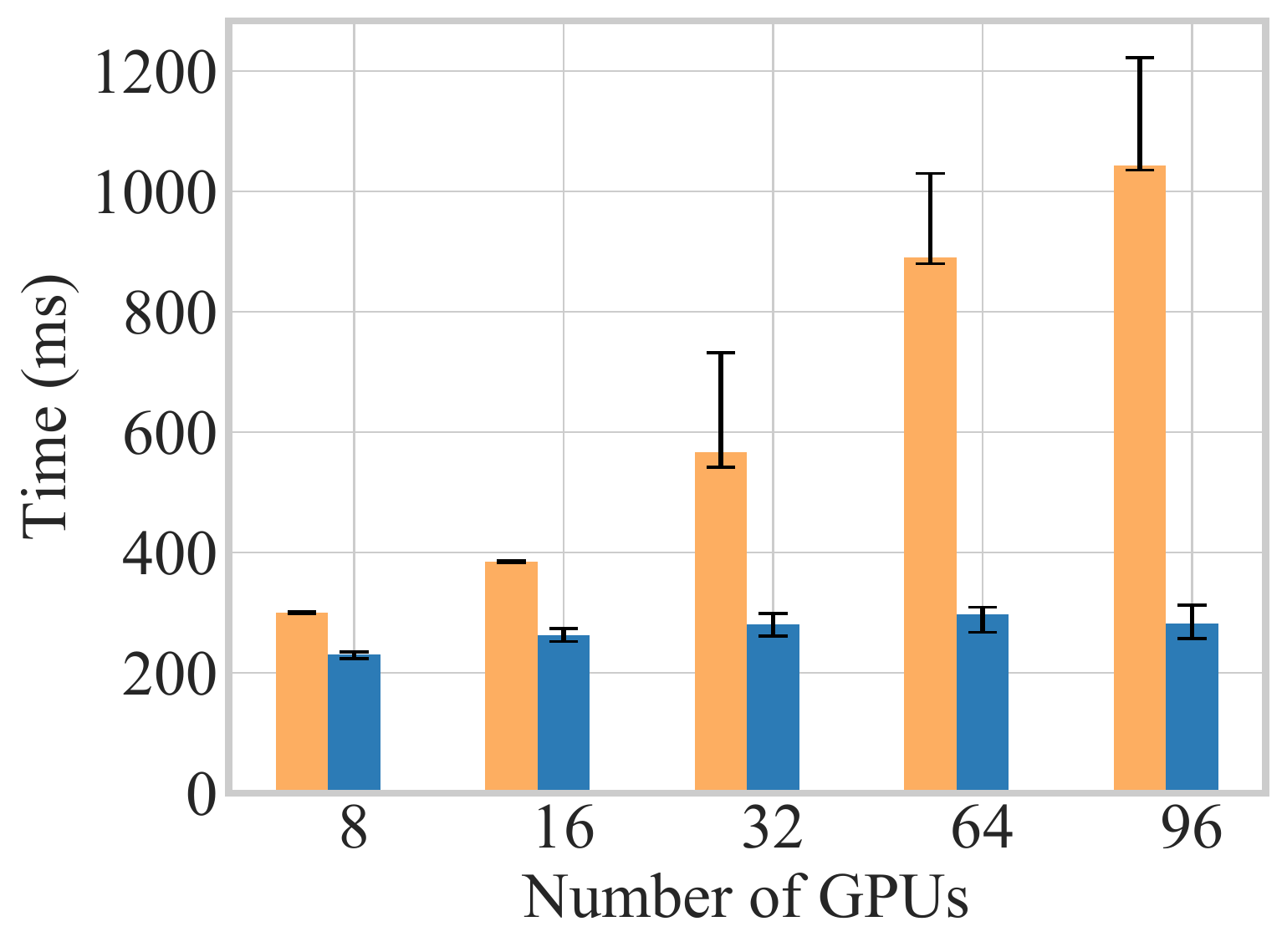}
    \vspace{-16pt}
    \caption{ResNet 101: Batch Size 64}
    \end{subfigure}
    \begin{subfigure}[b]{0.3\textwidth}
    \includegraphics[width=\textwidth]{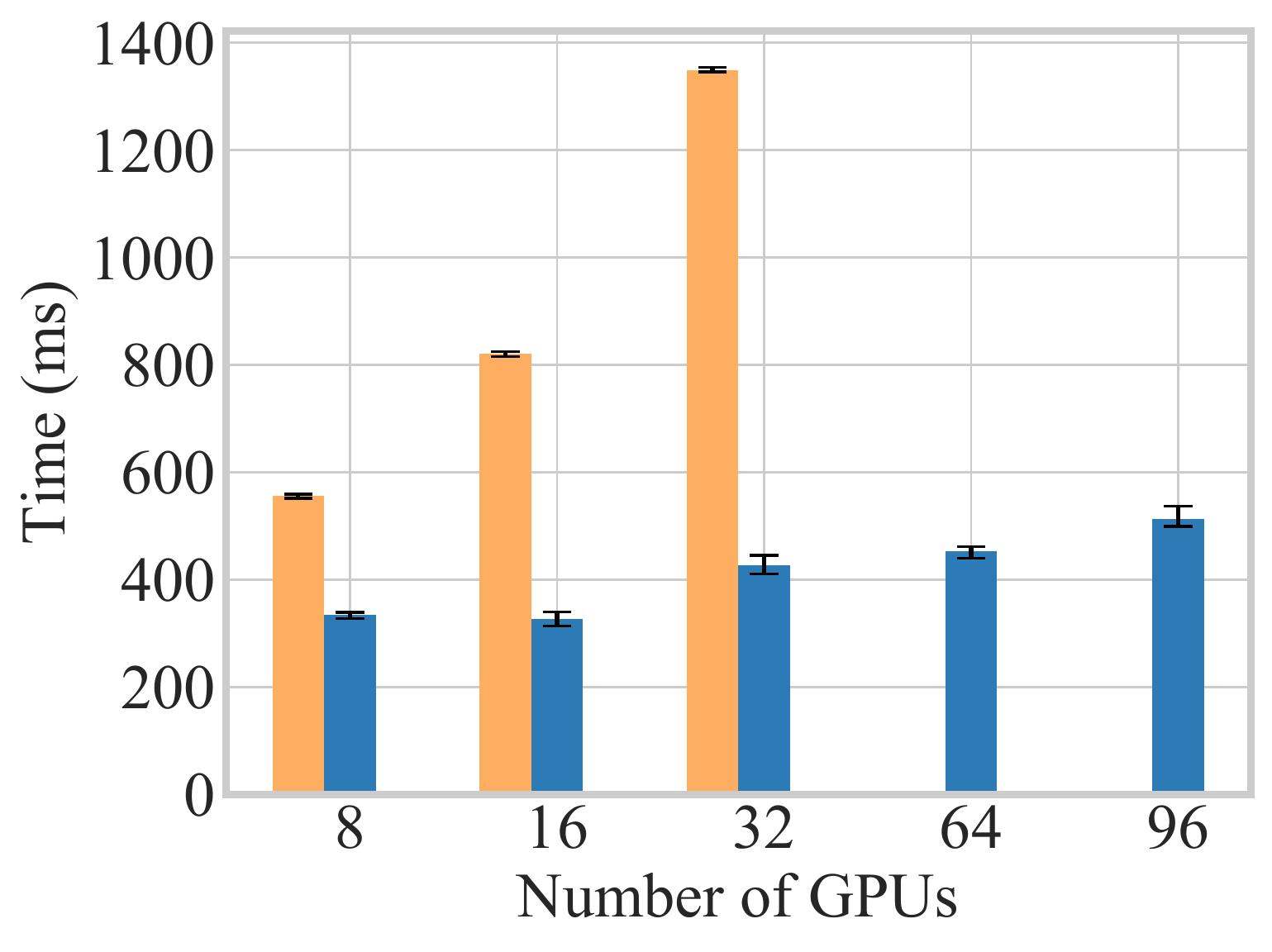}
    \vspace{-16pt}
    \caption{BERT: Batch Size 12}
    \end{subfigure}
    \end{center}
    \vspace{-10pt}
    \caption{\small{\textbf{Scalability of \signsgd:} Due to lack of support for {\it all-reduce} and linearly increasing decode time, across all three models, \signsgd{} performs considerably slower than syncSGD. For BERT we were not able to scale signSGD beyond 32 GPUs because we ran out of memory on a V100 GPU. This is due to the memory requirement increasing linearly with number of machines. }}
    \vspace{-0.0in}
    \label{fig:signsgd_allmodels_serial}
    \end{figure*}
We next analyse the performance of gradient compression methods against syncSGD. 

\vspace{-1mm}
\paragraph{PowerSGD.}\looseness=-1
\label{sec:powersgd_speedup}
We first study the scalability of PowerSGD when compared to syncSGD for ResNet-50, ResNet-101 , and $\text{BERT}_{\text{BASE}}$.  We use Rank-4, 8 and 16 as discussed previously.
As shown in Figure~\ref{fig:psgd_all_models} we can see that PowerSGD with Rank 4, 8, and 16 is \emph{slower} than syncSGD for ResNet-50 and ResNet-101 with batch size 64 (We investigate varying batch sizes in Section~\ref{sec:batch_size_vary}). 
This is primarily because syncSGD does not incur any overheads from compression and is able to overlap communication with computation.
On the other hand, for $\text{BERT}_{\text{BASE}}$, which is a much larger model(490MB), we see that for 96 GPUs, Rank-4 and Rank-8 are faster than syncSGD by around 18.8\% and 11.3\% respectively, while Rank-16 still takes longer than syncSGD.  

\vspace{-1mm}
\paragraph{\mstopk.}
\label{sec:topk_speedup}
Since the \mstopk~\cite{shi2021towards} operator is not compatible with {\it all-reduce} we use {\it all-gather} for communication. As shown in Figure~\ref{fig:mstopk_all_models}, only in 2 out of 15 different setups we observe a minuscule speedup (around 1.3\%) when compared against syncSGD, speedups are achieved when we are using \mstopk-0.01\%, i.e., when 99.9\% of the entries in the gradient are dropped. Also, due to high memory requirements for creating buffers for the all-gather primitive \mstopk{} does not scale beyond 32 GPUs for ResNet-101 and 16GPUs for BERT on a V100 GPU.

\begin{figure*}[t]
    \vspace{-10pt}
    \begin{center}
    \includegraphics[width=0.8\textwidth]{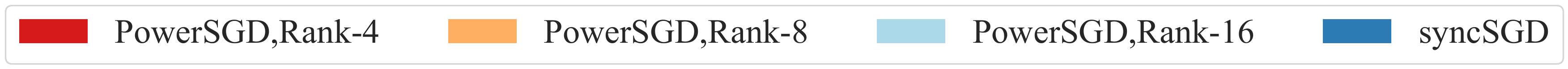}\\
    \vspace{-2pt}
    \begin{subfigure}[b]{0.32\textwidth}
    \includegraphics[width=\textwidth]{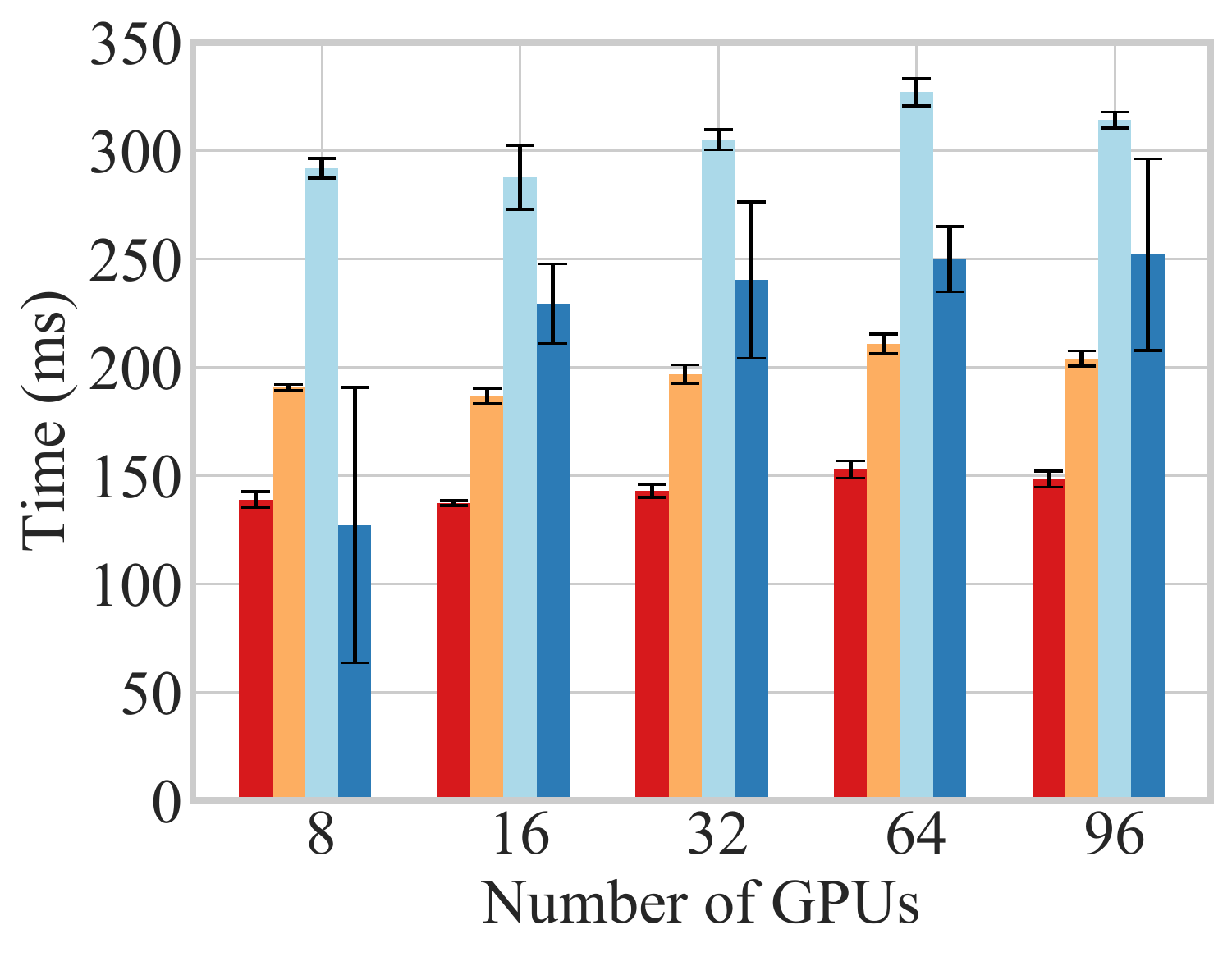}
    \vspace{-16pt}
    \caption{Resnet101: Batch Size 16}
    \end{subfigure}
    \begin{subfigure}[b]{0.32\textwidth}
    \includegraphics[width=\textwidth]{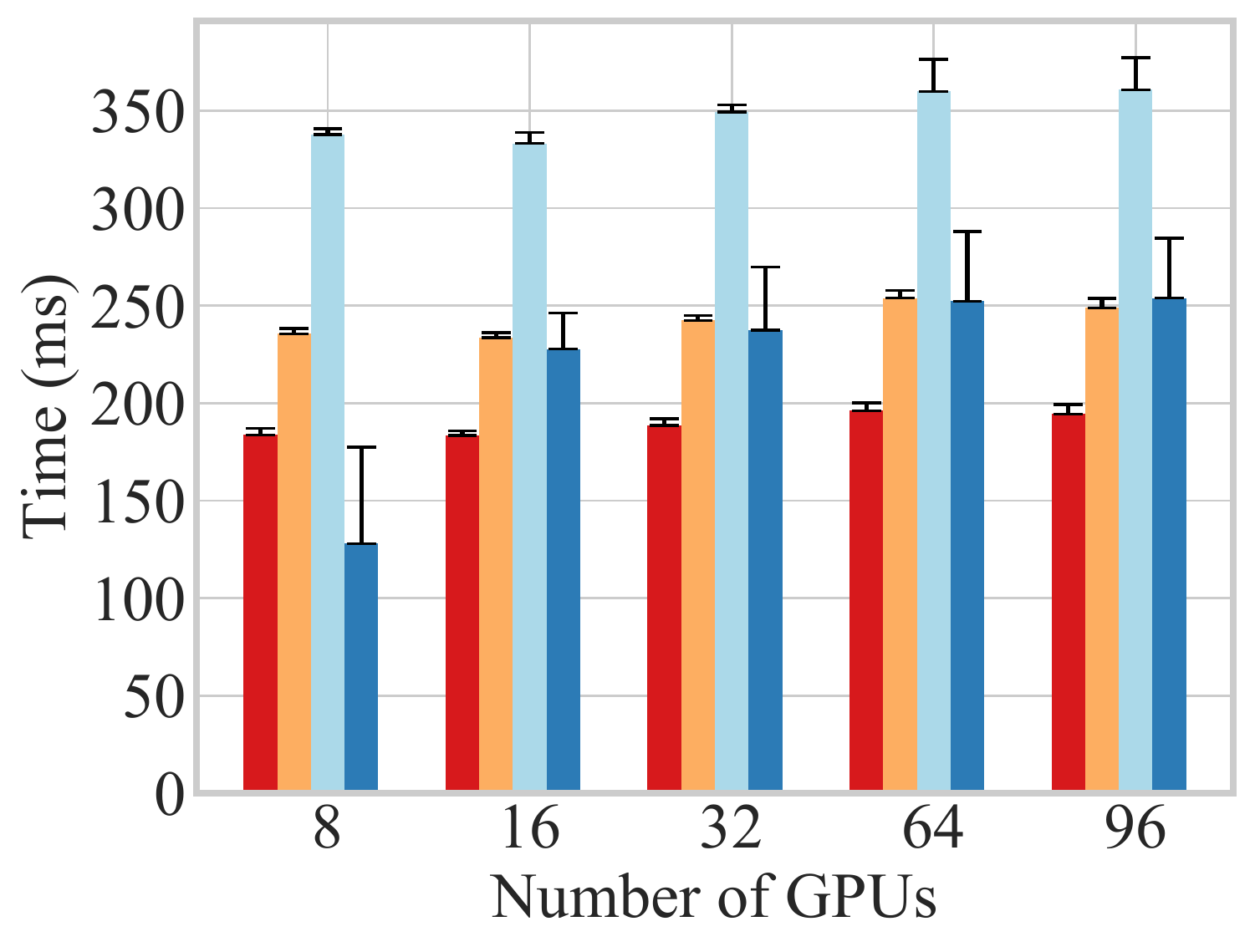}
    \vspace{-16pt}
    \caption{Resnet101: Batch Size 32}
    \end{subfigure}
    \begin{subfigure}[b]{0.32\textwidth}
    \includegraphics[width=\textwidth]{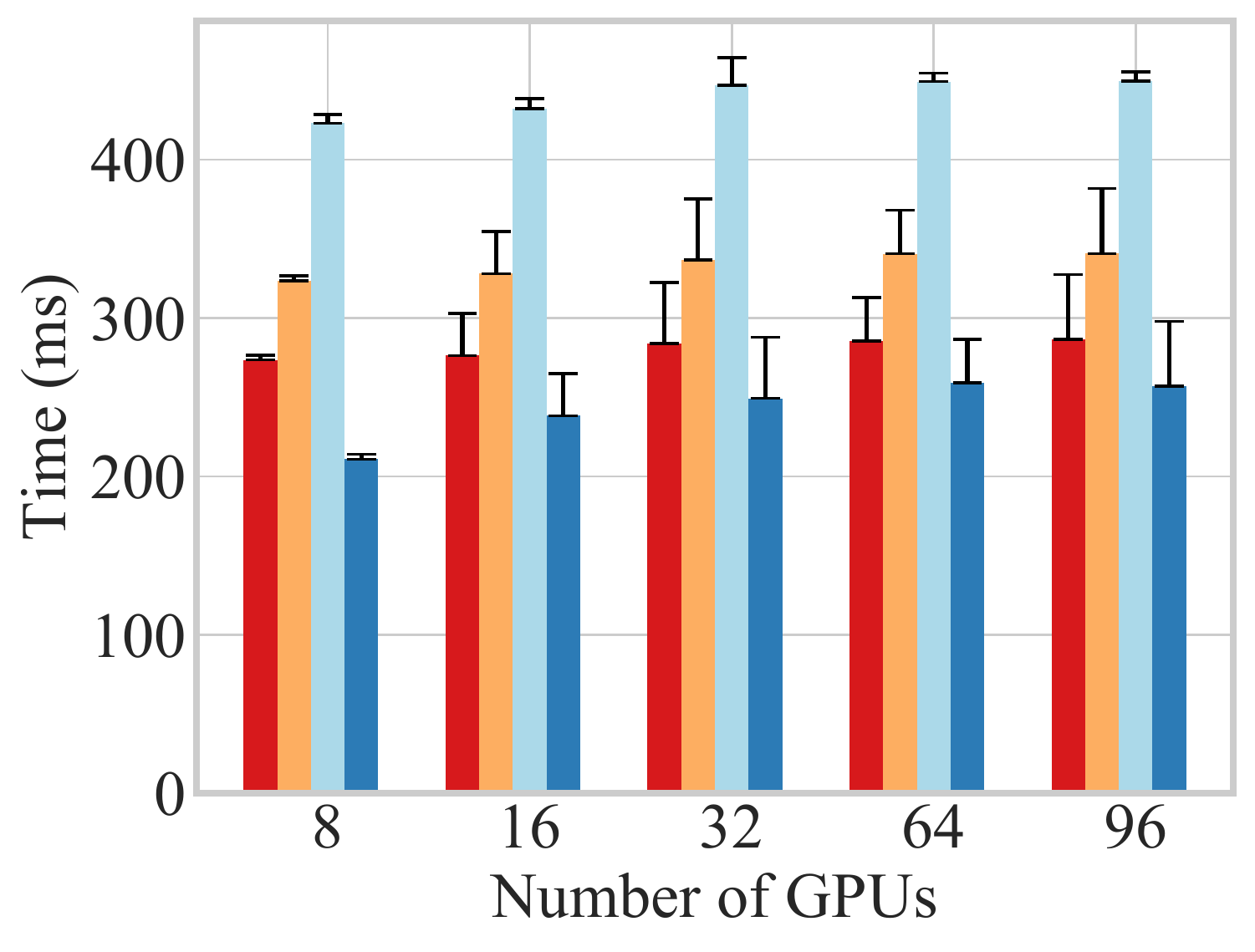}
    \vspace{-16pt}
    \caption{Resnet101: Batch Size 64}
    \end{subfigure}
    \end{center}
    \vspace{-11pt}
    \caption{\small{\textbf{Effect of varying batch size:} Here we compare \powersgd{} against $\text{Resnet101}$ on different batch sizes. We observe that large batch sizes provide more opportunity to syncSGD to hide the communication time, meanwhile at small batch sizes due to reduced computation time this overlap is not possible. Therefore  gradient compression methods become more useful at small batch sizes.}}
    \vspace{-18pt}
    \label{fig:resnet101_batchsize}
 \end{figure*}
\vspace{-1.5mm}
\paragraph{\signsgd{}.}
\label{sec:signsgd_speedup}
We study \signsgd{} with majority vote, where 1 bit is sent for each float (32 bit) leading to $32\times$  compression. Majority vote operation is not associative thus requiring use of all-gather. Figure~\ref{fig:signsgd_allmodels_serial},  shows that despite \signsgd{} being extremely quick to encode and decode, due to lack of comaptibility with {\it all reduce} communication time scales linearly. 
Further, due to overheads in creating buffers for the all-gather primitive we can not scale \signsgd{} on $\text{BERT}_{\text{BASE}}$ beyond 32 GPUs.\looseness=-1

\begin{figure}[t]
\begin{minipage}{0.43\textwidth}{\null
\captionof{table}{\small{\textbf{Encode \& Decode times for ResNet-50:} Even for a comparatively small network like ResNet-50, where time for backward pass is around 122ms, gradient compression methods have high overhead}}
\label{tab:encode_decode_time}
\resizebox{\linewidth}{!}{
\begin{tabular}{@{}cccc@{}}
\toprule
\begin{tabular}[c]{@{}l@{}}Compression \\ Method\end{tabular}       & \begin{tabular}[c]{@{}l@{}}Compression \\ Parameter\end{tabular} & \begin{tabular}[c]{@{}l@{}}Compression \\ Ratio\end{tabular} & $T_{encode-decode}$(ms) \\ \midrule
\multirow{3}{*}{\powersgd} & Rank-4               & $72\times$ & 45                      \\
                          & Rank-8                & $37\times$ & 64                      \\
                          & Rank-16               & $19\times$& 130                     \\ \midrule
\multirow{3}{*}{\mstopk}    & 1\%                  & $100\times$ &            103          \\
                          & .1\%                  & $1000\times$ &               104      \\ \midrule
\signsgd{}                  &             & $32\times$ & 16.34                   \\ \bottomrule
\end{tabular}}

}
\end{minipage}\quad
\begin{minipage}{0.5\textwidth}{\null
    \centering
    \includegraphics[width=\linewidth]{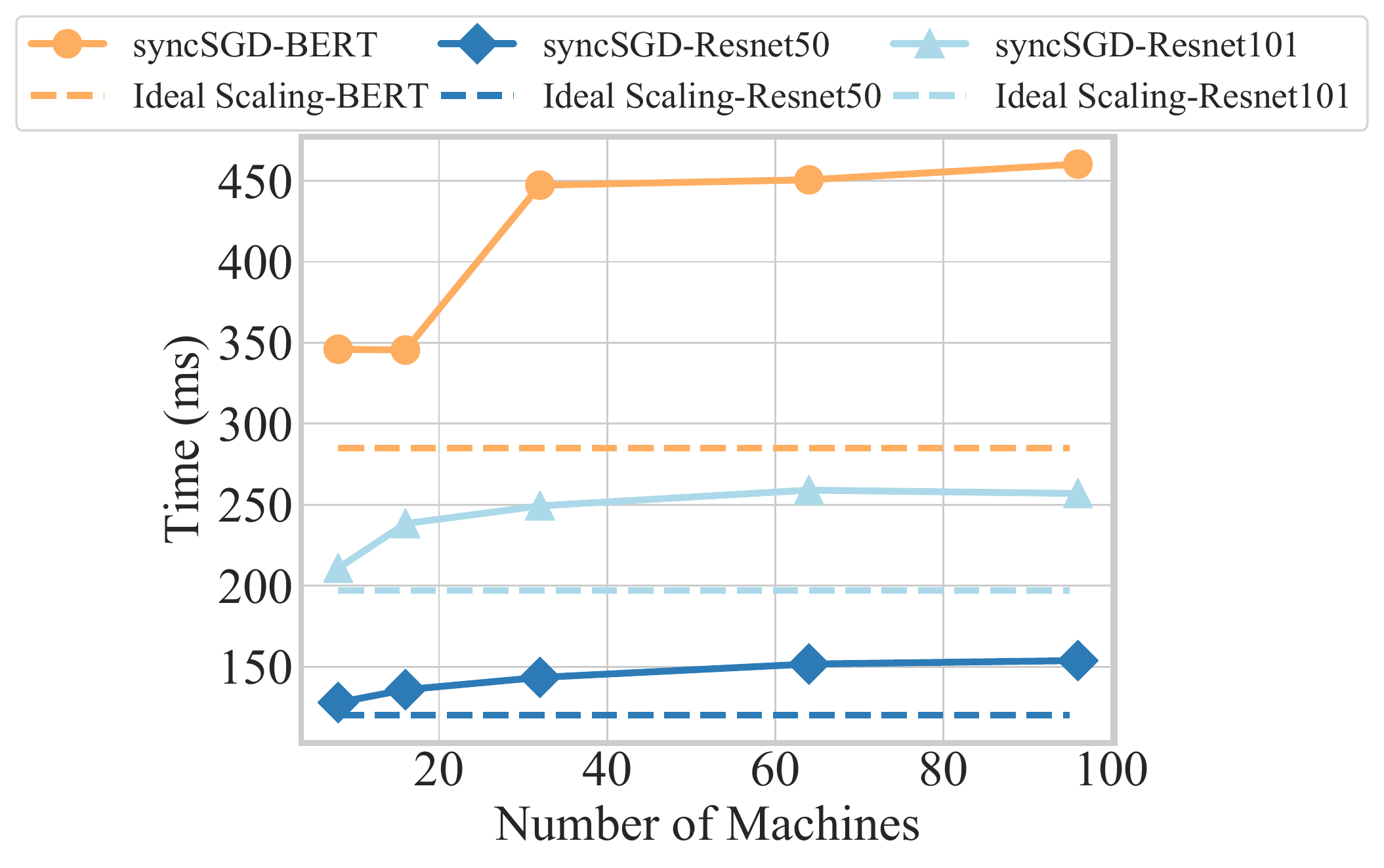}
    \captionof{figure}{\small{\textbf{Difference between linear scaling and observed performance:} We observe that the difference between linear scaling and syncSGD is less than 200 ms at 10Gpbs. This leaves little opportunity for gradient compression methods to provide speedups. }}%
    \label{fig:ideal_compare_main}
}
\end{minipage}
\vspace{-0.2in}
\end{figure}
\vspace{-2mm}
\paragraph{Why Doesn't  Gradient Compression Lead to Speedups?}\looseness=-1

There are three reasons for lack of speedups. First as stated in Section~\ref{sec:comm_comp_over}, compression methods are poor candidates for overlapping with gradient computation. 

Meanwhile vanilla syncSGD, as shown in Figure~\ref{fig:effect_overlap} is able to  benefit from overlapping communication and backward pass, which provides it a significant advantage. 

The second reason, as depicted in Table~\ref{tab:encode_decode_time}, is high overhead of compression. Due to system advances, we observe in Figure~\ref{fig:ideal_compare_main} that even for large models like $\text{BERT}_{\text{BASE}}$, for $96$ GPUs the difference between syncSGD and linear scaling is around 200ms. This indicates, that compression algorithms for being a viable alternative, need to be extremely fast and perform compression and communication in less than $200$ms even for such large models. 

Third reason for slowdown, as pointed out by prior works~\cite{vogels2019powersgd, cho2019gradzip} and experiments in previous section, is lack of compatibility with all-reduce. Compression methods which are compatible all-reduce like \powersgd{} are able to scale better.  For an operation to be compatible with all-reduce it must be associative, \ie the order of operations should not matter. However, Table~\ref{tab:methods} shows that several gradient compression methods are not compatible with all-reduce. In these cases, to perform gradient aggregation, the workers need to perform an all-gather operation. This can lead to high communication costs, leading to poor scalability as we increase the number of processors.\looseness=-1

\begin{remark}
Existing gradient compression methods provide limited benefits either due to encoding overheads or due to lack of compatibility with all-reduce across a range of models.
\end{remark}

\vspace{-1.5mm}
\subsection{Effect of Batch Size on Scalability}
\label{sec:batch_size_vary}
For analysing the effect of varying batch sizes, we compare PowerSGD against syncSGD given it is the most scalable method we encounter. In Figure~\ref{fig:resnet101_batchsize}, for ResNet-101, we find that the benefits of using PowerSGD with Rank-4 drops as the batch size increases. For instance, when using 96 GPUs, PowerSGD Rank-4 provides almost 42.5\% speedup when training using batch size 16.  This speedup drops to 25.7\% for batch size 32 and with batch size 64, we observe that PowerSGD Rank-4 is around 6.3\% slower than synSGD. In general, increasing batch size leads to an increase in the compute time providing more opportunity for syncSGD to overlap computation and communication.

\begin{remark}
Using large batch sizes often provides enough opportunity for syncSGD to overlap communication with communication thereby reducing the extent of benefits achieved from using gradient compression.  
\end{remark}

\section{Identifying Regimes of High Gradient Compression Utility}
In the previous section we looked at the performance of distributed training and gradient compression of popular models on currently available hardware. Next we study how to identify regimes, in terms of hardware or model characteristics, where gradient compression can provide significant gains \ie how will our above results change if we had 100Gbps bandwidth or with an $8\times$ faster GPU. To answer such questions we develop a performance model that can be used both by researchers and practitioners to reason about expected performance under different setups. 
\vspace{-1.5mm}
\subsection{Performance Model for Distributed Data Parallel}
\label{sec:performance_model}
Based on optimizations listed for syncSGD in \cite{li2020pytorch} we construct an analytical performance model. We assume the model being trained can be partitioned into $k$ buckets, where the first $k-1$ buckets are of size $b$ and the last bucket is of size $\hat{b}$, where $\hat{b} \leq b$. The total time observed for backward pass and gradient synchronization for synchronous SGD becomes:
{\small
\begin{equation*}
    T_{obs} \approx max(\gamma T_{comp}, (k-1)\times T_{comm}(b,p, BW))  + T_{comm}(\hat{b},p,BW)
\end{equation*}
}\noindent where $T_{obs}$ is the total time observed for backward pass and synchronization, $T_{comp}$ is the compute time for the backward pass on single machine, $(k-1)\times T_{comm}(b, p, BW)$ is the time required to communicate $k-1$ gradient buckets of size $b$ across $p$ GPUs at $BW$ bandwidth, and $T_{comm}(\hat{b}, p, BW)$ is the time to communicate the last bucket of size $\hat{b}$, which can not be overlapped with computation. Finally, $\gamma$ represents the factor of slowdown in backward pass due to overlap with communication. We observe $\gamma$ to between 1.04 to 1.1. In case of syncSGD when using ring-reduce, $T_{comm}(b,p, BW)$ becomes
{\small
\begin{equation}
\label{eq:ring_reduce}
T_{comm}(b,p, BW)= 2\alpha\times(p-1) +  2\times b \times \frac{(p-1)}{p\times BW}
\end{equation}
}\noindent
where $\alpha$ is the latency coefficient $b$ is the bucket size, $p$ is the number of GPUs and $BW$ is the bandwidth available. The performance model for gradient compression methods is in Appendix~\ref{app:grad_compression_perfomance_model}. 
\vspace{-1.5mm}

\begin{figure}[t]
    \centering
    \begin{minipage}[b]{0.3\textwidth}
    \includegraphics[width=0.98\linewidth,]{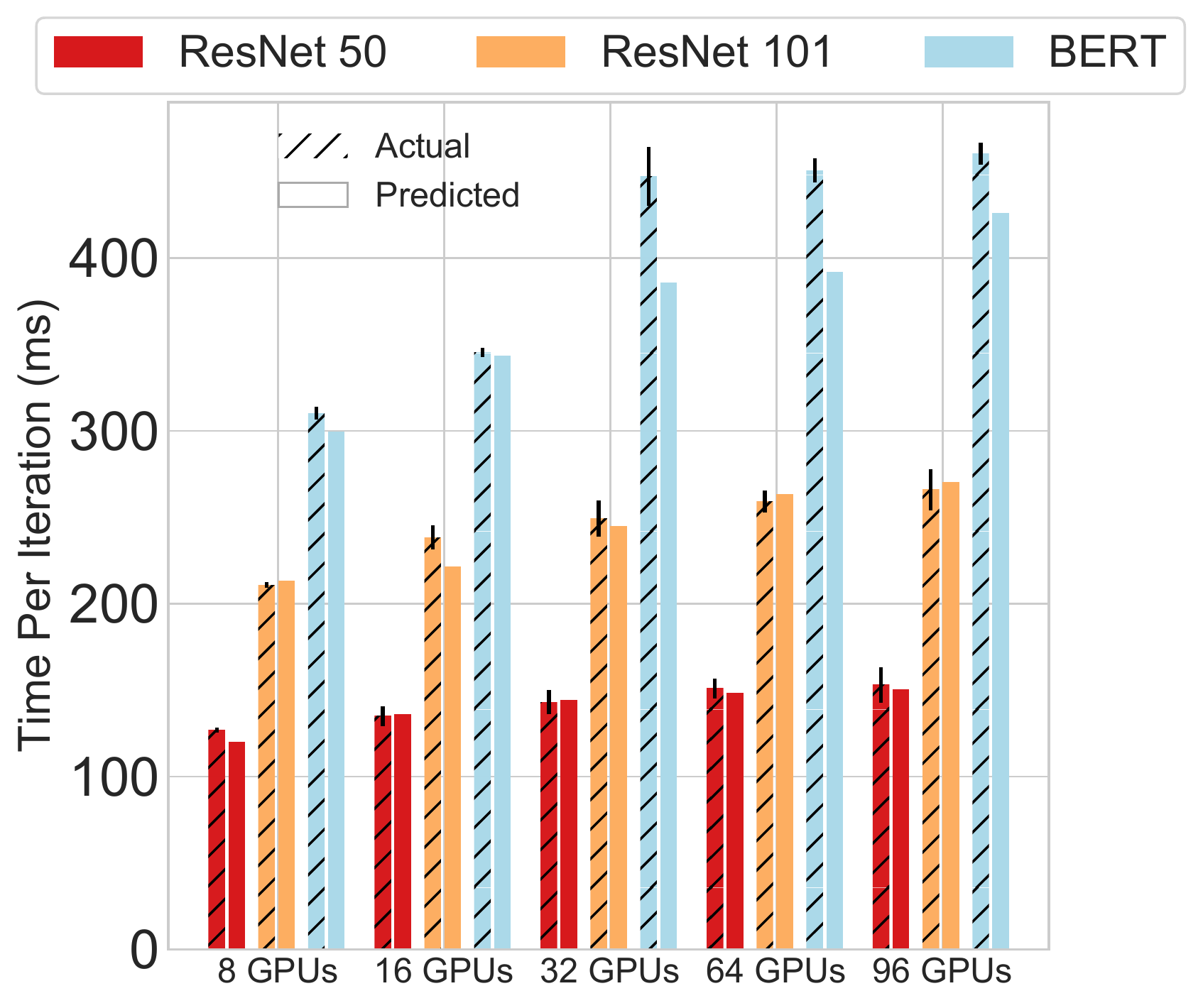}
    \vspace{-0.1in}
  \caption{ \small{\textbf{Verifying performance model for syncSGD: } Our performance model matches the actual performance for all three models across wide range of GPUs. The median difference between predictions and actual runtime is 1.8\%. Error bars show minimum and maximum value.}}
    \label{fig:perf_sync_verify_main}
  \end{minipage}\quad
    \begin{minipage}[b]{0.3\textwidth}
    \includegraphics[width=0.98\linewidth]{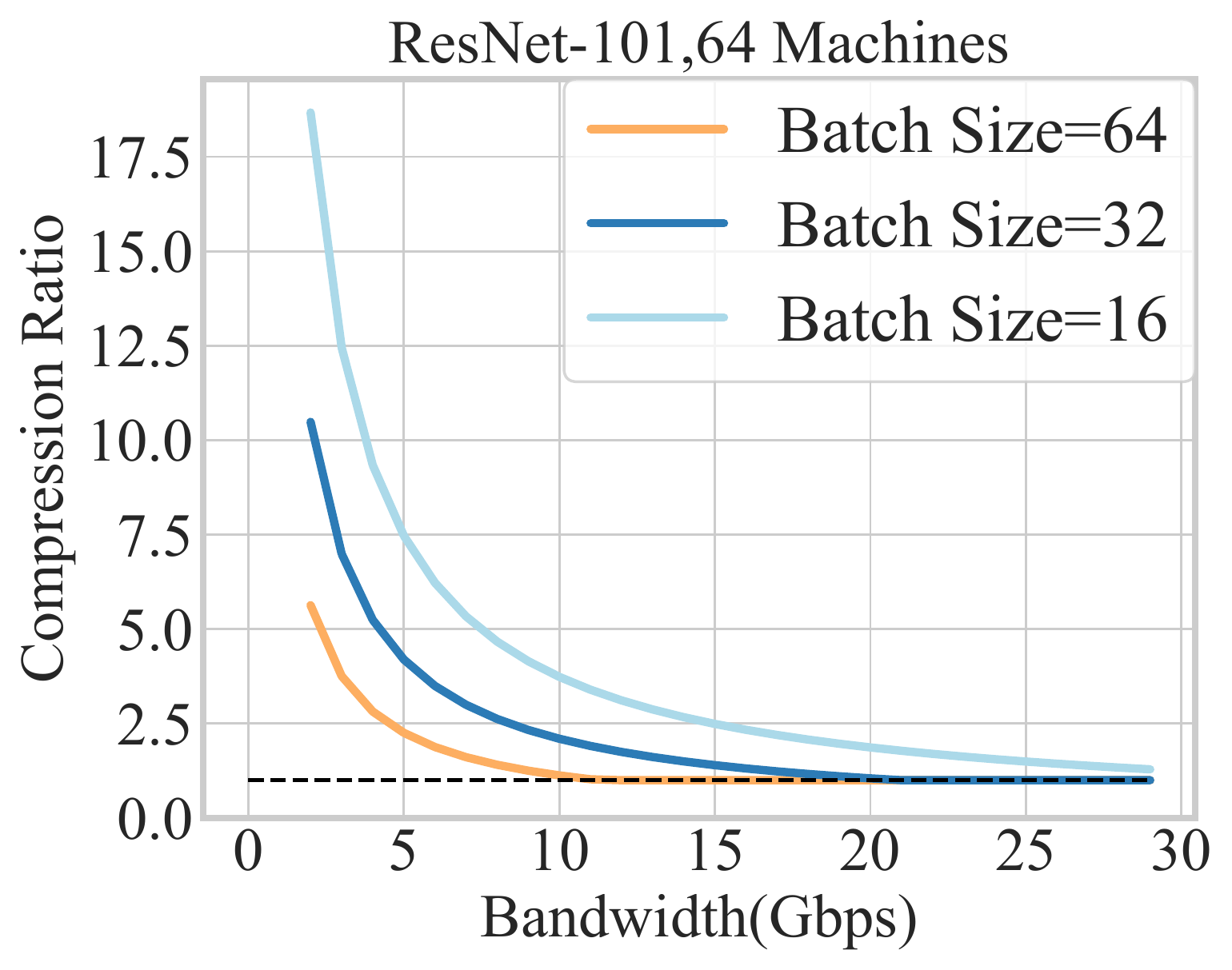}
    \vspace{-0.1in}
     \caption{\small{\textbf{Required gradient compression for near linear speedups (simulated):} Above figure is for ResNet-101 simulated for 64 machines. We observe that the required gradient compression for near linear scaling at 10 Gbps even for quite small batch sizes is around $4\times$.}}
       \label{fig:ideal_compression_resnet101}
    \end{minipage}\quad
    \begin{minipage}[b]{0.32\textwidth}
    \captionof{table}{\small{Compatibility of various gradient compression methods with all-reduce. Methods which are compatible scale better.}}
\label{tab:methods}
    \begin{tabular}[b]{@{}ll@{}}
\toprule
Compression Method        & All-reduce  \\ \midrule 
syncSGD                 & \cmark                        \\      
GradiVeq~\cite{yu2018gradiveq}                         & \cmark                      \\
\powersgd~\cite{vogels2019powersgd}                          & \cmark                      \\ 
Random-$k$~\cite{wangni2018gradient}      & \cmark                         \\
\topk~\cite{aji2017sparse} & \xmark \\
ATOMO~\cite{wang2018atomo}                          & \xmark                      \\
\signsgd~\cite{bernstein2018signsgd}                          & \xmark                          \\ 
TernGrad~\cite{wen2017terngrad}                   &          \xmark                  \\ 
QSGD~\cite{alistarh2017qsgd}                   &          \xmark                  \\ 
DGC~\cite{lin2017deep}                   &          \xmark                   \\ 
\bottomrule
\end{tabular}

    \end{minipage}
\vspace{-0.2in}
\end{figure}

\paragraph{Verifying Performance Model.} We first empirically verify our performance model using the same experimental setup as mentioned in Section~\ref{sec:eval_grad_compression}. As shown in Figure~\ref{fig:perf_sync_verify_main} we observe that our model very closely tracks the actual performance in all cases. The median difference between our prediction and actual runtime is 1.8\% and the maximum is 9.1\%. 
More details on verification and how we measure the values to input into the performance model can be found in Appendix~\ref{app:verification_pmodel}. 
\vspace{-1.5mm}
\paragraph{Limitations.} Currently, our performance model only supports the data-parallel setting and is not applicable on other forms of distributed training like model or pipeline parallelism, \ie we do not consider cases where the model can not fit in single GPU memory. Further, we do not account for asynchronous methods~\cite{dean2012large, grishchenko2018asynchronous, mota2013d}, \ie we assume that gradient synchronization is required after every iteration.
\vspace{-2mm}
\subsection{Insights from the Performance Model}
\paragraph{How Much Should We Compress?} Using the performance model we investigate how much compression will be required for linear scalability. In Figure~\ref{fig:ideal_compression_resnet101} we see that even at small batch sizes for ResNet-101 we need around $4\times$ compression for linear scalability, which is significantly smaller than what most gradient compression methods offer.
\vspace{-1.5mm}
\paragraph{Effect of Network Bandwidth on Gradient Compression.} Figure~\ref{fig:network_resnet101_bsize64_main} shows comparison between speedups for ResNet-101 when using syncSGD and \powersgd{} Rank-4 at different network bandwidths. In addition to estimating time taken with our performance model, we also use the TC command~\cite{tc} to limit bandwidth on a real cluster, thereby verifying our performance model (the markers represent measurements on hardware). From the figure, we see that gradient compression is very useful in low bandwidth settings ($\leq 8$ Gbps). Although low bandwidths are not common in data centers (10 Gbps is minimum with a V100 GPU on Amazon EC2), this shows that in certain cases like wide-area Federated Learning~\cite{bonawitz2019towards} gradient compression methods can be extremely useful. 
Several other insights and analysis from our performance model can be found in Appendix~\ref{app:analysis_pmodel}.
\vspace{-1.5mm}
\subsection{Takeaways for Practitioners and Researchers}
We have implemented our performance model in a simple tool that can simulate distributed training and thus help users reason about performance and expected speedups in different setups. We discuss some scenarios discovered using our tool, and the implications for practitioners and researchers.
\vspace{-1.5mm}
\paragraph{Extremely Large Models and Low Bandwidth.} Recently \powersgd{} was used by Ramesh et al.~\cite{ramesh2021zeroshot} to scale training of an extremely large model (12 billion parameters). However, the setup used was not the standard data parallel setup since the model did not fit on a single GPU. In cases where the model is extremely large, practitioners can plug in the model size into our performance model to calculate expected speedups from gradient compression.  
\vspace{-1.5mm}
\paragraph{Low Compute Density Workloads.} Highly scalable syncSGD implementations~\cite{li2020pytorch,sergeev2018horovod} rely on the overlap between communication and backward pass to provide high speedup. But if the compute density decreases without reduction in the number of parameters then overlap will reduce. An example of reduced compute density is a small batch size and we find gradient compression does indeed provide speedups for small batches. However, recent work has focused on increasing the batch size (memory permitting)~\cite{ devarakonda2017adabatch, yao2018large, you2019large} and designing algorithms to improve accuracy when using large batches.
\vspace{-1.5mm}
\paragraph{Focus on Compression Overhead.} Existing gradient compression methods focus heavily on amount of compression they provide. Our analysis shows that for linear scaling we do not need extremely high compression ratios. Instead the focus for ML researchers should be on reducing the encoding overhead. In Appendix~\ref{app:analysis_pmodel}, we show that reducing encode-decode time even at the expense of decreased compression ratio helps. As an extreme case, prior work on how to by-pass the gradient encoding and decoding step can potentially provide communication efficiency for free~\cite{wang2021pufferfish}.
\vspace{-1mm}
\paragraph{Using Auxiliary Hardware.}
In Section~\ref{sec:comm_comp_over} we showed that contention for compute resources inhibits overlapping compression with backward  pass. However, newer generation of network interface cards and switches support some basic arithmetic operations~\cite{sapio2019scaling}. If gradient compression methods are built using these rudimentary  operations, they can be offloaded to these auxiliary devices, which can enable overlapping compression with communication.
\vspace{-1.5mm}

\section{Conclusion}
In this work we study several gradient compression methods used to accelerate distributed ML training. We discover that existing gradient compression methods provide marginal speedups in a datacenter setup due to the overheads in compression. 
We develop a performance model that can help algorithm designers build scalable gradient compression algorithms. 
Our performance model also allows users to conduct what-if analyses and determine how much compression they need given a hardware setup. We believe this analysis provides the community clarity on the desirable properties for gradient compression and will lead to methods that can provide improved scalability in the future.


\bibliographystyle{unsrt}
\bibliography{ref}
\clearpage

%
\clearpage
\appendix
\section{Overlap gradient compression with computation}
\label{app:extra_overlap_results}
\begin{figure*}[b]
    \begin{center}
    \includegraphics[width=0.8\textwidth]{figs/resnet50_bsize64_real_bar_same_y_height_psgd_legend.pdf}\\
    \vspace{-2pt}
    \begin{subfigure}[b]{0.32\textwidth}
    \includegraphics[width=\textwidth]{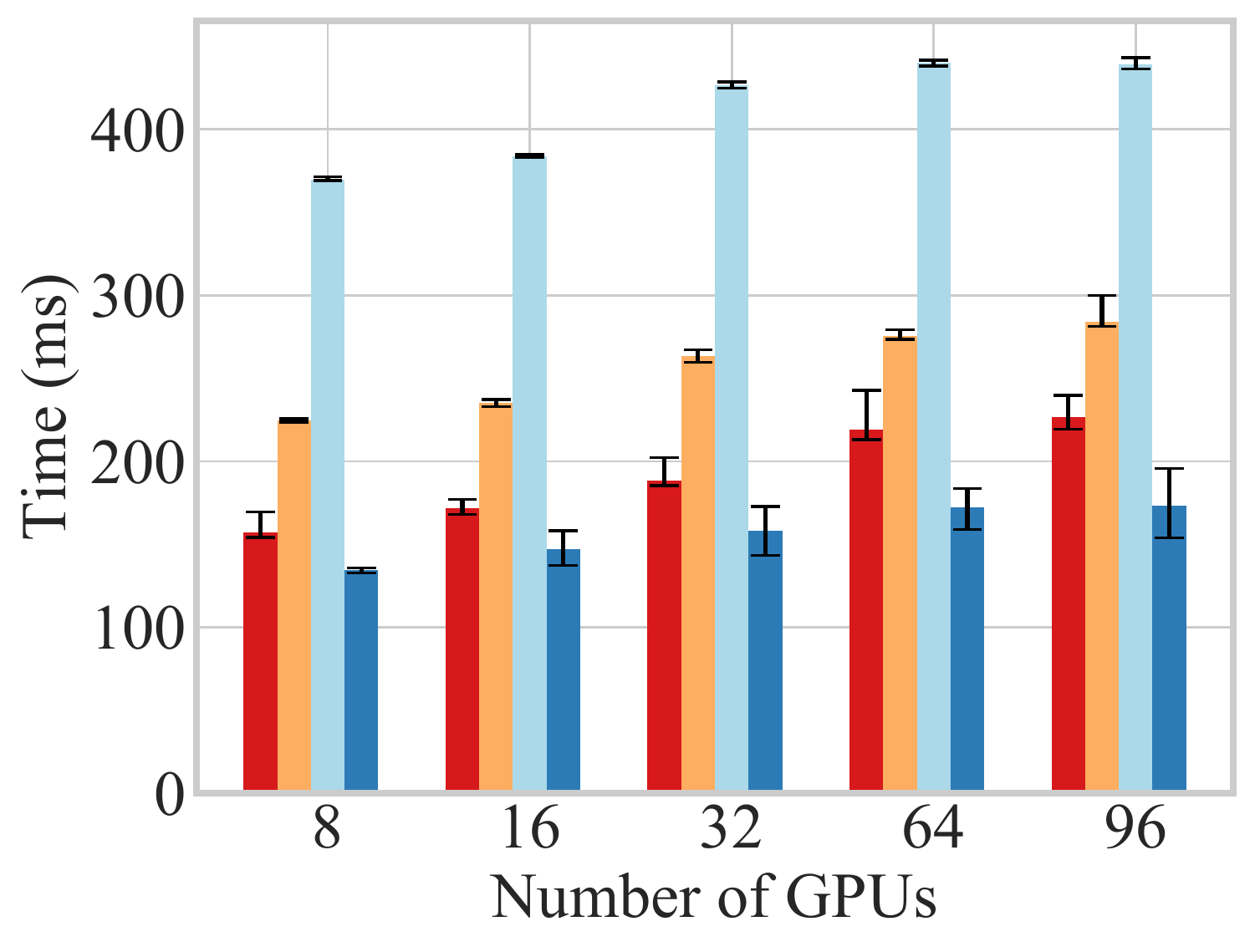}
    \caption{ResNet 50: Batch Size 64}
    \label{fig:psgd_resnet50_bsize64_new_api}
    \end{subfigure}
    \begin{subfigure}[b]{0.32\textwidth}
    \includegraphics[width=\textwidth]{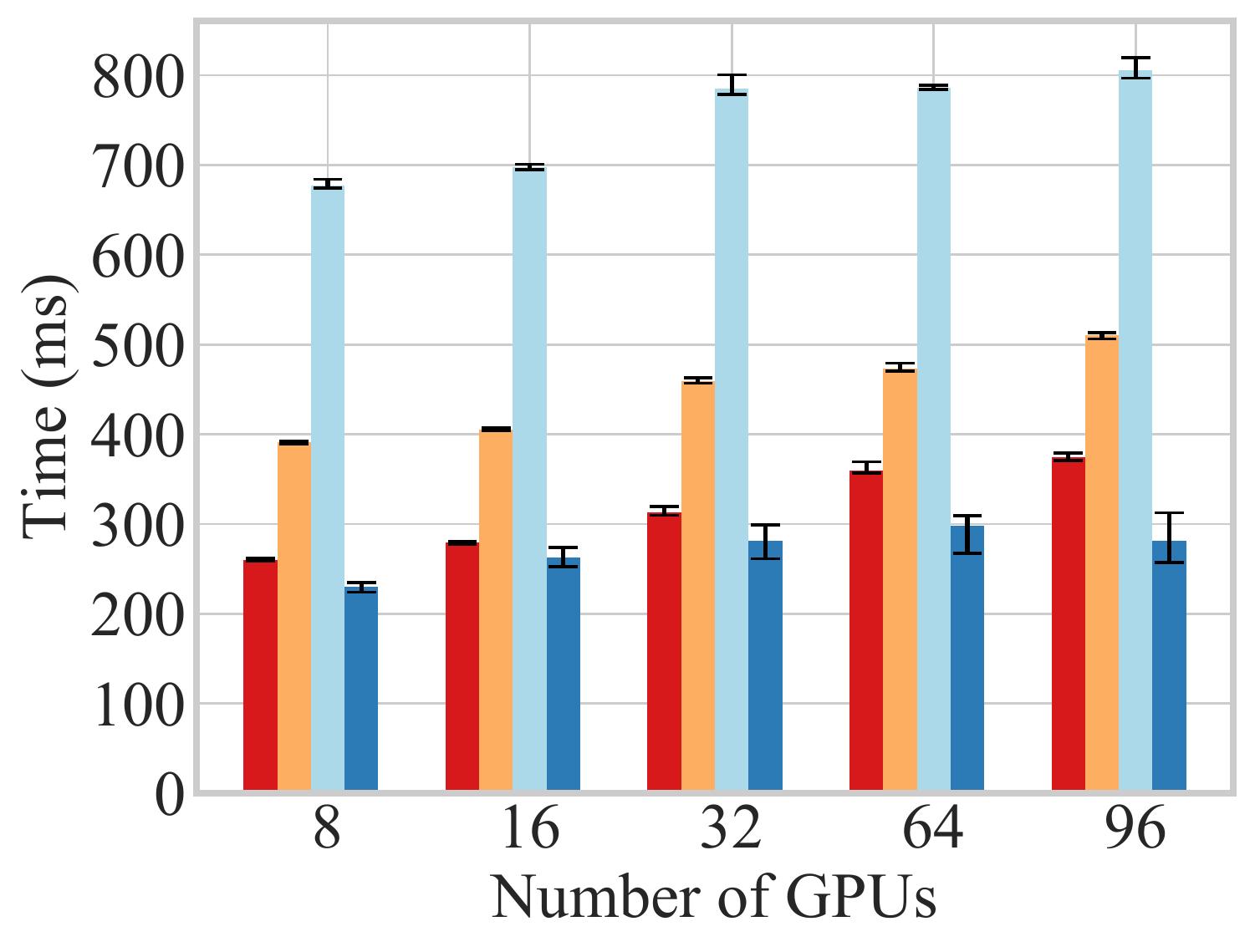}
    \caption{ResNet 101: Batch Size 64}
    \label{fig:psgd_resnet101_bsize64_new_api}
    \end{subfigure}
    \begin{subfigure}[b]{0.32\textwidth}
    \includegraphics[width=\textwidth]{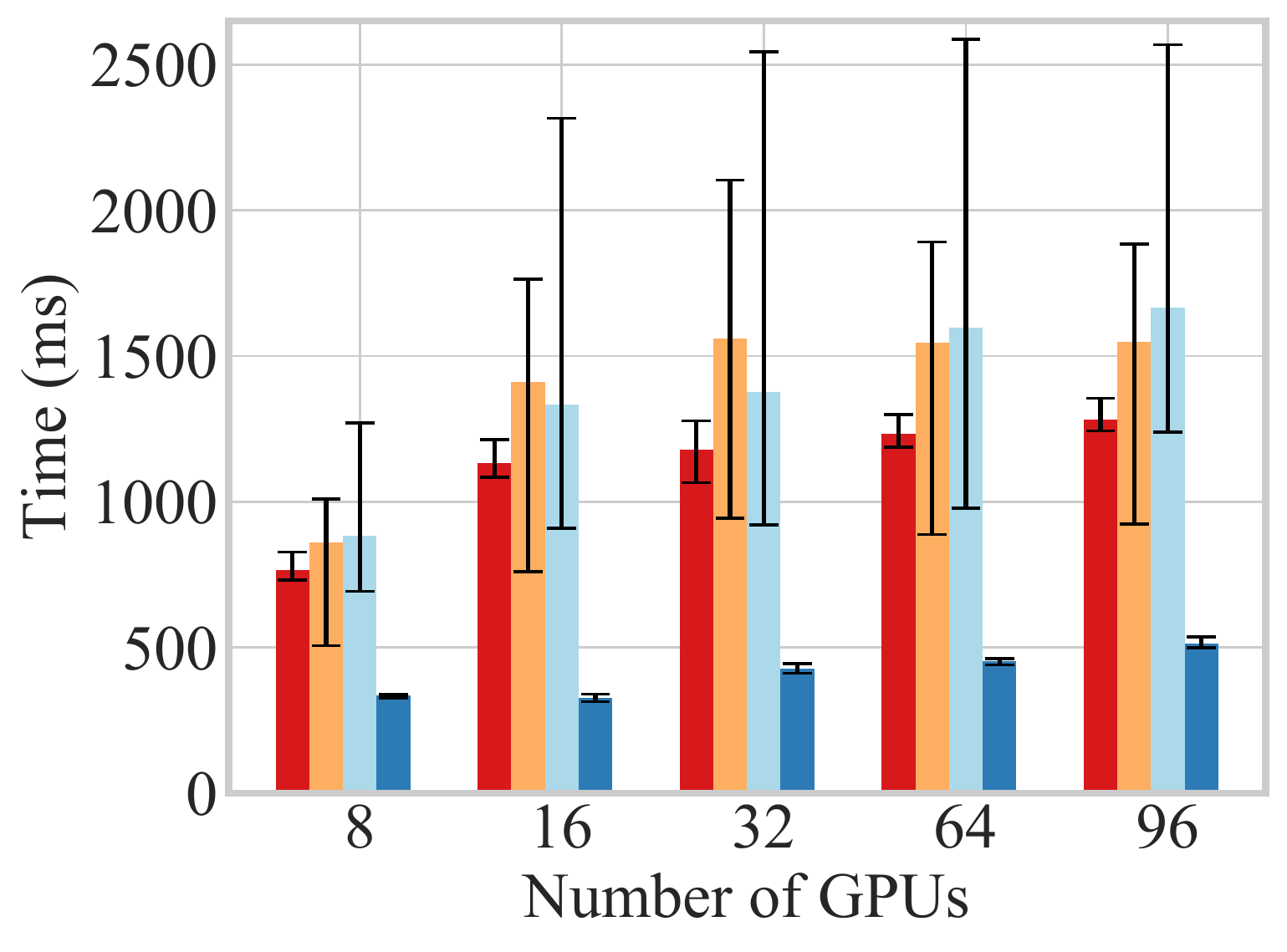}
    \caption{BERT: Batch Size 12}
    \label{fig:psgd_bert_bsize12_new_api}
    \end{subfigure}
    \end{center}
    \vspace{-0.1in}
    \caption{\small{\textbf{Scalability of PowerSGD with overlap:} When \powersgd{} is overlapped with backward we observe that it does not provide speedups in any of our experiments when compared against an optimized implementation of syncSGD. }}
    \label{fig:psgd_all_models_new_api}
    \vspace{-0.1in}
\end{figure*}

\begin{figure*}
    \begin{center}
    \includegraphics[width=0.6\textwidth]{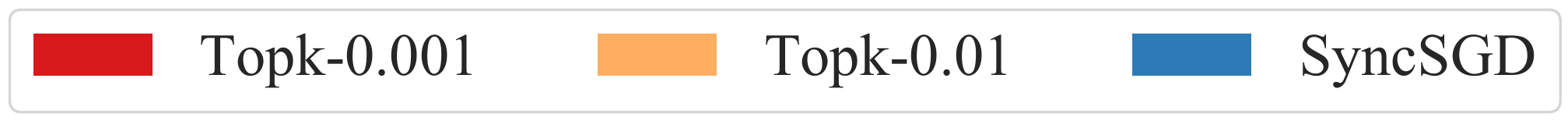}\\
    \vspace{-2pt}
    \begin{subfigure}[b]{0.32\textwidth}
    \includegraphics[width=\textwidth]{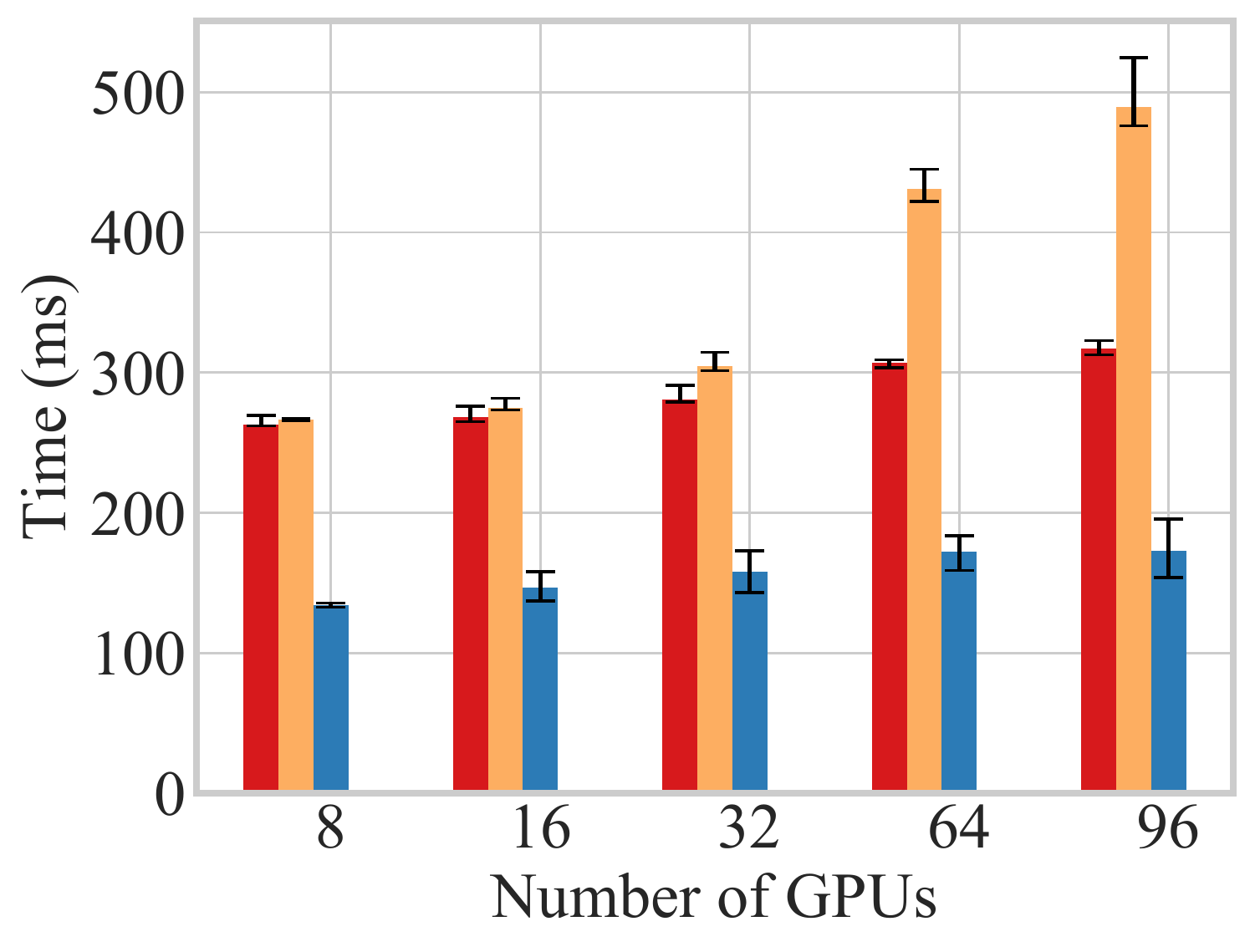}
    \caption{ResNet50: Batch Size 64}
     \label{fig:topk_resnet50_bsize64}  
    \end{subfigure}
    \begin{subfigure}[b]{0.32\textwidth}
    \includegraphics[width=\textwidth]{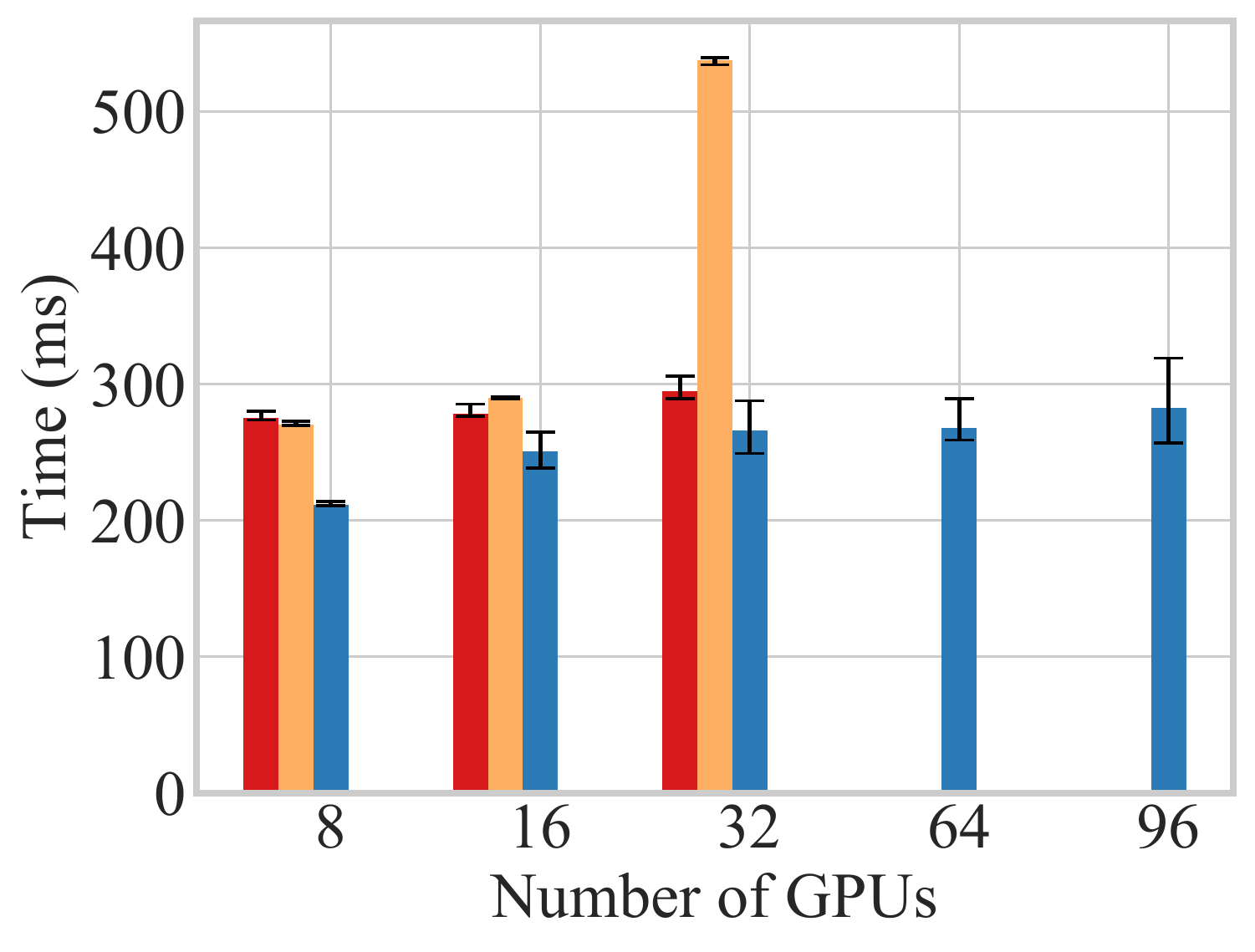}
    \caption{ResNet 101: Batch Size 64}
    \label{fig:topk_resnet101_bsize64}
    \end{subfigure}
    \begin{subfigure}[b]{0.32\textwidth}
    \includegraphics[width=\textwidth]{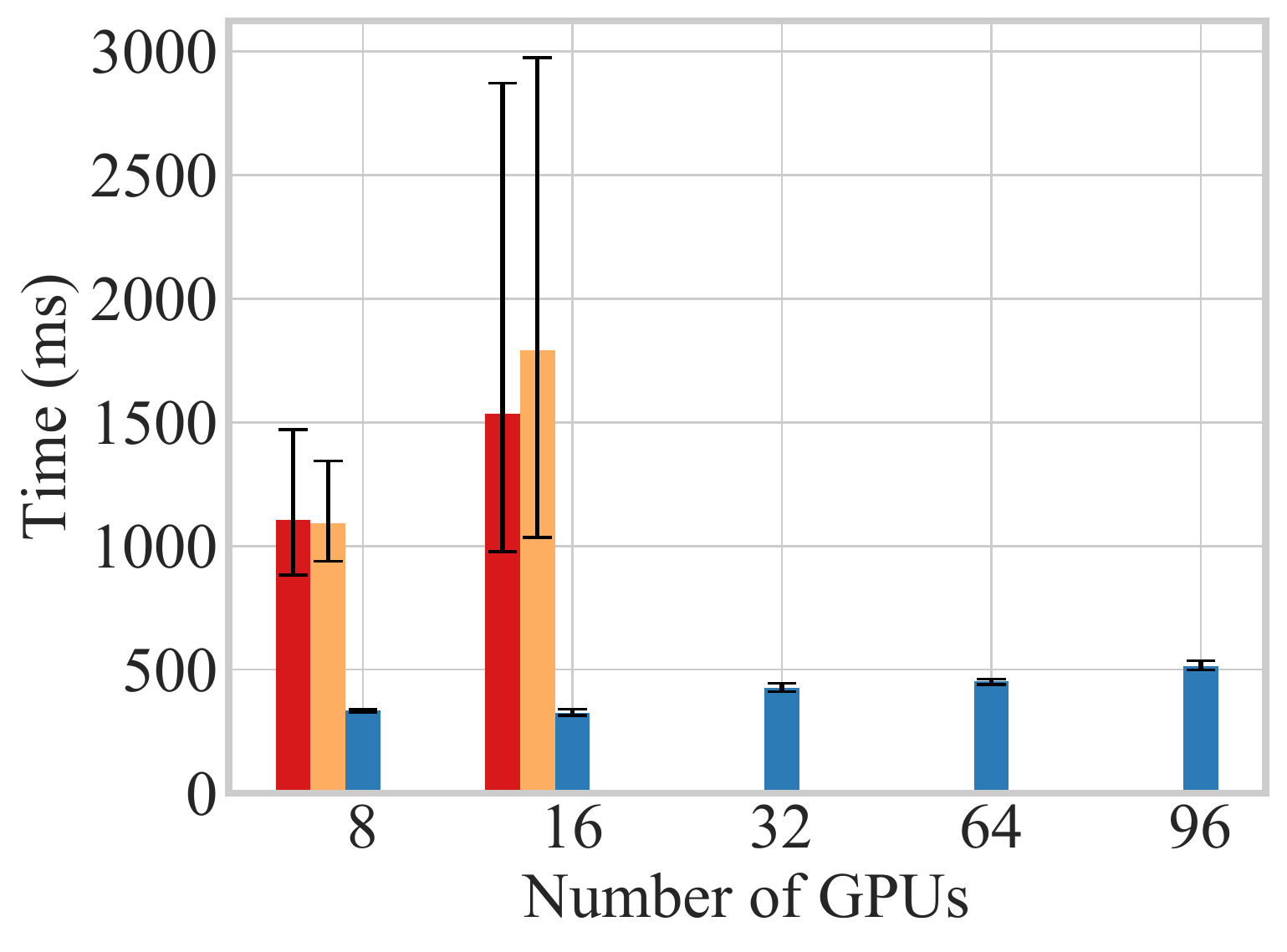}
    \caption{BERT: Batch Size 12}
    \label{fig:topk_bert_bsize12}
    \end{subfigure}
    \end{center}
    \vspace{-0.1in}
    \caption{\small{\textbf{Scalability of \mstopk{} with overlap:} Comparing the time taken for gradient computation and aggregation for \mstopk\ (with overlap) with syncSGD. For BERT and ResNet-101 we could not scale \mstopk\ beyond 16 and 32 GPUs respectively, due to memory requirement of \mstopk\ increasing linearly with number of machines and running out of available memory.}}
    \label{fig:topk_all_models_new_api}
    \vspace{-0.0in}
    \end{figure*}

\begin{figure*}[t]
    \begin{center}
    \includegraphics[width=0.3\textwidth]{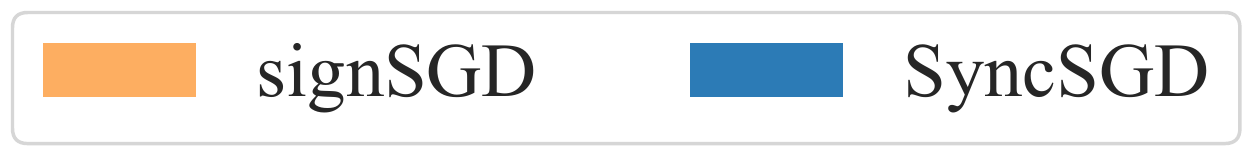}\\
    \vspace{-5pt}
    \begin{subfigure}[b]{0.3\textwidth}
    \includegraphics[width=\textwidth]{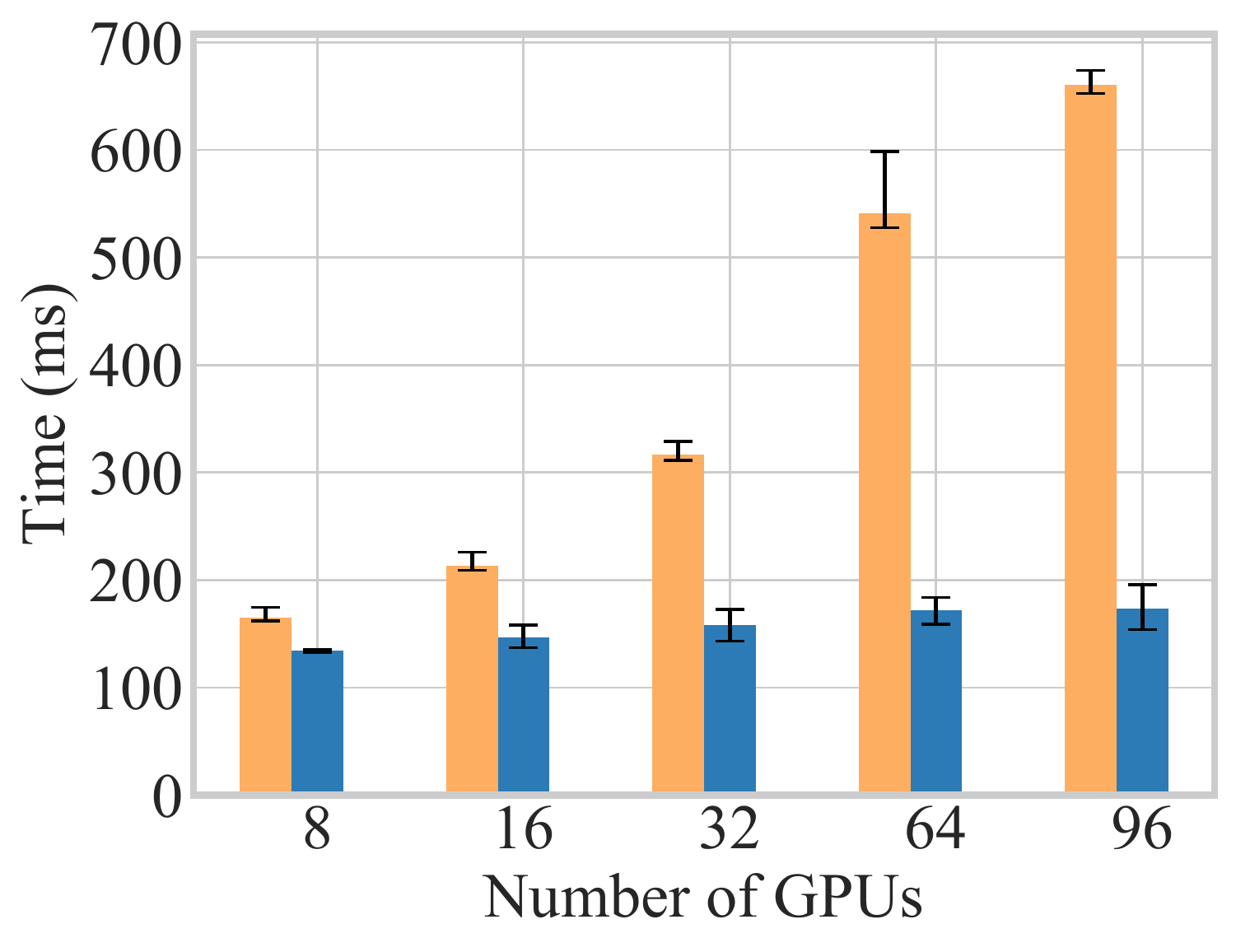}
    \caption{ResNet50: Batch Size 64}
     \label{fig:signsgd_resnet50_bsize64}
    \end{subfigure}
    \begin{subfigure}[b]{0.3\textwidth}
    \includegraphics[width=\textwidth]{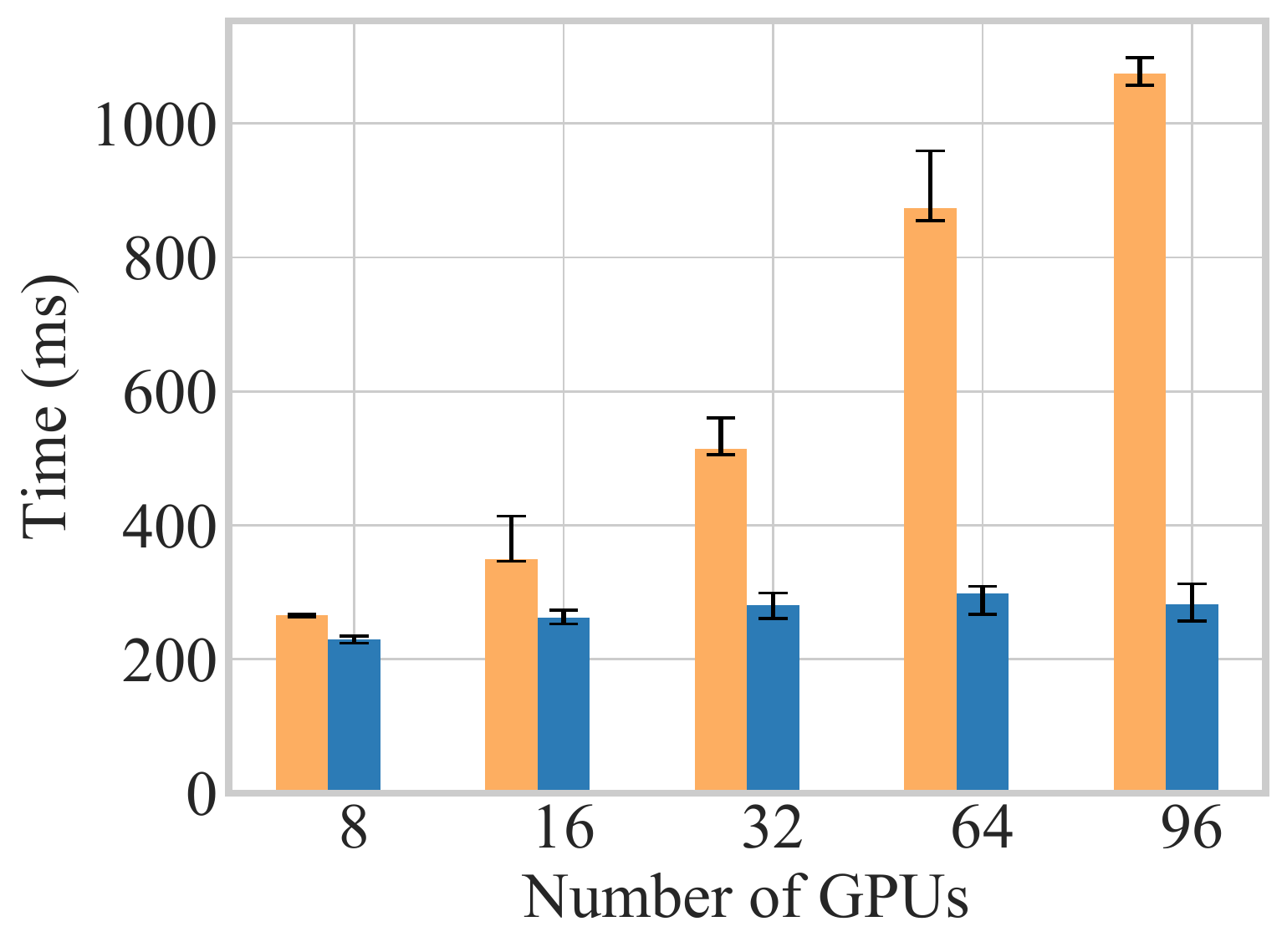}
    \caption{ResNet 101: Batch Size 64}
    \label{fig:signsgd_resnet101_bsize64}
    \end{subfigure}
    \begin{subfigure}[b]{0.3\textwidth}
    \includegraphics[width=\textwidth]{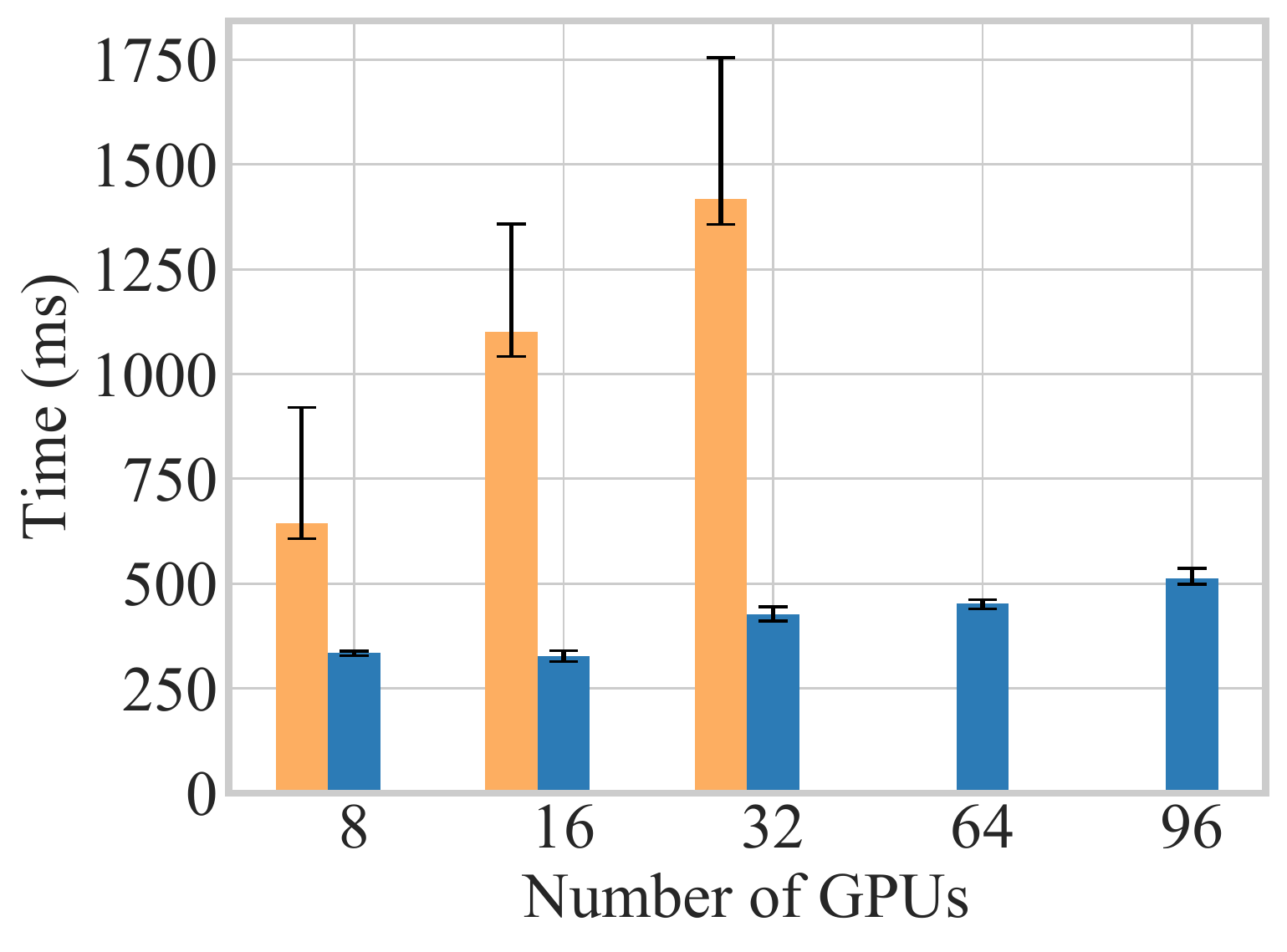}
    \caption{BERT: Batch Size 12}
    \label{fig:signsgd_bert_bsize12}
    \end{subfigure}
    \end{center}
    \vspace{-0.1in}
    \caption{\small{\textbf{Scalability of signSGD with overlap:} We compare the time taken for gradient computation and aggregation for signSGD with syncSGD. For BERT we could not scale signSGD beyond 32 GPUs, because the memory requirement of signSGD increase linearly with number of machines and for BERT we ran out available memory. }}
    \vspace{-0.0in}
    \label{fig:signsgd_allmodels_new_api}
    \end{figure*}

In this section, we include additional results in which we consider overlapping gradient compression with gradient computation. \powersgd{} was recently implemented with overlap in PyTorch v1.8~\cite{pytorchpsgd}. For integrating \signsgd{} and \mstopk{} we used the recently introduced DDP Communication hook~\cite{pytorchcommhooks} interface. The DDP Communication hook interface was recently added in PyTorch v1.8. Comparing Figure~\ref{fig:psgd_all_models} with Figure~\ref{fig:psgd_all_models_new_api}, Figure~\ref{fig:mstopk_all_models} with Figure~\ref{fig:topk_all_models_new_api} and Figure~\ref{fig:signsgd_allmodels_serial} with Figure~\ref{fig:signsgd_allmodels_new_api}  we observe that overlapping gradient compression with gradient computation is slower compared to performing gradient compression post gradient computation. Therefore to consider the best case for gradient compression, in the main paper we only consider gradient compression being performed post backward pass. As discussed in the main paper this phenomenon can be primarily attributed to both compression and backward pass being compute intensive and thus competing for the same resources on the GPU, leading to an overall slowdown.

\section{Performance model for gradient compression}
\label{app:grad_compression_perfomance_model}
In Section~\ref{sec:performance_model} we described our performance model for syncSGD with system optimizations. Here we describe our performance model for gradient compression.

From the perspective of performance, the scalability of a compression method depends on two main factors i) can the aggregation be performed using { \it all-reduce} ii) the encode decode time for compression. ~\Cref{tab:methods} classifies a number of gradient compression methods based on compatibility with  {\it all-reduce}. Ideally for high scalability we would like the method to be both {\it all-reduce} compatible and have low encode-decode time.

In Section~\ref{sec:overlap_failure} and Appendix~\ref{app:extra_overlap_results} we have shown that the best case from the perspective of runtime will be performing gradient compression post backward pass. Based on this finding,  a generic performance model will be 
\begin{equation*}
\begin{split}
T_{obs} \approx T_{comp}+ T_{encode-decode}  + T_{comm}(\hat{b},p, BW)
\end{split}
\end{equation*}
where $T_{comp}$ is the time required for gradient computation, $T_{encode-decode}$ is the overhead of compressing and decompressing the gradients. Since after compression gradients are extremely small they are then sent in a single bucket, $T_{comm}(\hat{b},p,BW)$ is the time required to communicate compressed gradients of size $\hat{b}$,  across $p$ GPUs at $BW$ bandwidth.
We now derive specific performance models for studying gradient compression schemes from the generic model stated above. 
\paragraph{PowerSGD.} \powersgd{} requires sending two low rank  matrices, $P$ and $Q$. But $T_{encode-decode}$ as stated in Table~\ref{tab:encode_decode_time} has high overhead. The performance model becomes-
\begin{equation*}
\begin{split}
T_{obs} \approx T_{comp} + T_{encode-decode} + T_{comm}(P,p, BW) + T_{comm}(Q,p, BW)
\end{split}
\end{equation*}
Where $p$ is the number of GPUs, and $T_{comm}$ is calculated using Equation~\ref{eq:ring_reduce}.  

\paragraph{\mstopk.}
For \mstopk\, the output of compression is the \topk\%\ gradient values ($\hat{g}$) and their corresponding indices ($\hat{i}$). Further, \topk\ operator is not compatible with {\it all-reduce}, therefore we need to use {\it all-gather} collective, thus $T_{comm}$ will be calculated from  $$T_{comm}(\hat{g},p,BW) = \frac{\hat{g} \times (p-1)}{BW}$$
where $\hat{g}$ is the gradient size, $p$ is the number of GPUs. A similar calculation applies to $\hat{i}$ the indices. Overall the performance model becomes.
\begin{equation*}
\begin{split}
T_{obs} \approx T_{comp} + T_{encode-decode}+ T_{comm}(\hat{g},p, BW) + T_{comm}(\hat{i}, p, BW)
\end{split}
\end{equation*}

\paragraph{\signsgd.}
SignSGD, only sends 1bit for each 32bit leading to around $32\times$ gradient compression. However SignSGD is not compatible with all-reduce leading to a performance model as follows:
\begin{equation*}
\begin{split}
T_{obs} \approx T_{comp} + T_{encode-decode} +T_{comm}(\hat{g},p, BW) 
\end{split}
\end{equation*}
where $T_{comm}(\hat{g}, p, BW) = \frac{\hat{g}\times (p-1)}{BW}$ and $\hat{g} = \frac{g}{32}$. For \signsgd{} we only consider {\it all-gather} collective, \ie each node receives the encoded gradients from all other nodes. Prior work~\cite{vogels2019powersgd} has observed that using {\it all-gather}  collective performs better than just using {\it gather} collective, due to lack of support in NCCL library.

\begin{figure*}[t]
    \begin{center}
    \includegraphics[width=0.5\textwidth]{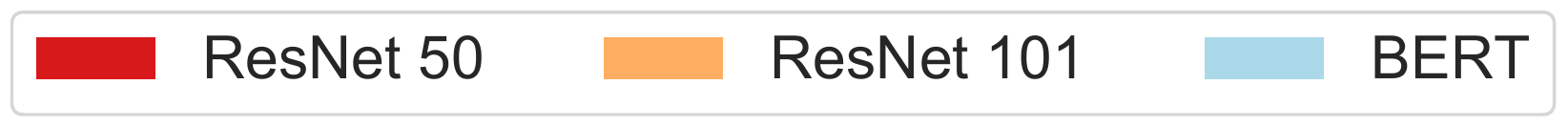}\\
    \vspace{-2pt}
    \begin{subfigure}[b]{0.3\textwidth}
    \includegraphics[width=\textwidth]{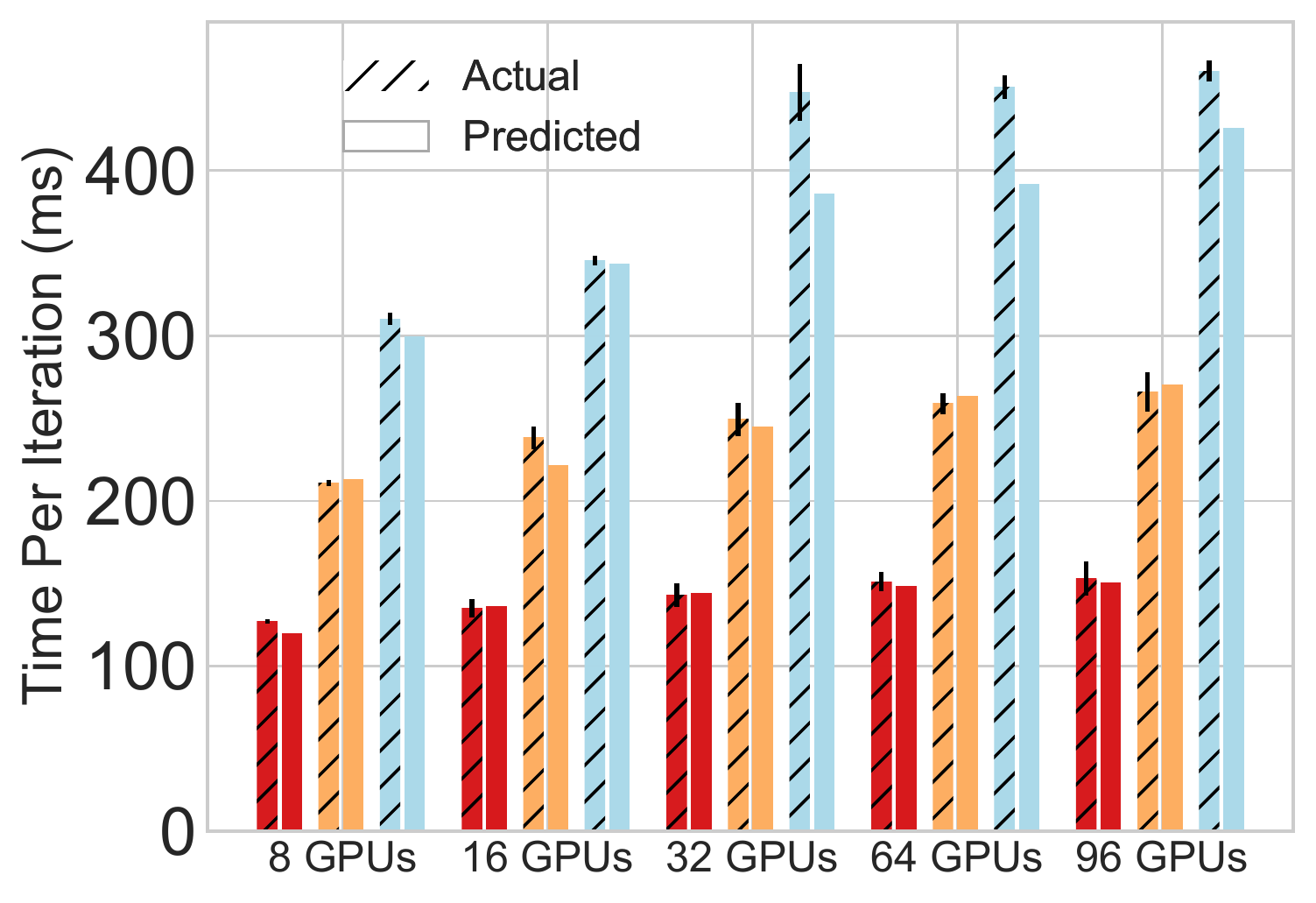}
    \caption{syncSGD}
    \label{fig:perf_sync_verify}
    \end{subfigure}
    \begin{subfigure}[b]{0.3\textwidth}
    \includegraphics[width=\textwidth]{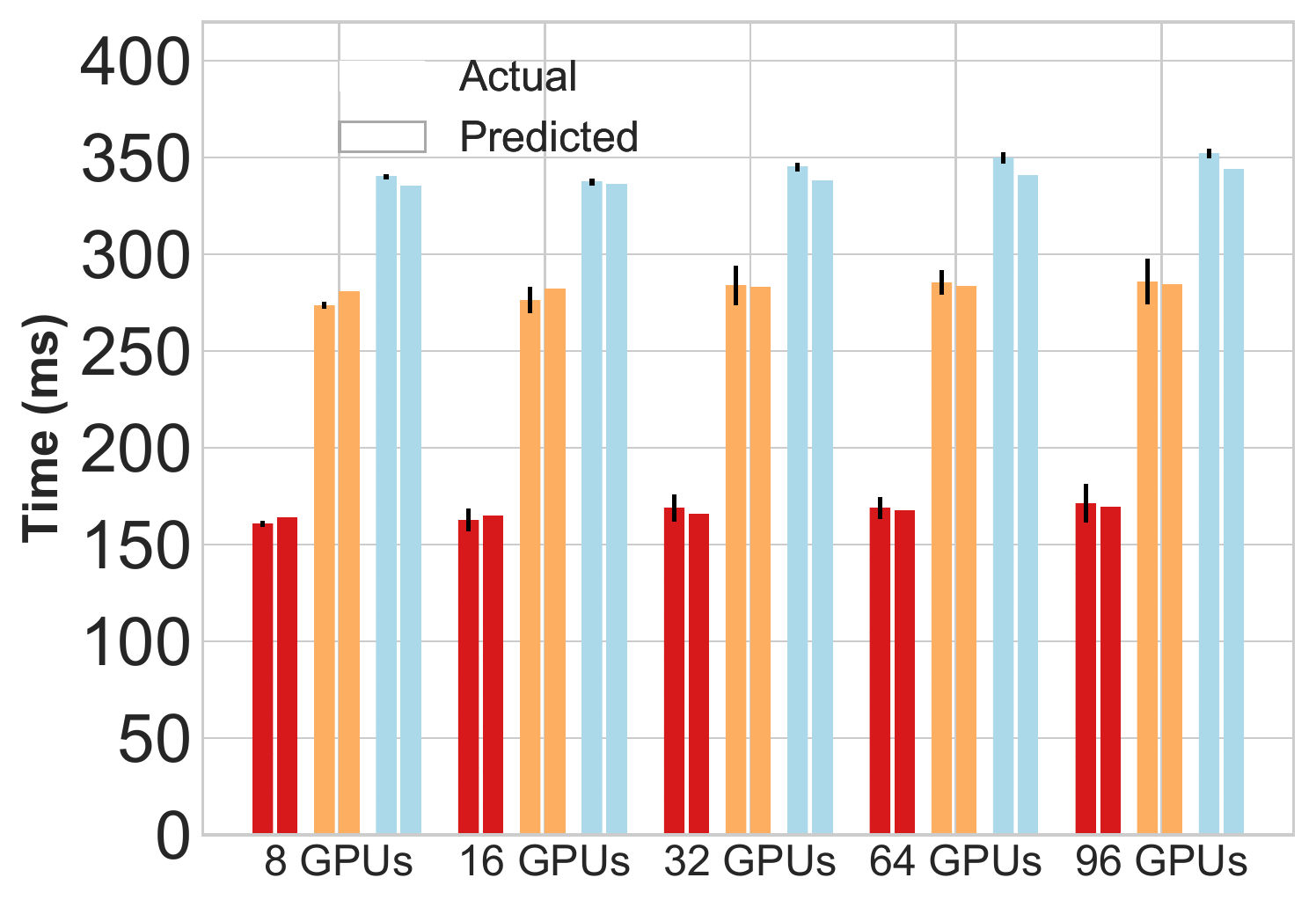}
    \caption{\powersgd{}}
     \label{fig:perf_compare_resnet50_ssgd}
    \end{subfigure}
    \begin{subfigure}[b]{0.3\textwidth}
    \includegraphics[width=\textwidth]{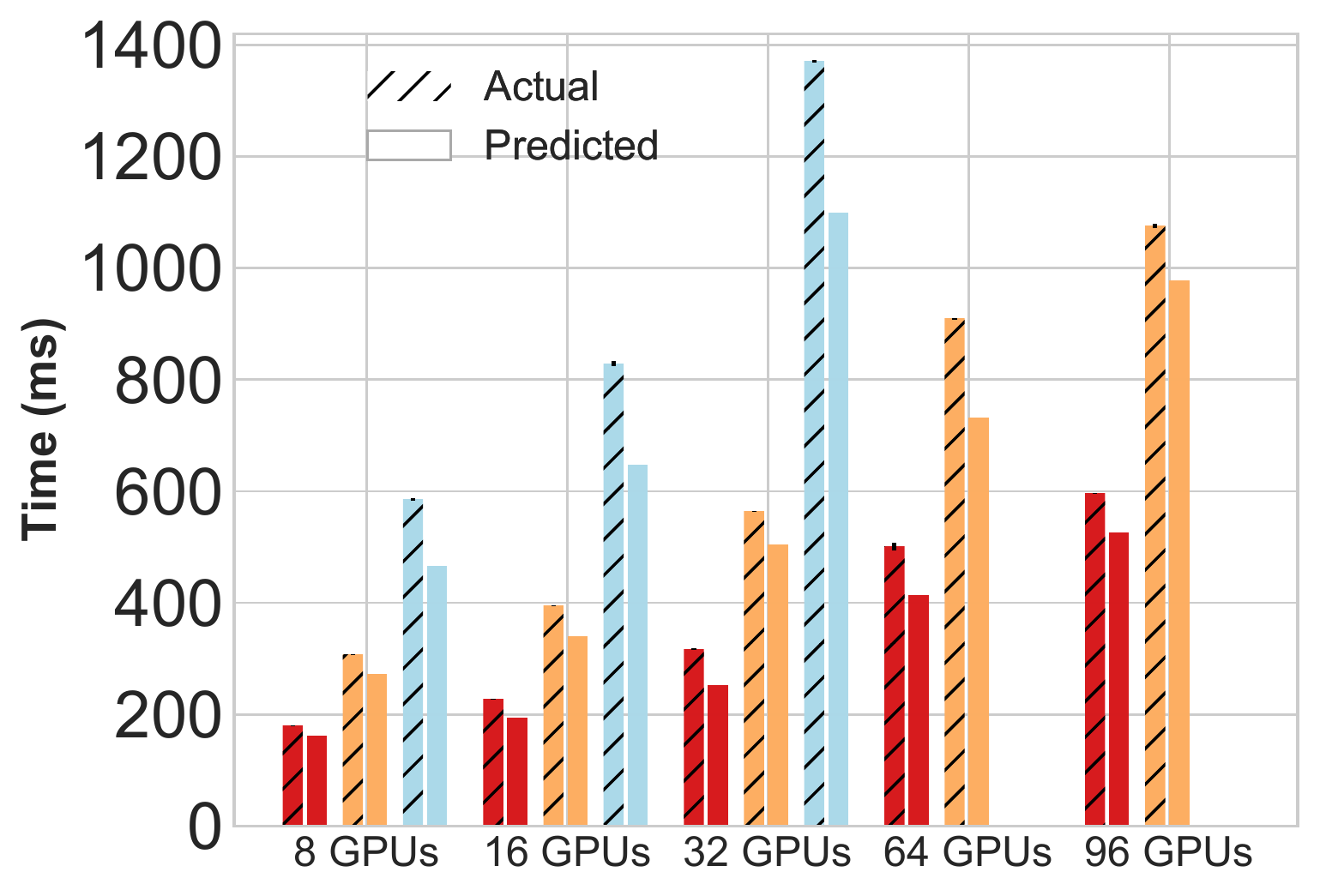}
    \caption{\signsgd{}}
    \label{fig:perf_compare_vgg19_ssgd}
    \end{subfigure}
    \end{center}
    \vspace{-14pt}
    \caption{ \small{\textbf{Evaluating our performance model on actual hardware: }We evaluate our performance model on AWS on p3.8xlarge instance. We observe that our performance model quite closely tracks the actual performance of both syncSGD implementation of PyTorch as well as performance of gradient compression methods. Before all experiments we calculated the available pairwise bandwidth using iperf3\cite{iperf3}, and calculate the latency term by performing all reduce based on the vector of size equivalent to number of machines. For BERT we could not scale signSGD beyond 32 GPUs, because signSGD's memory requirement increase linearly with number of machines and for BERT we ran out available memory. }}
    \label{fig:eval_perf_model_bar}
    \vspace{-0.1in}
\end{figure*}

\section{Verifying the performance model}
\label{app:verification_pmodel}
In this section we describe how we verify our performance model and calculate the values required for using our analytical performance model.

In case of syncSGD the backward pass and gradient synchronization are overlapped, therefore it is not easy to segregate the time spent in communication and time spent in computation. First we calculate just the time taken for backward pass on a single machine this forms $T_{comp}$ in the performance model. To calculate $\gamma$, we run distributed training but with Nsight Systems profiling switched on. From Nsight systems we track kernels launched during backward pass and find how long does it takes for the compute phase of backward pass. The ratio between the two allows us to calculate $\gamma$. For all our experiments we disable NCCL auto tuning and forced it to use ring algorithm by setting the \textsc{NCCL\_TREE\_THRESHOLD=0}. To calculate $T_{encode-decode}$ we calculate the time required for compression and decompression for each iteration and plug it in the model. Before each run we calculate available bandwidth between each pair of instances using iperf3~\cite{iperf3} and take the minimum of these values as $BW$. For calculating $\alpha$ we perform ring-reduce on a small tensor and divide the obtained value by $(p-1)$ where $p$ is the number of GPUs. 

Figure~\ref{fig:eval_perf_model_bar} shows that our model closely tracks the experiments performed on real hardware. In case of syncSGD and \powersgd{} (schemes using all-reduce) we observe the maximum deviation from actual experiments to be around 9.1\%.  In case of \signsgd{} the maximum deviation observed is 19.1\%,
the reason for high difference for \signsgd{} is that {\it all-gather} collective has an all to all pattern which causes degraded network performance due to widely reported issues of incast~\cite{chen2009understanding, alizadeh2010data}. In future a utility which can simulate the traffic pattern of {\it all-gather} collective and provide us more accurate measurements of the effective bandwidth available during all-to-all communications can be helpful in providing better estimates of per iteration time.  

\paragraph{Using the Performance Model.} To use the performance model, similar to verification we calculate $T_{comp}$, the time for backward pass on a single machine for a given batch size and model. It depends on hardware, computation requirements of the model and the batch size used for training. 
For gradient compression methods we also calculate $T_{encode-decode}$ for \signsgd{}, \topk{} and \powersgd{}. We only include the computation time and disregard the time for extracting gradients, or copying back the decompressed gradients to the model. As these timings can be improved with tighter integration with the training frameworks. For this calculation we run each experiment 60 times and discard the first 10, we assign the mean of remaining 50 as  $T_{encode-decode}$. Table~\ref{tab:encode_decode_time} shows the times for $T_{comp}$ and $T_{encode-decode}$ for ResNet-50 when using V100 GPU on AWS. Thus without running large scale experiments, practitioners and researchers can utilize our performance model to predict speedups when performing distributed training with and without using gradient compression.

\section{What-If Analysis}
\label{app:analysis_pmodel}
\begin{figure*}[t]
    \begin{center}
    \begin{subfigure}[b]{0.32\textwidth}
    \includegraphics[width=\textwidth]{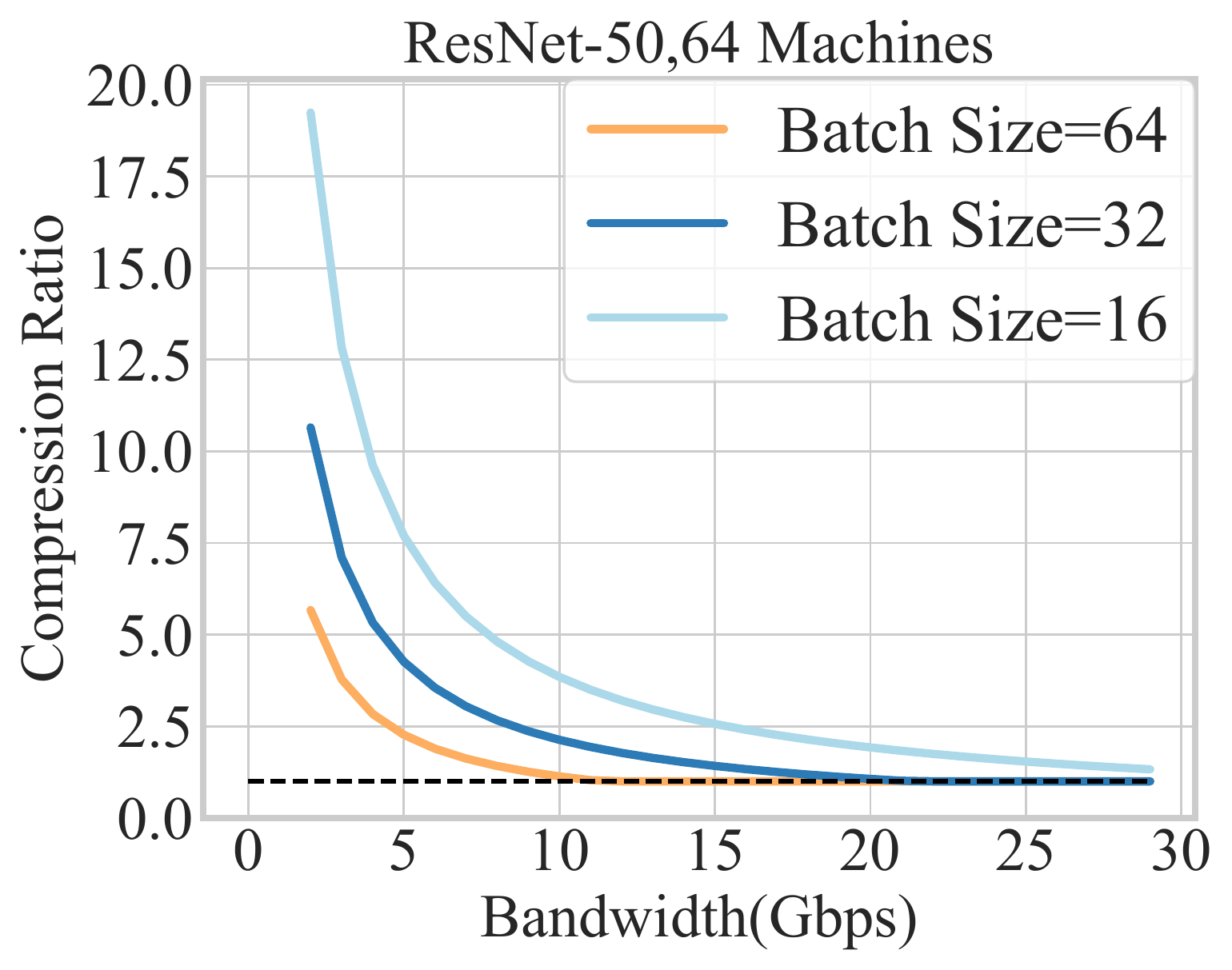}
    \caption{ResNet50: 64 GPUs}
     \label{fig:ideal_compression_resnet50_app}
    \end{subfigure}
    \begin{subfigure}[b]{0.32\textwidth}
    \includegraphics[width=\textwidth]{figs/resnet101_ideal_compression.pdf}
    \caption{ResNet101: 64 GPUs}
    \label{fig:ideal_compression_resnet101_app}
    \end{subfigure}
    \begin{subfigure}[b]{0.32\textwidth}
    \includegraphics[width=\textwidth]{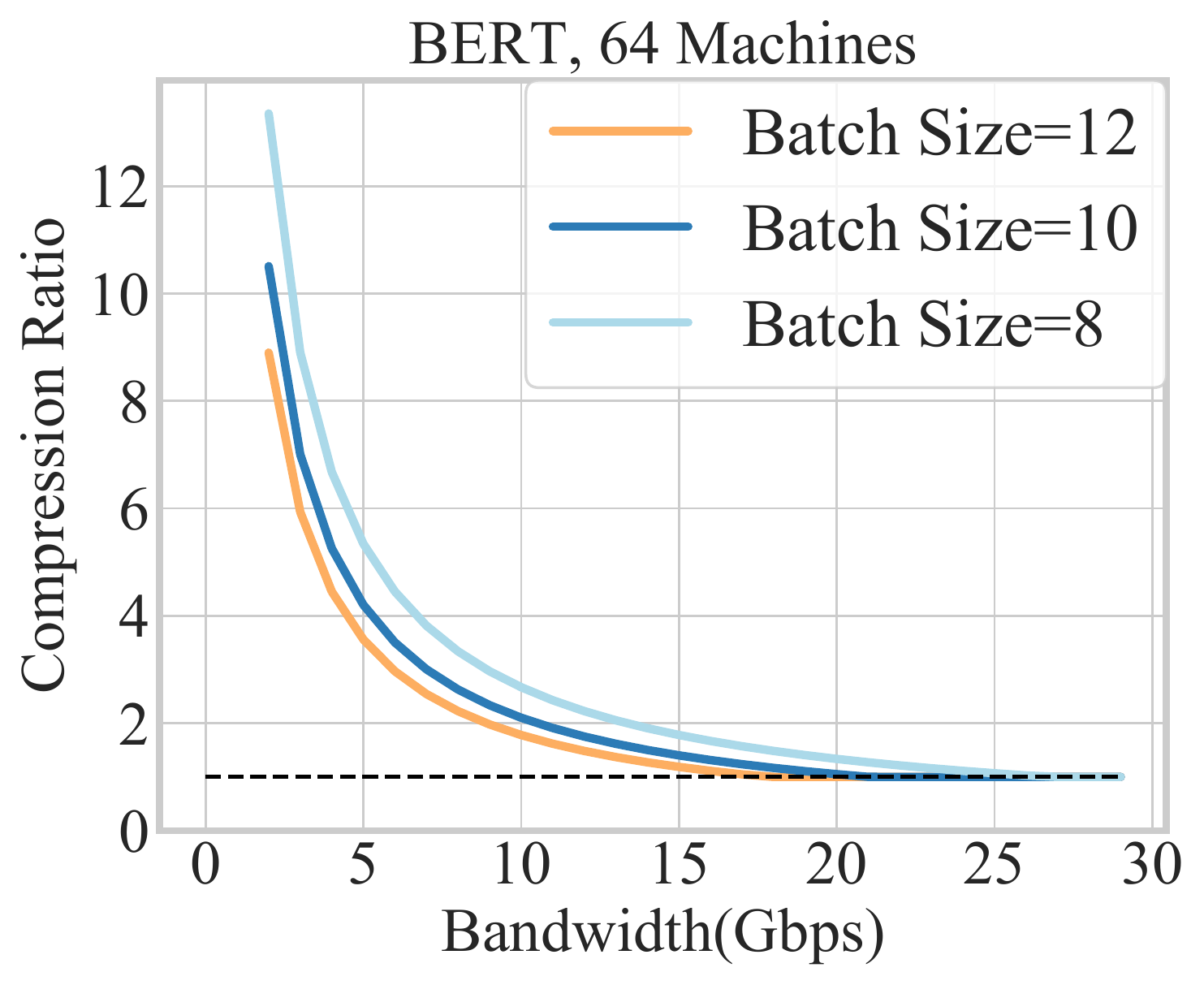}
    \caption{BERT: 64 GPUs}
    \label{fig:ideal_compression_BERT_app}
    \end{subfigure}
    \end{center}
    \vspace{-0.1in}
    \caption{\small{\textbf{Required gradient compression for near optimal speedups (simulated):} We observe that the required gradient compression for near optimal scaling is quite small. At 10 Gbps even for quite small batch sizes we need less than $4\times$ gradient compression, which is quite small compared to what popular gradient compression methods. }}
    \label{fig:ideal_compression}
    \vspace{-0.2in}
\end{figure*}

Our performance model also allows us to consider several what-if scenarios. To understand how and where gradient compression methods will be useful, we can vary several factors like compute availability, encode-decode time, network bandwidth etc. Based on our results in Section~\ref{sec:powersgd_speedup} which show that \powersgd{} Rank-4 is the most scalable compression scheme, we use PowerSGD with Rank-4 as the baseline for these what-if analyses. 

\paragraph{Required Compression for linear scaling.} Existing gradient compression methods provide massive amount of compression which often leads to poor accuracy. Using our performance model we study the amount of gradient compression required for linear scaling. Figure~\ref{fig:ideal_compression} shows that in most common models at 10 Gbps we do not need compression greater than $4\times$. This shows that focus of gradient compression should be to reduce the overheads of compression rather than providing very high compression rates.

\paragraph{Effect of Network Bandwidth}

\begin{figure*}[t]
    \begin{center}
    \includegraphics[width=\textwidth]{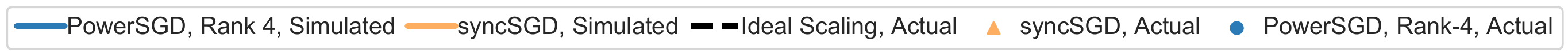}\\
    \vspace{-2pt}
    \begin{subfigure}[b]{0.32\textwidth}
    \includegraphics[width=\textwidth]{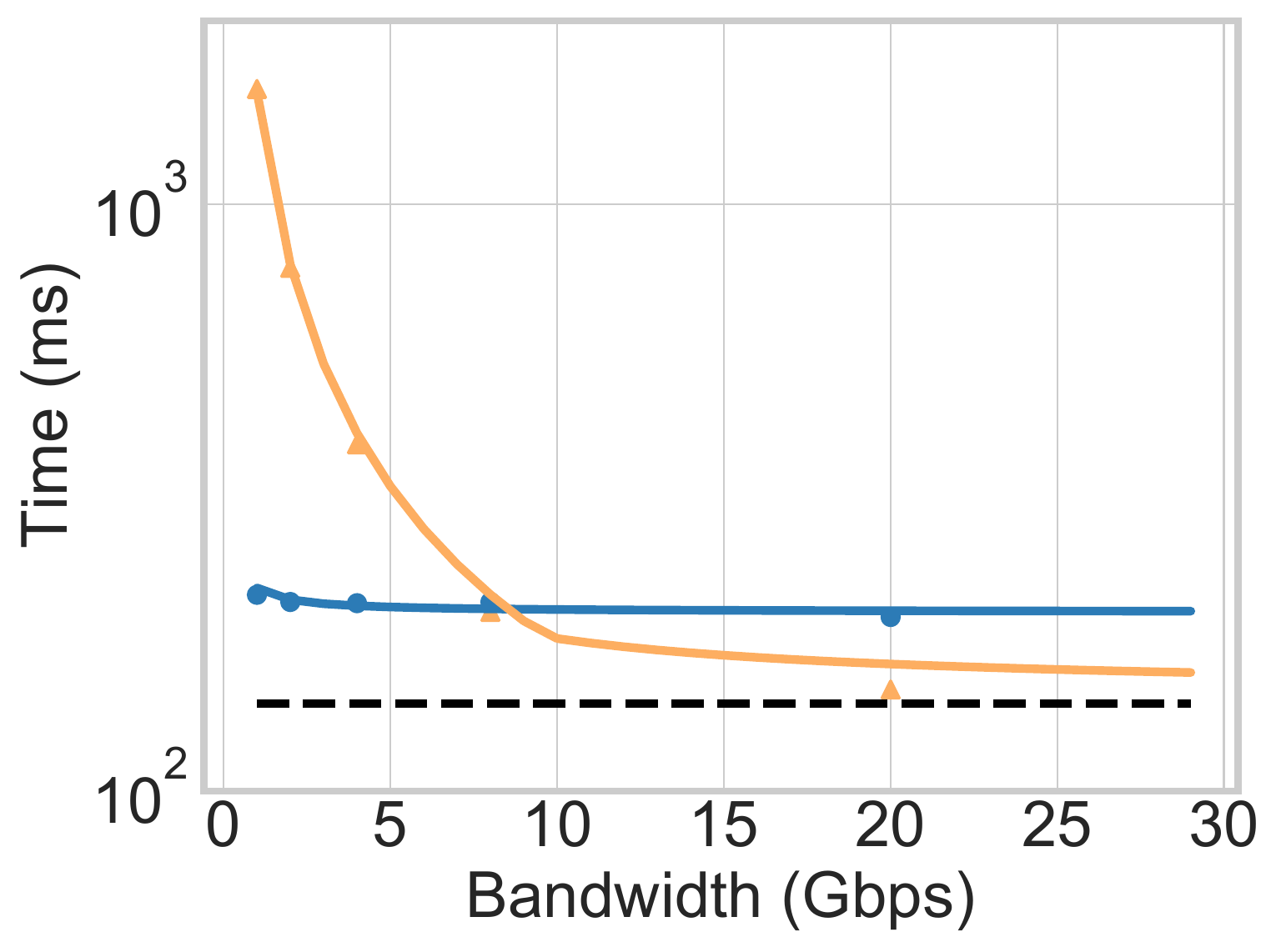}
    \caption{ResNet50: Batch Size 64}
     \label{fig:network_resnet_bsize64}
    \end{subfigure}
    \begin{subfigure}[b]{0.32\textwidth}
    \includegraphics[width=\textwidth]{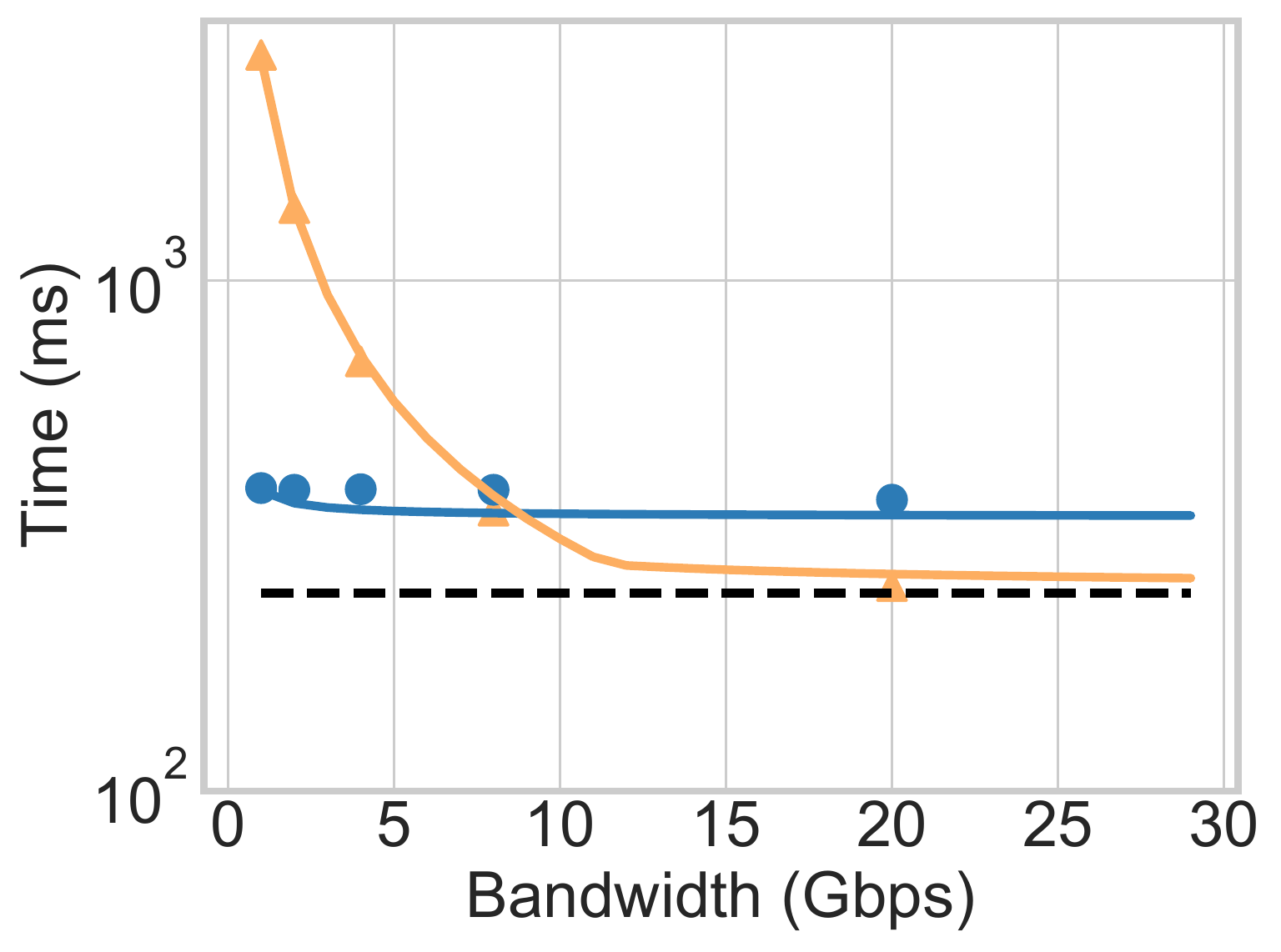}
    \caption{ResNet101: Batch Size 64}
    \label{fig:network_resnet101_bsize64}
    \end{subfigure}
    \begin{subfigure}[b]{0.32\textwidth}
    \includegraphics[width=\textwidth]{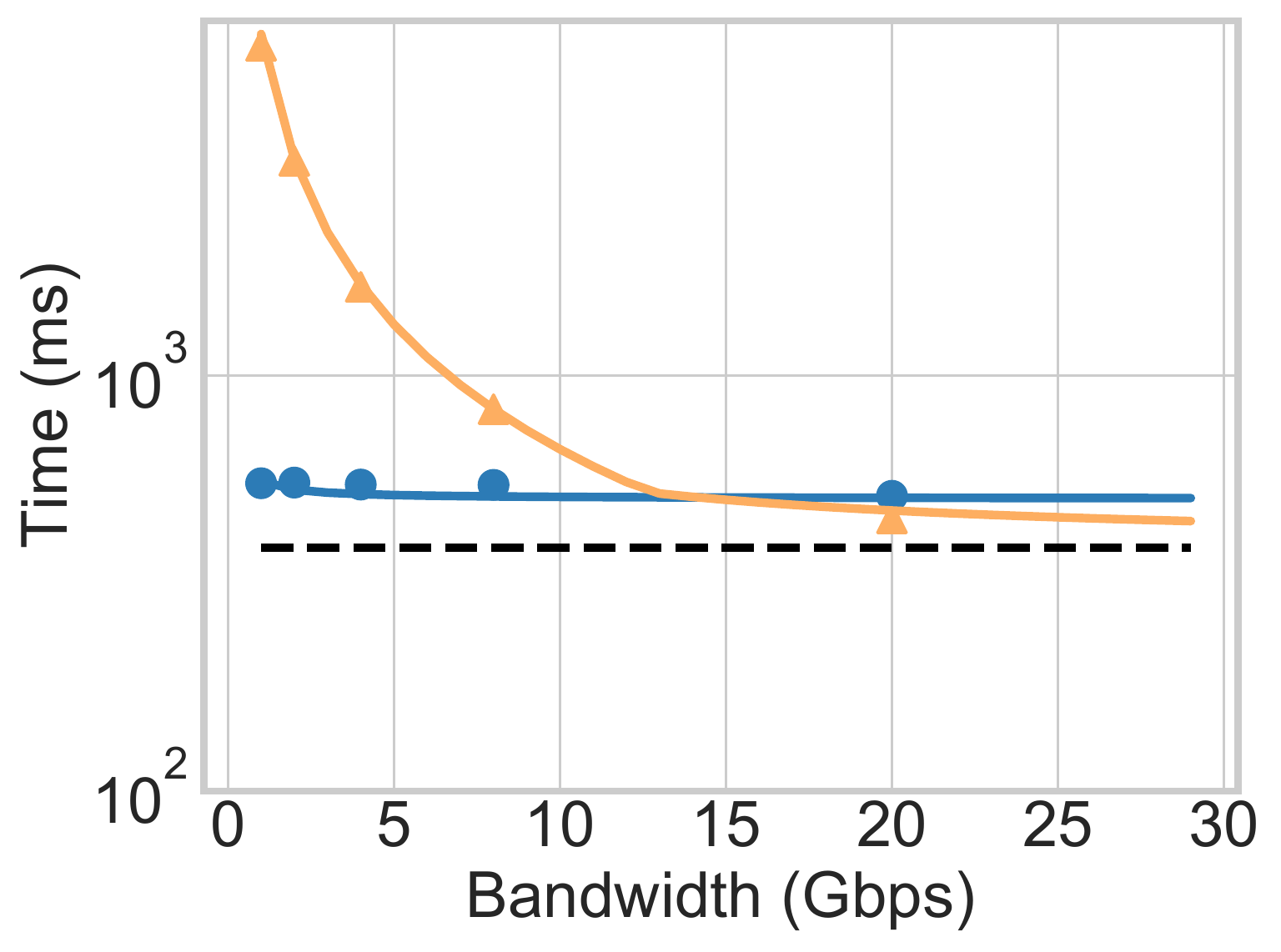}
    \caption{BERT: Batch Size 12}
    \label{fig:network_bert_bsize12}
    \end{subfigure}
    \end{center}
    \vspace{-0.1in}
    \caption{\small{\textbf{Evaluating effect of network bandwidth on training (simulated):}  We vary bandwidth availability and analyse the performance of synchronous SGD vs PowerSGD Rank 4. We observe that as bandwidth increase significantly it helps synchronous SGD since it has a larger communication overhead. Moreover we observe the PowerSGD provides massive gains at extremely low bandwidth (1Gbps) but as bandwidth scales we see PowerSGD gets bounded by compute availability. The markers are values from actual experiments, this also shows how close our performance model is to actual measurement.  }}
     \label{fig:network_change}
    \vspace{-0.1in}
\end{figure*}
In Figure~\ref{fig:network_change} we vary network bandwidth available from 1Gbps to 30Gbps and see how this changes the speedup offered by PowerSGD.
We see that, for example, in the case of Resnet-50, PowerSGD offers considerable speedup at low network bandwidths (1-7 Gbps) but becomes slower than synchronous SGD when bandwidth available becomes $>9Gbps$.
This is due to the fact that syncSGD benefits more from availability of higher bandwidth since it communicates significantly more while PowerSGD is still limited by extra time spent in the encode-decode step. For BERT which is a communication heavy network, PowerSGD becomes slower than syncSGD at around 15Gbps. In Figure~\ref{fig:network_change} the markers represent values from actual experiments. To perform these experiments we used the {\it tc} command in linux to modify the available bandwidth. For experiments with bandwidth less than 10Gbps we used {\it p3.8xlarge} instances which provide a maximum of 10Gbps bandwidth. And for 20 Gbps experiment we used {\it p3.16xlarge} instance which provides 25 Gbps bandwidth. The markers are extremely close to the values from our analytical performance model thus verifying that our performance model can indeed be useful in several settings. 
\paragraph{Effect of faster compute}
Next we analyze how the effect of gradient compression changes when newer hardware with higher compute capabilities arrive in future.

\begin{figure*}[t]
    \begin{center}
    \includegraphics[width=0.3\textwidth]{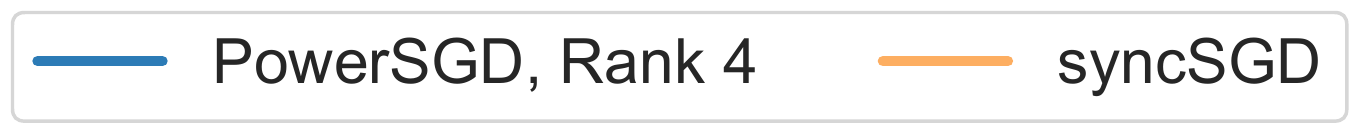}\\
    \vspace{-2pt}
    \begin{subfigure}[b]{0.32\textwidth}
        \includegraphics[width=\textwidth]{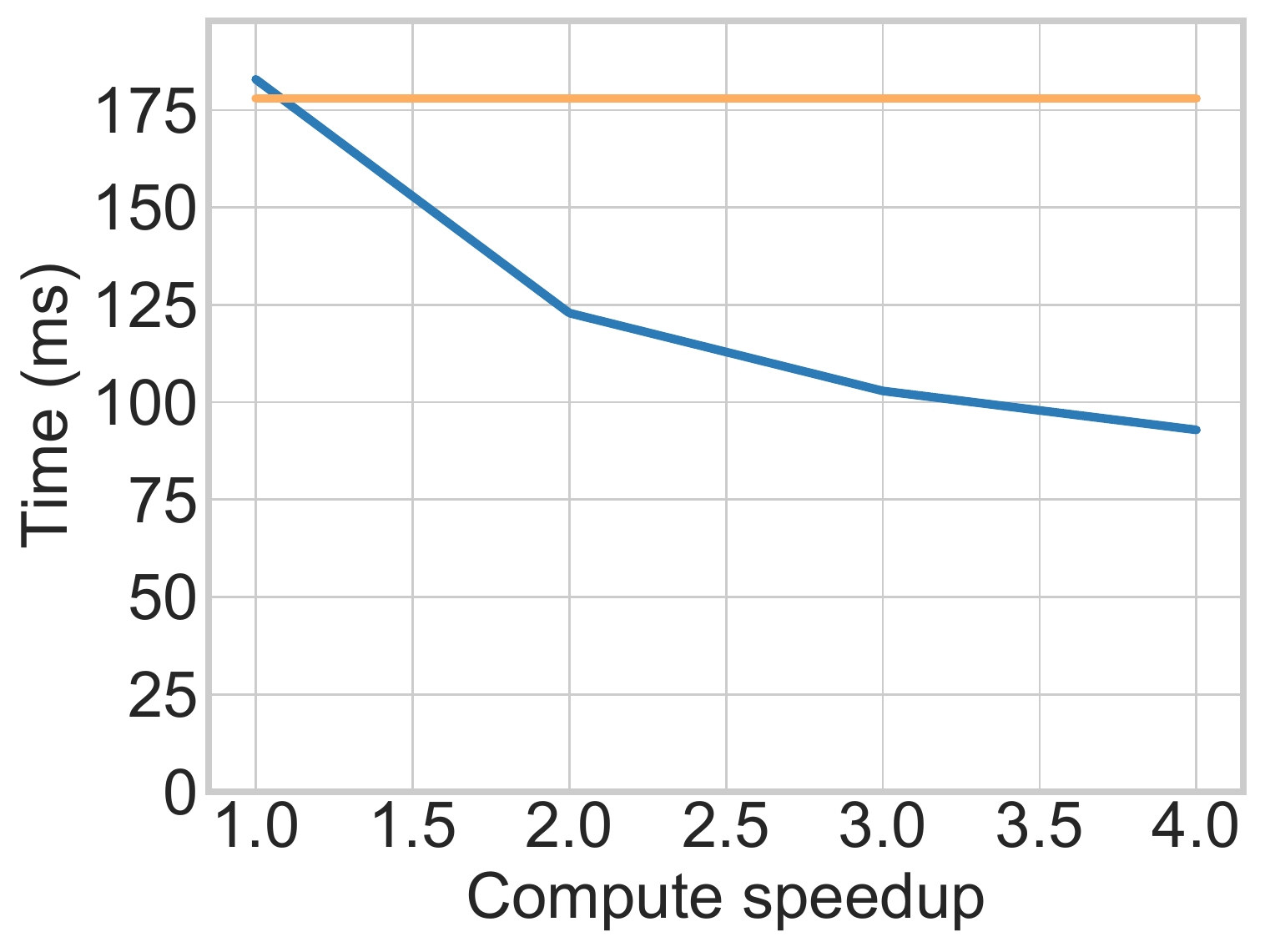}
    \caption{ResNet50: Batch Size 64}
     \label{fig:speedup_resnet50_bsize64}
    \end{subfigure}
    \begin{subfigure}[b]{0.32\textwidth}
    \includegraphics[width=\textwidth]{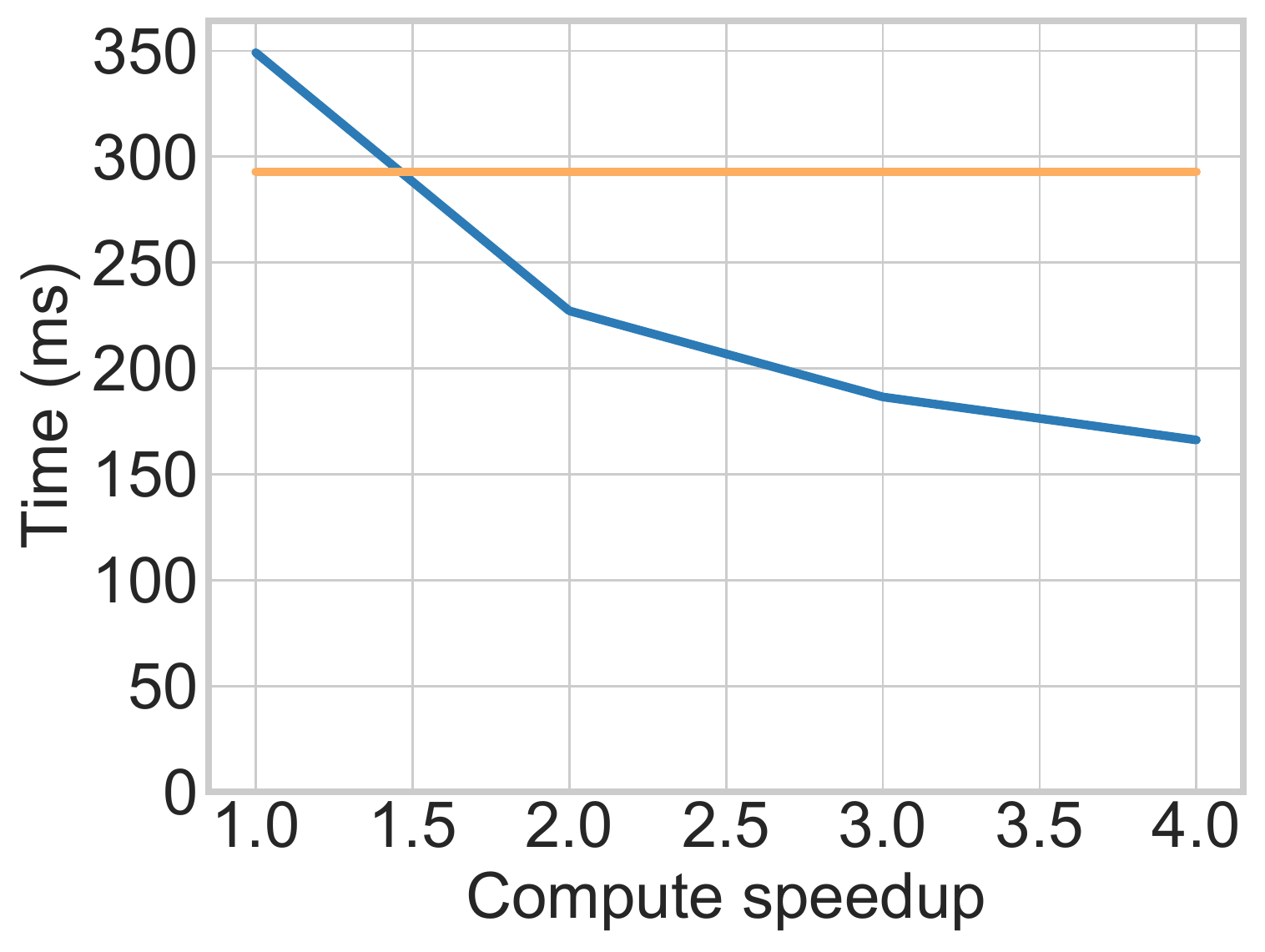}
    \caption{ResNet101: Batch Size 64}
    \label{fig:speedup_resnet101_bsize64}
    \end{subfigure}
    \begin{subfigure}[b]{0.32\textwidth}
    \includegraphics[width=\textwidth]{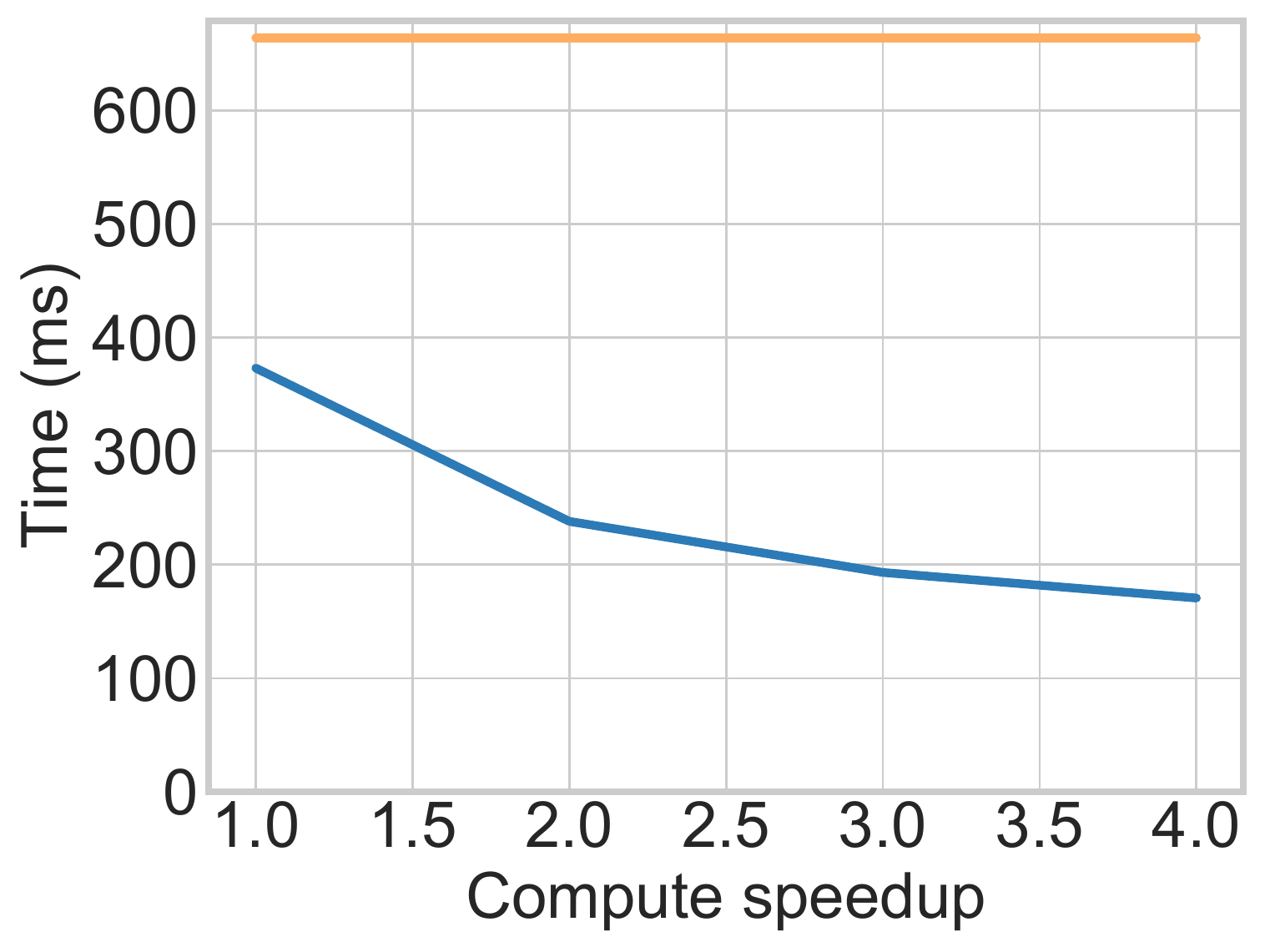}
    \caption{BERT: Batch Size 12}
    \label{fig:speedup_bert_bsize12}
    \end{subfigure}
    \end{center}
    \vspace{-0.1in}
    \caption{\small{\textbf{Evaluating effect of compute speedup on training time (simulated):}Assuming network capacity remains at 10Gigabit but compute capabilities go up, we observe in that case PowerSGD will end up providing significant benefit, meanwhile synchronous SGD will end up being communication bound and will not be able to utilize increased compute. Showing that if compute capabilities increase drastically but network bandwidth remains stagnant, gradient compression methods will become useful. }}
    \label{fig:speedup_change}
    \vspace{-0.1in}
\end{figure*}

 In Figure~\ref{fig:speedup_change}, we plot the effect of compute capabilities improving by up to $4\times$, while network bandwidth remains constant at 10 Gbps. We can see that for Resnet-50, PowerSGD with Rank-4 can provide 1.75x speedup if the compute becomes around 3.5x faster. 

 There are two reasons for this, (i) As compute gets faster, the encode-decode time also reduces by the same factor, (ii) with a faster backward pass, there is less opportunity for synchronous SGD to overlap computation with communication, making it communication bound. 

 \begin{figure*}[t]
    \begin{center}
    \begin{subfigure}[b]{0.32\textwidth}
    \includegraphics[width=0.95\textwidth]{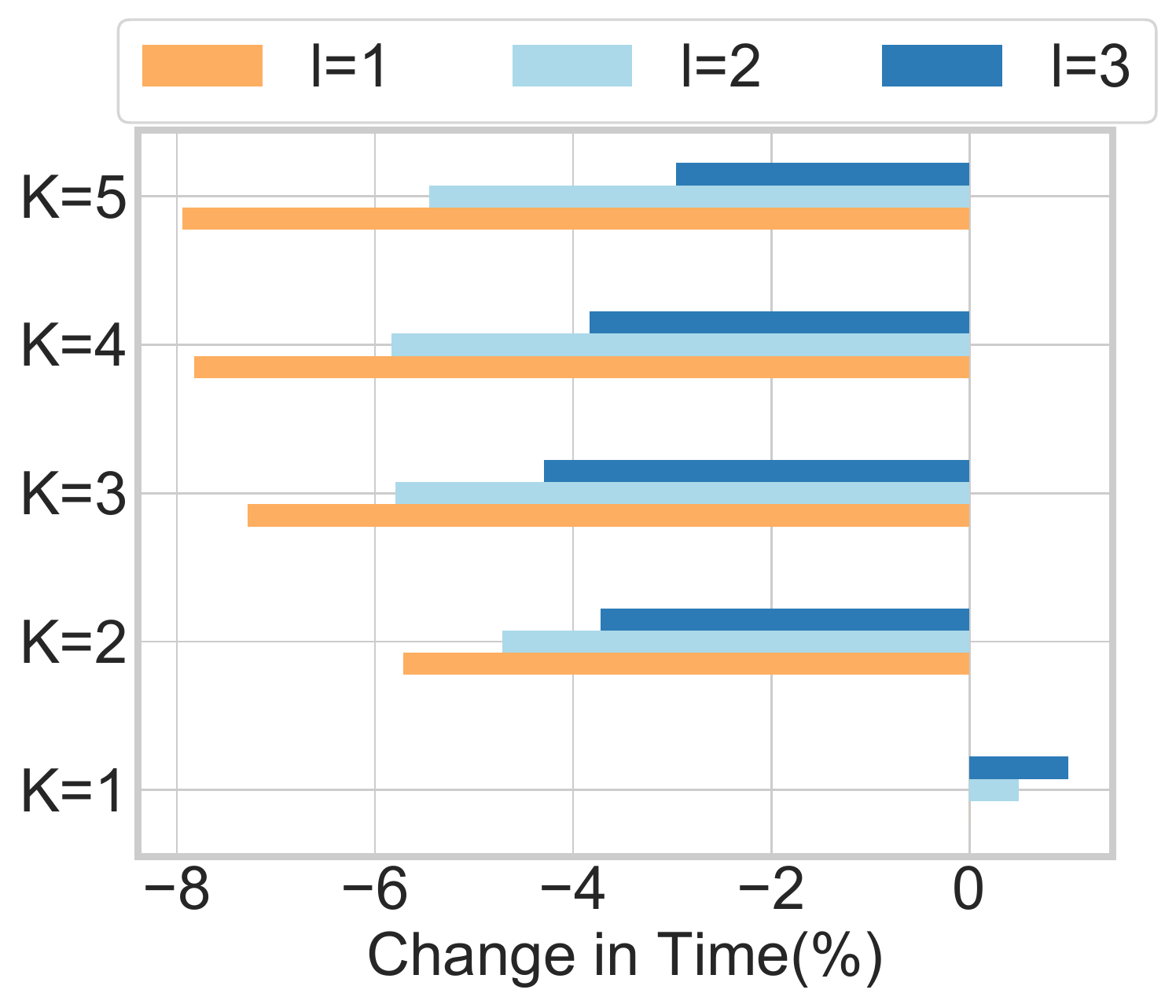}
    \caption{ResNet50: Batch Size 64}
     \label{fig:encode_compression_resnet50_bsize16}
    \end{subfigure}
    \begin{subfigure}[b]{0.32\textwidth}
    \includegraphics[width=0.95\textwidth]{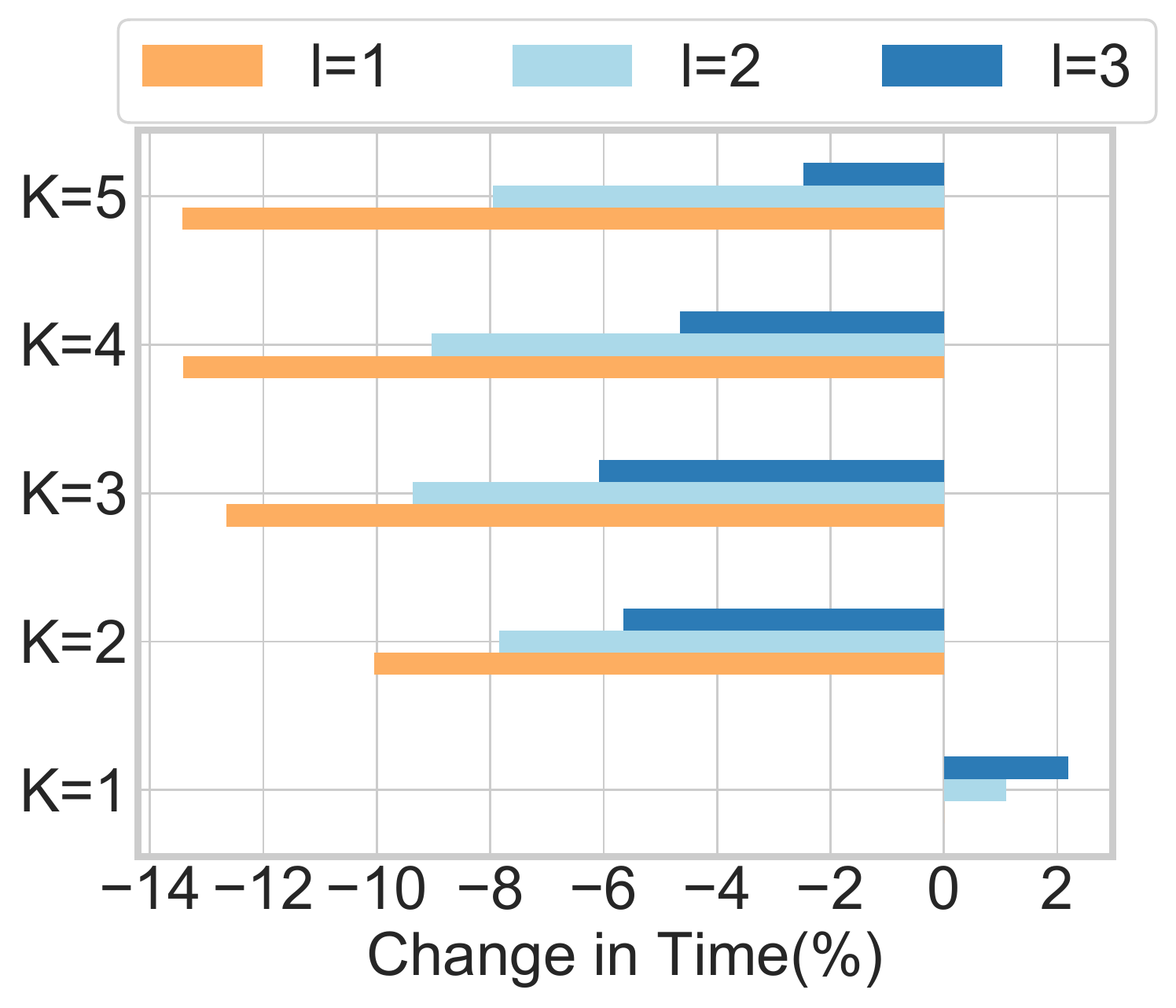}
    \caption{ResNet101: Batch Size 64}
    \label{fig:encode_compression_resnet50_bsize64}
    \end{subfigure}
    \begin{subfigure}[b]{0.32\textwidth}
    \includegraphics[width=0.95\textwidth]{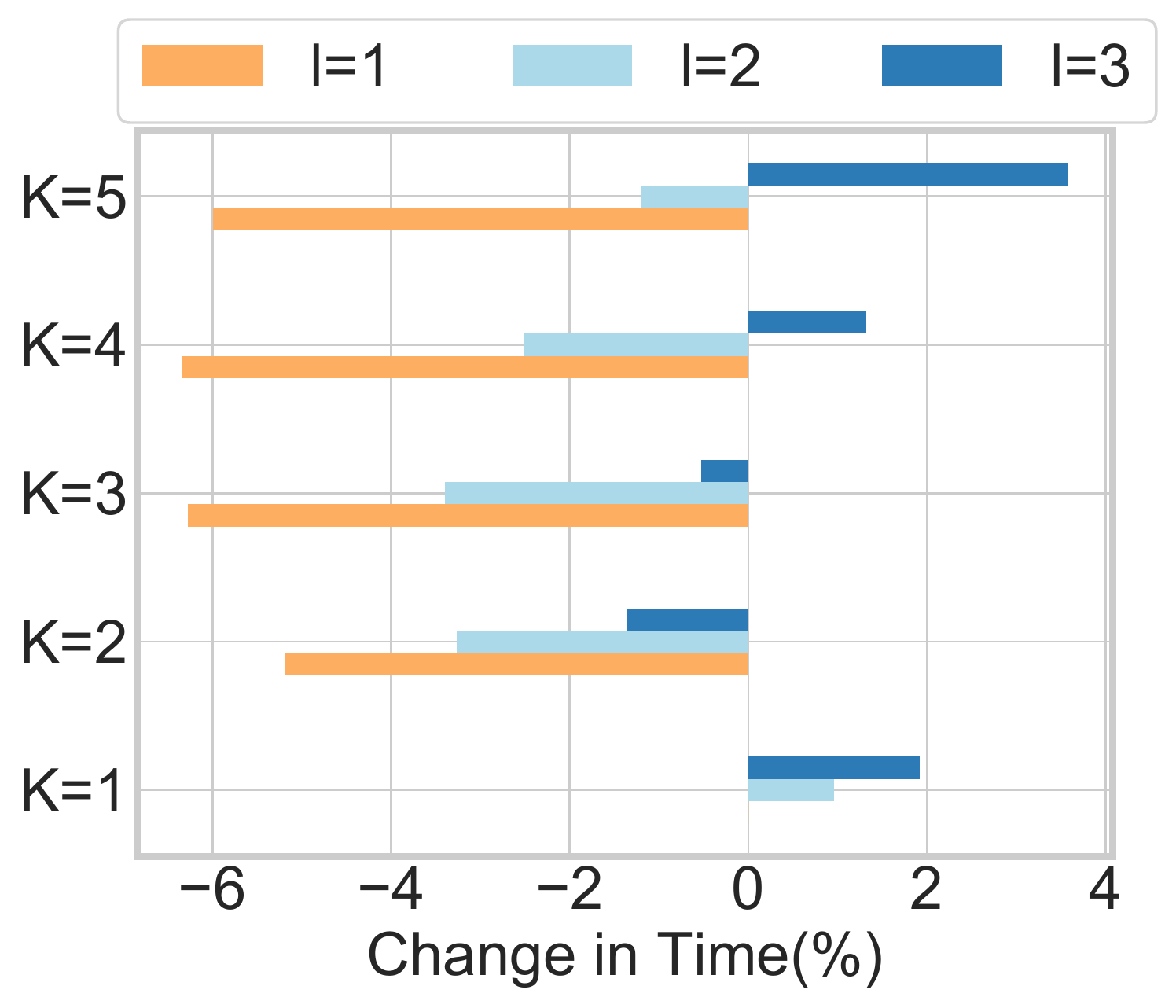}
    \caption{BERT: Batch Size 12}
    \label{fig:encode_compression_resnet50_bsize128}
    \end{subfigure}
    \end{center}
    \vspace{-0.1in}
    \caption{\small{\textbf{Varying encoding-decoding time and compression (simulated) :} We observe that reducing encode-decode time even if it leads to reduced gradient compression is very useful and can make methods like PowerSGD more viable.}}
    \label{fig:encode_decode_time}
    \vspace{-0.1in}
\end{figure*}

\paragraph{Tradeoff between encode-decode time and compression ratio }

Finally, we explore the tradeoff between the effect of reducing encode-decode time, while simultaneously decreasing the compression ratios by similar proportions. For this we consider a hypothetical gradient compression scheme in which if we decrease encode-decode time by a factor $k$ the size of gradients communicated increases by $lk$.  
For example, if say $k=2$ and $l=2$ then a 2x decrease in encode-decode time would be accompanied by a 4x increase in size of gradients. This setup is to study what would happen if we had compression schemes that offered a variety of trade-off points.
We vary $k$ from 1 to 4 in increments of 1 and try 1,2 and 3 as values of $l$. Using PowerSGD with Rank-4 as the baseline, we see in Figure~\ref{fig:encode_decode_time} that any reduction in encode-decode time even at the expense of increased communication helps.

\section{Implementation Details}
Since we are comparing time, we did our best to use the most optimized implementations. For \powersgd{} without overlap, we used the author provided code which is JIT-optimized. For \powersgd{} with overlap we used the one supported in PyTorch natively~\cite{pytorchpsgd}. For \signsgd{} we used the author provided C++ library which packs signs into bitmaps an operation which is not natively supported by PyTorch. We implemented \mstopk{} using vector instructions thus avoiding expensive for loops. For communication we only used the highly optimized NCCL communication library. For overlapping gradient compression with computation we used the communication hook~\cite{pytorchcommhooks} interface provided in PyTorch v1.8. To code is available at \url{https://github.com/uw-mad-dash/GradCompressionUtility.git}.
\end{document}